\documentclass[11pt]{article}

\title{Testing Endogeneity of Spatial Weights Matrices in Spatial Dynamic Panel Data Models}
\author{Jieun Lee\\Department of Economics, \\ University of Illinois Urbana-Champaign, \\ \textit{jieun2@illinois.edu}}

\usepackage{amsmath}
\usepackage{amsthm}
\usepackage{mathtools}
\usepackage{undertilde}
\usepackage{amssymb}
\usepackage{mathrsfs} 
\usepackage{setspace}
\usepackage{multirow}
\usepackage{caption}
\DeclareCaptionLabelFormat{AppendixTables}{A.#2}
\usepackage{subcaption}
\usepackage{graphics}
\usepackage{lscape}
\usepackage{changepage}

\newenvironment{indentblock}{\begin{adjustwidth}{\parindent}{}\hspace{-\parindent}}{\end{adjustwidth}}

\newcommand*{\field}[1]{\mathbb{#1}}%

\usepackage{graphicx}
\makeatletter
\newcommand{\distas}[1]{\mathbin{\overset{#1}{\kern\z@\sim}}}%
\newsavebox{\mybox}\newsavebox{\mysim}
\newcommand{\distras}[1]{%
	\savebox{\mybox}{\hbox{\kern3pt$\scriptstyle#1$\kern3pt}}%
	\savebox{\mysim}{\hbox{$\sim$}}%
	\mathbin{\overset{#1}{\kern\z@\resizebox{\wd\mybox}{\ht\mysim}{$\sim$}}}%
}
\makeatother

\usepackage{amsmath,amssymb}
\makeatletter
\newsavebox\myboxA
\newsavebox\myboxB
\newlength\mylenA
\newcommand*\xoverline[2][0.75]{%
	\sbox{\myboxA}{$\m@th#2$}%
	\setbox\myboxB\null
	\ht\myboxB=\ht\myboxA%
	\dp\myboxB=\dp\myboxA%
	\wd\myboxB=#1\wd\myboxA
	\sbox\myboxB{$\m@th\overline{\copy\myboxB}$}
	\setlength\mylenA{\the\wd\myboxA}
	\addtolength\mylenA{-\the\wd\myboxB}%
	\ifdim\wd\myboxB<\wd\myboxA%
	\rlap{\hskip 0.5\mylenA\usebox\myboxB}{\usebox\myboxA}%
	\else
	\hskip -0.5\mylenA\rlap{\usebox\myboxA}{\hskip 0.5\mylenA\usebox\myboxB}%
	\fi}
\makeatother

\theoremstyle{definition}
\newtheorem{definition}{Definition}
\newtheorem{assumption}{Assumption}

\newtheorem{proposition}{Proposition}
\newtheorem{cor}{Corollary}
\newtheorem{remark}{Remark}

\begin{document} 
	\onehalfspacing
	
	\date{}
	\maketitle
	
	\vspace{-7mm}
	\begin{abstract}
		I propose Robust Rao's Score (RS) test statistic to determine endogeneity of spatial weights matrices in a spatial dynamic panel data (SDPD) model (Qu, Lee, and Yu, 2017). I firstly introduce the \textit{bias-corrected} score function since the score function is not centered around zero due to the two-way fixed effects. I further adjust score functions to rectify the over-rejection of the null hypothesis under a presence of local misspecification in contemporaneous dependence over space, dependence over time, or spatial time dependence. I then derive the explicit forms of our test statistic. A Monte Carlo simulation supports the analytics and shows nice finite sample properties. Finally, an empirical illustration is provided using data from Penn World Table version 6.1. \\
		
	\end{abstract}
	
	\newpage
	\section{Introduction}
	\onehalfspacing

	In a connected world, it is natural that an outcome of interest arises as a complex of interactions among spatial units in some effective area. In this spirit, Spatial econometrics analyzes the effects of spatial interactions among spatial units on their final economic outcome. 
	
	An ordinary factor considered to affect such interactions is the physical distances that represent the adjacency with its neighborhood or farness where the interactions are expected to decay at some specific rate as the distances get far. Thus, an intuitive use in spatial weights matrices ($W$) has been found in the predetermined geography, wherefore exogenous $W$ have been conventional in spatial econometrics (Moran, 1950; Cliff and Ord, 1972; Anselin, 1988, 2001; Ertur and Koch, 2007; Kelejian and Prucha, 2010; Elhorst, 2010, 2014; Lee and Yu, 2010; Dogan and Taspinar, 2013). 
	
	Empirical literature, however, assert that economic activities such as international trade also carry out knowledge spillover effects (Frankel and Rose, 1998; Baxter and Kouparitas, 2005; Ditzen, 2018). Inspired by this, a special $W$ of the bilateral trade flow is used (Ertur and Koch, 2011; Ho, Wang, and Yu, 2013). For more examples where economic factors form $W$, see Conley and Ligon (2002), Conley and Topa (2002), Parent and LeSage (2008), and Skevas, Skevas, and Cabrera (2021).

	$W$ can be \textit{endogenous}, however, if the space is intrinsically physical or economic (Pinkse and Slade, 2010). This is because economic variables are broadly and closely connected one another and thus the random shock in one variable might be correlated to that in the outcome of interest. If they are significantly dependent each other, then the exogenous assumption on $W$ may not be valid any more and inferences from the ordinary spatial autoregressive (SAR) estimators and test statistics will be invalid.
	
	To tackle this problem, Qu and Lee (2015) modeled the source of endogeneity of $W$ and proposed the maximum likelihood estimator (MLE) that controls endogeneity of $W$, which is known to be consistent and asymptotically normal distributed in cross section (Qu and Lee, 2015). In a spatial dynamic panel data (SDPD) model, the score function is not centered around zero. Also, the ML estimator has a bias with an order of $O(max\{n^{-1},T^{-1}\})$, even with large $n$ and large $T$, due to the two-way fixed effects (Yu, Jong, and Lee, 2008; Lee and Yu, 2010; Qu, Lee, and Yu, 2017). This can be rectified by the bias correction method (Yu, Jong, and Lee, 2008; Lee and Yu, 2010; Qu, Lee, and Yu, 2017; Bera, Dogan, Taspinar, and Leiluo, 2019). I thus firstly introduce the \textit{bias-corrected} score function centered around zero. The initial condition issue with finite $T$ is not of concern here with large $T$ setting. 
	
	I further adjust score functions to avoid the over-rejection of the null hypothesis under presences of local misspecification (Davidson and Mackinon, 1987; Saikkonen, 1989) in contemporaneous dependence over space, dependence over time, or spatial time dependence. This adjustment is in agreement with the existing Robust Rao's Score (RS) test literature: see Bera and Yoon (1993); Dogan, Taspinar, and Bera (2018); Bera, Dogan, and Taspinar (2018, 2019); Bera, Dogan, Taspinar, and Leiluo (2019); Bera, Bilias, Yoon, Taspinar, and Dogan (2020). The RS test is \textit{robust} in the sense that its asymptotic distribution is a central chi-square distribution regardless of local parametric misspecifications. A Monte Carlo simulation shows nice finite sample properties in size and power. 
	
	Another advantage of the RS test is that it is \textit{computationally efficient} since it only requires the restricted MLE under the null where the parameters above are assumed to be zero, reducing the spatial dynamic panel data models to the simple fixed-effects model. As illustrated in Section 6, the elapsed time (in seconds) for the Robust RS test is less than that for Conditional Lagrange Multiplier test (Qu and Lee, 2015; Cheng and Fei Lee, 2017).
	
	Literature on testing and model specifications have been mostly considered for cross-sectional spatial models (For example, see Anselin, 1988, 2001; Kelejian and Robinson, 1992; Anselin et al., 1996; Baltagi and Li, 2001; Yang, 2010; Bera et al., 2018, 2019; Dogan et al., 2018), whereas only few studies are found for the spatial static panel models (Baltagi et al., 2003, 2007, 2009; Baltagi and Liu, 2008; Debarsy and Ertur, 2010; Baltagi and Yang, 2013) and for the SDPD models (Yang, 2016; Taspinar et al., 2017; Bera et al., 2019). The majority of these studies lie on testing the presence of spatial dependence. To my best knowledge, testing endogeneity of $W$ has been developed only in Bera et al. (2018) in the cross-sectional spatial models and no such test has been developed yet in the SDPD models. A valid score test to test the endogeneity of $W$ is developed in this paper resolving two challenges of uncentered score function and presence of local misspecifications in the SDPD models. 
	
	The rest of the paper is organized as follows. In Section 2, I review the model specification and adopt the assumptions in Qu, Lee, and Yu (2017). In Section 3, I review the ML estimation which maximizes the concentrated log-likelihood. In Section 4, I develop the robust Rao's score test for testing endogeneity of $W$. In Section 5, I derive the explicit forms of the test statistic. In Section 6, I conduct a Monte Carlo simulation to explore its finite sample properties and an empirical illustration is provided using Penn World Table version 6.1. In Section 7, I conclude. The proofs of Propositions are provided in the Appendix.

	\section{Model specification}
	Following Jenish and Prucha (2009, 2012), consider a generalized spatial processes on a possibly unevenly spaced lattice. Let $\vert\vert Z \vert\vert_p=[E|Z|^p]^{1/p}$ when the absolute $\text{$p^{th}$}$ moment exists.
	
	\begin{definition}
		Let $Z=\{Z_{\ell,L}: \ell \in D_L, L \geq 1\}$ be a random field with $||Z_{\ell, L}||_{p} < \infty, p \geq 1$ and $\epsilon=\{\epsilon_{\ell,L}: \ell \in D_L, L \geq 1\}$ be another random field, where $|D_L| \rightarrow \infty$ as $L \rightarrow \infty,$ and let $d=\{d_{\ell,L}: \ell \in D_L, L \geq 1\}$ be an array of finite positive constants. Then the random field $Z$ is said to be \textit{$L_p(d)$-near-epoch dependent (NED)} on the random field $\epsilon$ if
		\begin{equation*}
		||Z_{\ell,L}-E(Z_{\ell,L}|\mathcal{F}_{\ell,L}(s))||_p \leq d_{\ell,L}\Psi(s) 
		\end{equation*} 
		\noindent for some sequence $\Psi(s) \geq 0$ with $\lim\limits_{s \rightarrow \infty}\Psi(s)=0$, where $\mathcal{F}_{\ell,L}(s)=\sigma(\epsilon_{j,L}: j\in D_L, \rho(\ell,j) \leq s)$ is the $\sigma$-field generated by $\epsilon_{j,L}$ within distance $s$ from $\ell.$ Here $\ell$ is an index as well as a location for simplification.
	\end{definition}
	
	I adopt the following assumptions as in Qu, Lee, and Yu (2017) on the spatial setting of observations. 
	\begin{assumption} 
		For a sample with $n$ units over $T$ periods, observations are located on a (possibly) unevenly spaced lattice $D \subset \mathbb{R}^{d+1}$, $d \geq 1$ and it is infinitely countable. The location $\ell: I \times T \rightarrow D_L \subset D$ is a mapping of individual $i \in I=\{1, \dots, n\}$ and time $t \in \{1,\dots,T\}$ to its location $\ell(i,t) \in D_L \subset \mathbb{R}^{d+1}$, $L=nT.$ For each spatial unit $i$, the location $\ell(i,t)$ is always one unit apart from $\ell(i,t-1)$ with respect to the time dimension. For a fixed $t=1,\dots,T$, any two elements in $D$ are separated by at least $\rho_{ct}>0$ distance from each other, i.e., for any $\ell(i,t), \ell(j,t)\in D$, $\rho_{ij,t} \geq \rho_{ct}$, where $\rho_{ij,t}$ is the distance between $\ell(i,t)$ and $\ell(j,t)$ for a fixed $t$. 
	\end{assumption}
	
	\noindent Assumption 1 allows the asymptotic inference on increasing domain under the space-time NED. Without loss of generality, assume $\rho_{ct}=1$ for all $t=1,\dots,T$, which means their least physical distance is one unit apart given each period $t.$
	
	Let $\{(\epsilon_{l(i,t)}, v_{l(i,t)}): l(i,t) \in D_L, i \in \field{N}, t \in \field{T}\}$ be a random field of error terms defined on a probability space $(\Omega, \mathcal{F}, P)$, where $D_L \subset D$ is a finite set and $D$ satisfies Assumption 1. To simplify the notation, let $(\epsilon_{l(i,t)}, v_{l(i,t)})$ be denoted by $(\epsilon_{it}, v_{it})$, where $\epsilon_{it}$ is the $p \times 1$ column vector for $t=1,\dots,T$ formulated from the $i^{th}$ row of $n \times p$ matrix $\varepsilon_{nt}$, and $v_{it}$ is the $i^{th}$ element of $n \times 1$ vector $V_{nt}$, for all $t = 1, \cdots, T.$ A SDPD model with individual and time fixed effects for $n$ cross-sectional units can be specified as
	\begin{align}
		\label{eq1}
		\begin{split}
		Y_{nt}&= \lambda_0 W_{nt} Y_{nt} + \rho_0 W_{n,t-1}Y_{n,t-1} + \gamma_0 Y_{n,t-1}  \\
		&\hspace{5mm}+ X_{1nt}\beta_0 + c_{n10} + \alpha_{t10}1_n + V_{nt}, \quad t=1,\dots,T, \\
		\end{split}
	\end{align}

	\noindent where $Y_{nt}=(y_{1t},\dots,y_{nt})'$ is a $n \times 1$ vector of observations on the dependent variable, $W_{nt}=(w_{ij})_{t}$ is a $n \times n$ spatial weights matrix with zero diagonal elements, $\lambda_0$ denotes the autoregressive parameter, $X_{1nt}$ is a $n \times k_1$ matrix of deterministic explanatory variables that are bounded in absolute value, and $\beta$ is a $k_1 \times 1$ vector of parameters. $c_{n10}$ is a $n \times 1$ column vector of individual fixed effects and $\alpha_{t10}$ is the $t^{th}$ element of $T \times 1$ time fixed effects vector $\alpha_{T10}$, respectively, and $V_{nt}=(v_{1t}, \cdots, v_{nt})'$ is a $n \times 1$ vector of error terms with zero mean and variance $\sigma_0^2$. I also generalize $W_{nt}$ to be time varying and endogenous by modeling a construction of $W_{nt}$ as a bounded function of $Z_{nt}$ such that
	\begin{align}
		\label{eq2}
		\begin{split}
		&(W_{nt})_{ij}= w_{ij,nt}=h(z_{i,nt},z_{j,nt}), \\ 
		&Z_{nt}= Z_{n,t-1}\kappa_0 + X_{2nt}\Gamma_0 + c_{n20} + 1_n \alpha_{t20}' + \varepsilon_{nt},
		\end{split}
	\end{align}

	\noindent where $h_{ij,t}(\cdot)$ is a nonnegative and uniformly bounded function and $Z_{nt}=(z_{1t}, \cdots, z_{nt})'$ is a $n \times p$ matrix of dependent variables with $z_{it}=(z_{1it}, \cdots, z_{p it})'$ being a $p \times 1$ vector and $\kappa_0$ is its associated parameter. $X_{2nt}$ is a $n \times k_2$ matrix of deterministic explanatory variables whose elements are bounded in absolute value, $\Gamma_0$ is a $k_2 \times p$ matrix of associated coefficients, and $\varepsilon_{nt}=(\epsilon_{1t}, \cdots, \epsilon_{nt})'$ is a $n \times p$ matrix of errors, where $\epsilon_{it}=(\epsilon_{it,1}, \cdots, \epsilon_{it,p})'$ is a $p \times 1$ vector. $c_{n20}$ is again a $n \times 1$ column vector of individual fixed effects and $\alpha_{t20}$ is the $t^{th}$ element of $T \times 1$ time fixed effects vector $\alpha_{T20}$, respectively. $1_n$ is a $n \times 1$ vector of ones. The initial values in $Y_{n0}$ are assumed to be observable. 
	
	\noindent For the SDPD model in (1), it might be convenient to adopt the subscript $n$ to simplify notation. Due to non-linearity in $W^{'}_{nt}s$, one may work on the random fields for $\{y_{it}\}$ and $\{(v_{it},\epsilon'_{it})\}$. The settings are similar as in Jenish and Prucha (2012) so that one may apply the spatial-time LLN. At time $t=0,$ the $n$ units are located in the Euclidean space $\mathbb{R}^{d+1}$. At time $t=1$, the $n$ units shift vertically upward to an affine plane parallel to $\mathbb{R}^d$ planes. As time passes, it keeps shifting upward each time by one time unit. For the variables in past periods (history), one may shift the plane $\mathbb{R}^d$ at $t=0$ downward to an affine plane at $t=-1$, and so on. A sample of individuals with $n$ units over $T$ periods are located in the $\mathbb{R}^{d+1}$ space. Using the maximum metric (Jenish and Prucha, 2012)
	\begin{equation*} ||\ell(i,t)-\ell(j,\tau)||_{\infty}=\max\{|t-\tau|,||\ell(i,0)-\ell(j,0)||_{\infty}\}, \end{equation*}
	each observation indexed by $(i,t)$ will be located at $\ell(i,t)\in\mathbb{R}^{d+1}$ where $\ell(i,0)$ is the physical location of an individual at time 0. Any two individuals located in the $\mathbb{R}^d$ plane at $t=0$ are assumed at least one unit apart. For each spatial unit $i,$ the location $\ell(i,t)$ is always one unit apart from $\ell(i,t-1)$ with respect to the time dimension. There are $L$ locations in $\mathbb{R}^{d+1}$. 
	For the spatial weight matrix, $w_{ij,nt} \neq 0$ if an individual at spatial unit $i$ at time $t$ directly links to  $j$ at time $t$; $w_{ij,n,t-1}\neq 0$ if $i$ links to $j$ at time $t-1$. Any individual at time $t$ does not directly link to any future loads from anyone including itself $t+\tau$ for $\tau \geq 1,$ and not the past $t-\tau$ with $\tau \geq 2$ but indirect links are allowed at $t-\tau$ for $\tau \geq 2$. Note that $i$ indirectly links to itself in past periods. This is a mapping of $(i,t)$ to a location $\ell$ in $D_L$, i.e., $\ell=\ell(i,t) \in D_L$ with $L=nT.$ Since $y_\ell$ corresponds to a $y_{it}$, one may define the spatial NED process with a base $\sigma$-field generated by $\{v_{it},\epsilon_{it}'\}$ as follows:
	\begin{equation*} \mathcal{F}_{\ell,L}(s)=\sigma((v_{j,\tau},\epsilon_{j,\tau})'): \ell(j,\tau) \in D_L, ||\ell(j,\tau)-\ell(i,t)||_\infty \leq s).\end{equation*}
	
	\noindent The $Y=\{y_{it}\}$ is a NED if $\vert\vert y_{\ell,L}-E(y_{\ell,L}\vert\mathcal{F}_{\ell,L}(s))\vert\vert_p \leq d_{\ell,L}\Psi(s)$ for some $\Psi(s),$ which goes to zero as $s \rightarrow \infty.$ 
	
	\begin{assumption} 
		The error terms $v_{it}$ and $\epsilon_{it}$ are \textit{iid} and jointly normal distributed $(v_{it},\epsilon_{it}')' \distas{iid} N(0,\Sigma_{v\epsilon 0})$, where $\Sigma_{v\epsilon 0}=\begin{pmatrix*} \sigma_{v0}^2 & \sigma_{v\epsilon 0}' \\ \sigma_{v\epsilon 0} & \Sigma_{\epsilon 0} \end{pmatrix*}$, $\sigma_{v0}^2$ is a scalar variance of $v_{it}$, $\Sigma_{\varepsilon 0}$ is a $p \times p$ covariance matrix of $\epsilon_{it}=(\epsilon_{it,1}, \cdots, \epsilon_{it,p})'$, and $\sigma_{v\epsilon 0}$ is a $p \times 1$ covariance matrix between $v_{it}$ and $\epsilon_{it}=(\epsilon_{it,1}, \cdots, \epsilon_{it,p})'.$
	\end{assumption}
	
	\noindent Note that since $W_{nt}$ is a function of $\varepsilon_{nt}$, $W_{nt}$ is endogenous in the estimation equation if $\sigma_{v\epsilon 0} \neq 0$, which leads to a biased estimator due to the endogeneity. From the joint distribution of $(v_{it},\epsilon_{it}')',$ one has $E(v_{it}|\epsilon_{it})=\epsilon_{it}'\delta_0$ where $\delta_0=\Sigma_{\epsilon 0}^{-1}\sigma_{v\epsilon 0}$ and $Var(v_{it}|\epsilon_{it})=\sigma_{\xi 0}^2$ where $\sigma_{\xi 0}^2=\sigma_{v0}^2-\sigma_{v \epsilon 0}'\Sigma_{\epsilon 0}^{-1}\sigma_{v \epsilon 0}.$ Let $\xi_{nt}=V_{nt}-\epsilon_{nt}\delta_0.$ Since the expectation of $\xi_{nt}$ conditional on $\varepsilon_{nt}$ is zero, $\xi_{nt}$ is uncorrelated with $\varepsilon_{nt}.$ Hence $\eqref{eq1}$ can be represented as
	\begin{align}
		\label{eq3}
		\begin{split}
		Y_{nt}&=\lambda_0 W_{nt}Y_{nt} + \rho_0 W_{n,t-1}Y_{n,t-1} + \gamma_0 Y_{n,t-1} + X_{1nt}\beta_0 \\
		&\hspace{3mm} +c_{n10}+\alpha_{t10}1_n \\
		&\hspace{3mm} +(Z_{nt}-Z_{n,t-1}\kappa_0-X_{2nt}\Gamma_0-c_{n20}-1_n\alpha_{t20}')\delta_0+\xi_{nt},
		\end{split}
	\end{align}
	
	\noindent where $E(\xi_{nt}|\varepsilon_{nt})=0$ and $Var(\xi_{nt}|\varepsilon_{nt})=\sigma_{\xi 0}^2 I_n,$ and the elements of $\xi_{nt}$ are iid across $i$ and $t$. Note that $(Z_{nt}-Z_{n,t-1}\kappa_0-X_{2nt}\Gamma_0-c_{n20}-1_n\alpha_{t20}')$ are control variables in $Y_{nt}$ to control the endogeneity of $W_{nt}$.
	
	\noindent I additionally assume the followings.
	
	\begin{assumption} 
		$n$ is an increasing function of $T$, and $T$ goes to infinity. 
	\end{assumption}
	
	\begin{assumption} 
		Let $X_{nt}$ denote the collection of distinct columns in $X_{1nt}$ and $X_{2nt}.$ Elements of $X_{nt}$ are nonstochastic and bounded, and \\$\lim\limits_{n,T \rightarrow \infty} \frac{1}{nT}\sum\limits_{t=1}^T \tilde{X}_{nt}'J_n\tilde{X}_{nt}$ exists and is nonsingular, where $\tilde{X}_{nt}=X_{nt}-\frac{1}{T}\sum\limits_{t=1}^T X_{nt}$ and $J_n=I_n-\frac{1}{n}1_n 1_n'.$ Furthermore, $\lim\limits_{n, T \rightarrow \infty} \frac{1}{nT}\sum\limits_{t=1}^T E(\tilde{Z}_{n,t-1}^{(-1)}J_n \tilde{Z}_{n,t-1}^{(-1)})$ exists and is nonsingular, where $\tilde{Z}_{n,t-1}^{(-1)}=Z_{n,t-1}-\frac{1}{T}\sum\limits_{t=1}^T Z_{n,t-1}.$
	\end{assumption}
	
	\begin{assumption}
		(i) The spatial weight $w_{ij,t} \geq 0$ and $w_{ii,t}=0$ for all $i, j,$ and $t$; (ii) For those nonzero weights between $i \neq j$, $w_{ij,t}=h(z_{it},z_{jt})\cdot I(\rho_{ij} \leq \rho_c)$ or the row-normalized version $w_{ij,t}=h(z_{it},z_{jt})\cdot I(\rho_{ij} \leq \rho_c)/\sum\limits_{\rho_{ik} \leq \rho_c} h_{ik}(z_{it},z_{kt})$. (iii) For two different periods $t$ and $t'$, the Lipschitz condition holds such that 
		\begin{equation*}
		|h(z_{it'},z_{jt'})-h(z_{it},z_{jt})| \leq c_0 (|z_{it'}-z_{it}| + |z_{jt'}-z_{jt}|).
		\end{equation*}
	\end{assumption}
	
	\begin{assumption} 
		$\sup\limits_{n,t} ||W_{nt}||_{\infty} \leq C_w$, $||\gamma_0||_1 < 1$, and $|\lambda_0|C_w + |\rho_0| + |\gamma_0|C_w < 1$, where $C_w$ is a finite constant.
	\end{assumption}
	
	\begin{assumption}
		The vector of parameters $\theta=(\lambda, \gamma, \rho, \beta', \delta, vec'(\kappa,\Gamma), \alpha, \sigma_\xi^2)'$ is in the interior of a compact set $\Theta$ and the true parameter vector $\theta_0$ is in the interior of $\Theta,$ where $vec(\cdot)$ denotes the matrix operator that stacks columns of a given matrix.
	\end{assumption}
	
	\noindent Assumption 3 requires large $n$ and large $T$ case. As $T$ is large under assumption 3, the initial condition problem would not be an issue here. Assumption 4 excludes an issue of multicollinearity. Also $X_{1nt}$ and $X_{2nt}$ are allowed to overlap and have the common variables because identification is not of interest for our purpose. Assumption 5 imposes features of $W_{nt}$ such that: (i) allows time-varying $W_{nt}$ while the physical distance is fixed over time; For technical purpose, (ii) states that two units are not considered spatially connected if their exogenous distance exceeds a certain threshold, even though their economic/social factors are. This is a popular setting in empirical studies; (iii) is a condition on $h(\cdot)$ so that $z_{it}$ and $z_{jt}$ determine the time NED property for $w_{ij,t}$. Assumption 6 guarantees the stability of the dynamic process by controlling the magnitude of spatial interaction of $W_{nt}$ matrix, which is defined in Section 3. The parameter space is characterized under Assumption 7. \\

	\section{ML estimation}
	Following Qu, Lee, and Yu (2017), one may put the model $\eqref{eq1}$ into a big matrix form. Denote

	\begin{align*}
	& Y_L=\begin{pmatrix} Y_{n1} \\ Y_{n2} \\ \vdots \\ Y_{nT} \end{pmatrix},  \hspace{14mm} \ell_0(\gamma_0, \rho_0)=\begin{pmatrix} \gamma_0 Y_{n0} + \rho_0W_{n0}Y_{n0} \\ 0 \\ \vdots \\ 0 \end{pmatrix}, \\
	& \alpha_{1L} = \begin{pmatrix} \alpha_{11} \\ \alpha_{21} \\ \vdots \\ \alpha_{T1} \end{pmatrix} \otimes 1_n,  \hspace{5mm} \alpha_{2L} = \begin{pmatrix} \alpha_{12}' \\ \alpha_{22}' \\ \vdots \\ \alpha_{T2}' \end{pmatrix} \otimes 1_n,  
	\end{align*}

	\noindent and $X_{1L}, \epsilon_L, \xi_L, Z_L$ and $X_{2L}$ are defined similarly. Also, denote
	\begin{equation*}
	W_L(\eta)=\lambda W_{1L} + \gamma W_{2L} + \rho W_{3L},
	\end{equation*}
	
	\noindent with $W_{1L}=\begin{pmatrix} W_{n1} & 0 & \cdots & 0 \\ 0 & W_{n2} & \cdots & 0 \\ \vdots & \vdots& \ddots & \vdots \\ 0 & 0 & 0 & W_{nT}\end{pmatrix},$ $W_{2L}=\begin{pmatrix} 0 & 0 & \cdots & 0 \\ I_n & 0 & \cdots & 0 \\ 0 & \ddots & \ddots & 0 \\ 0 & 0 & I_n & 0 \end{pmatrix},$ \\ $W_{3L}=\begin{pmatrix} 0 & 0 & \cdots & 0 \\ W_{n1} & 0 & \cdots &0  \\ \ddots & \ddots & \ddots & \vdots\\ 0 & 0 & W_{n,T-1} & 0\end{pmatrix},$ \\ 
	
	\noindent so that
	\begin{equation*}
	W_L(\eta_0) =\begin{pmatrix}
	\lambda_0 W_{n1} & 0 & 0 & \cdots & 0 \\ \gamma_0 I_n+\rho_0 W_{n1} & \lambda_0 W_{n2} & 0 & \cdots & 0 \\ \vdots & \vdots & \vdots & \ddots & \vdots \\ 0 & 0 & \cdots & \gamma_0 I_n + \rho_0 W_{n,T-1} & \lambda_0 W_{nT}
	\end{pmatrix},
	\end{equation*}
	
	\noindent where $\eta=(\lambda, \gamma, \rho)'$ and $\eta_0$ is its true value. Therefore, the model in $\eqref{eq1}$ and in  $\eqref{eq2}$ can be written in the matrix form as:
	\begin{equation*}
	Y_L = W_L(\eta_0)Y_L + X_{1L}\beta_0 + \ell_0(\gamma_0, \rho_0) + c_{1L0} + \alpha_{1L0} + \epsilon_L \delta_0 + \xi_L,
	\end{equation*}
	\begin{equation*}
	Z_L = Z_{L,-1}\kappa_0 + X_{2L}\Gamma_0 + c_{2L0} + \alpha_{2L0} + \epsilon_L, 
	\end{equation*}
	
	\noindent where $c_{1L} = 1_T \otimes c_{1n}$, $c_{2L}=1_T \otimes c_{2n}$. From Assumption 2, $V_L=\varepsilon_L \delta_0 + \xi_L$ where $\xi_L \sim N(0,\sigma_{\xi_0}^2 I_n)$ conditional on $\varepsilon_L$.
	Hence,
	\begin{align}
		\label{eq4}
		\begin{split}
		Y_L &= W_L(\eta_0)Y_L + X_{1L}\beta_0 + \ell_0(\gamma_0,\rho_0) + c_{1L0} + \alpha_{1L0} \\ 
		& \hspace{25mm} +(Z_L-Z_{L,-1}\kappa_0 - X_{2L}\Gamma_0 - c_{2L0} -\alpha_{2L0})\delta_0 + \xi_L \\
		&= W_L(\eta_0)Y_L + X_{1L}\beta_0 + \ell_0(\gamma_0,\rho_0) + \xoverline{c_{1L0}} + \xoverline{\alpha_{1L0}} \\ 
		& \hspace{25mm}+(Z_L-Z_{L,-1}\kappa_0-X_{2L}\Gamma_0)\delta_0 + \xi_L,
		\end{split}
	\end{align}

	\noindent where $\xoverline{c_{1L0}}=c_{1L0}-c_{2L0}\delta_0$ and $\xoverline{\alpha_{1L0}}=\alpha_{1L0}-\alpha_{2L0}\delta_0.$ \\

	For asymptotic analysis, it is useful to have the likelihood function presented in matrix form involving $W_L({\eta}).$  Let $\theta=(\lambda,\phi_1',\delta',\phi_2',\alpha',\sigma_{\xi}^2)',$ where $\phi_1=(\gamma,\rho,\beta')'$, $\phi_2=vec(\Phi_2)$ with $\Phi_2=(\kappa',\Gamma')',$ and $\alpha$ is the $J \times 1$ column vector of distinct elements in $\Sigma_\epsilon.$ Corresponding to $\phi_1$ and $\phi_2$, I define $R_{nt}=[Y_{n,t-1}, W_{n,t-1}Y_{n,t-1}, X_{1nt}]$, $K_{nt}=[Z_{n,t-1},X_{2nt}]$ and their matrix form $R_L$ and $K_L$ with row dimension $L=nT.$  I denote $S_L(\eta)=I_L-W_L(\eta)$, $S_L=I-W_L(\eta_0)$ and correspondingly $G_{jL}(\eta)=W_{jL}S_L^{-1}(\eta)$ and $G_{jL}=W_{jL}S_{L}^{-1}$ for $j=1, 2, 3$, where $I_L$ is the $L \times L$ identity matrix. I also denote $Q_{1L}(\theta_0,c_{1L0},\alpha_{1L0})=G_{1L}(X_{1L}\beta_0+\ell_0(\gamma_0,\rho_0)+\varepsilon_L\delta_0+c_{1L0}+\alpha_{1L0})$.  One may derive the log likelihood function from Assumption 2 as it only imposes assumptions only on the first and second moments but third and fourth ones. From Assumption 2 and equation $\eqref{eq4}$, the log likelihood function in the matrix form is
	\begin{align}
		\label{eq5}
		\begin{split}
		&\ln L_L(\theta,\xoverline{c_{1L}}, \xoverline{\alpha_{1L}}, c_{2L}, \alpha_{2L})\\
		&=-\frac{nT}{2}\ln 2\pi + \ln\vert S_L(\eta)\vert -\frac{nT}{2}\ln \sigma_{\xi}^2 - \frac{nT}{2}\ln\vert \Sigma_\epsilon\vert  \\
		& \hspace{4mm}-\frac{1}{2}vec'(Z_L-K_L\Phi_2-c_{2L}-\alpha_{2L})(\Sigma_\varepsilon^{-1} \otimes I_L) vec(Z_L-K_L\Phi_2-c_{2L}-\alpha_{2L}) \\
		& \hspace{4mm}-\frac{1}{2\sigma^2_{\xi}}[Y_L-W_L(\eta)Y_L-X_{1L}\beta-(Z_L-K_L\Phi_2)\delta-\ell_0(\gamma,\rho)-\xoverline{c_{1L}}-\xoverline{\alpha_{1L}}]' \\
		&\hspace{15mm}\times [Y_L-W_L(\eta)Y_L-X_{1L}\beta-(Z_L-K_L\Phi_2)\delta-\ell_0(\gamma,\rho)-\xoverline{c_{1L}}-\xoverline{\alpha_{1L}}].
		\end{split}
	\end{align}

	As in Lee and Yu (2010), I use the two orthogonal projectors for taking time deviation and cross sectional deviation from their means, $J_n = I_n-\frac{1}{n}1_n 1_n'$ and $J_T = I_T - \frac{1}{T}1_T 1_T'$. Let $J_L=J_T \otimes J_n$. The concentrated log-likelihood function is then
	\begin{align}
		\label{eq6}
		\begin{split}
		\ln L_L^c(\theta)&=-\frac{nT}{2}\ln2\pi + \ln\vert S_L(\eta)\vert -\frac{nT}{2}\ln\sigma_\xi^2 - \frac{nT}{2}\ln\vert \Sigma_\epsilon\vert  \\
		&\hspace{3mm} -\frac{1}{2}vec'(Z_L-K_L\Phi_2)(\Sigma_\varepsilon^{-1}\otimes J_L)vec(Z_L-K_L\Phi_2) \\
		&\hspace{3mm} -\frac{1}{2\sigma_\xi^2}[S_L(\eta)Y_L -X_{1L}\beta_0 - (Z_L-K_L\Phi_2)\delta -\ell_0(\gamma,\rho)]'  \\
		&\hspace{3mm} \cdot J_L \cdot [S_L(\eta)Y_L -X_{1L}\beta_0 - (Z_L-K_L\Phi_2)\delta -\ell_0(\gamma,\rho)].
		\end{split}
	\end{align}
	
	\noindent The robust RS test statistics needs the asymptotic distribution of score functions derived as follows (Qu, Lee, and Yu, 2017): 
	\begin{align*}
	\frac{\partial lnL_L^c(\theta)}{\partial \theta}&
	=\begin{bmatrix}
	\frac{1}{\sigma^2_{\xi 0}}(W_{1L}Y_L)'J_L\xi_L -tr(G_{1L})\\
	\frac{1}{\sigma^2_{\xi 0}}R_L'J_L\xi_L \\
	\frac{1}{\sigma^2_{\xi 0}}\varepsilon_L'J_L\xi_L \\
	\Sigma_{\varepsilon 0}^{-1} \otimes (K_L'J_L)vec(\varepsilon_L) - \frac{1}{\sigma^2_{\xi 0}}\delta_0 \otimes K_L'J_L \xi_L \\
	-\frac{nT}{2\sigma^2_{\xi 0}} + \frac{1}{2\sigma^2_{\xi 0}}\xi_L'J_L\xi_L \\
	-\frac{nT}{2}\frac{\partial \ln\vert\Sigma_{\varepsilon 0}\vert}{\partial \alpha} -\frac{1}{2}\frac{\partial}{\partial \alpha}(tr[\Sigma_{\varepsilon 0}^{-1}\varepsilon_L'J_L \varepsilon_L])\end{bmatrix}	_.
	\end{align*}
	
	\noindent Due to the two way fixed effects, the score function of $lnL_L^c(\theta)$ can be decomposed of the unbiased score function and bias term (Qu, Lee, and Yu, 2017) as follows:
	
	\begin{equation}
	\frac{\partial \ln L_L^c(\theta_0)}{\partial \theta}=\frac{\partial \ln L_{1,L}^c(\theta_0)}{\partial \theta}+\Delta_L,	\label{eq7}
	\end{equation}
	
	\noindent where
	\begin{equation*}
	\frac{\partial \ln L_{1,L}^c(\theta_0)}{\partial \theta}=\begin{bmatrix} 
	\frac{1}{\sigma^2_{\xi 0}}(W_{1L}Y_L)'J_L\xi_L - tr(G_{1L}J_L) \\
	\frac{1}{\sigma^2_{\xi 0}}R_L'J_L\xi_L - [tr(G_{2L}J_L),tr(G_{3L}J_L),0]' \\
	\frac{1}{\sigma^2_{\xi 0}}\varepsilon_L'J_L\xi_L \\
	\Sigma_{\varepsilon 0}^{-1}\otimes K_L'J_Lvec(\varepsilon_L)+ vec\left(\frac{n-1}{T}\left(\sum\limits_{t=1}^{T-1}\sum\limits_{h=1}^{T-t}\kappa'^{(h-1)}\right)\Sigma_{\varepsilon 0}^{-1}\right)-\frac{1}{\sigma_{\xi 0}^2}\delta_0 \otimes K_L'J_L\xi_L \\
	-\frac{(n-1)(T-1)}{2\sigma^2_{\xi 0}} + \frac{1}{2\sigma^2_{\xi 0}}\xi_L'J_L\xi_L \\
	-\frac{(n-1)(T-1)}{2}\frac{\partial \ln\vert\Sigma_{\varepsilon 0}\vert}{\partial  \alpha} - \frac{1}{2}\frac{\partial}{\partial \alpha}(tr[\Sigma_{\varepsilon 0}^{-1}\varepsilon_L'J_L\varepsilon_L]) \\
	\end{bmatrix}
	\end{equation*}
	
	\noindent and
	\begin{equation*}
	\Delta_L = E\left(\frac{\partial lnL_L^c(\theta_0)}{\partial \theta}\right) \begin{bmatrix}
	tr(W_{1L}(I_L-W_L)^{-1}J_L) - tr(G_{1L}) \\
	[tr(G_{2L}J_L),tr(G_{3L}J_L),0]' \\
	0_{p \times 1} \\
	-vec\left(\frac{n-1}{T}\left(\sum\limits_{t=1}^{T-1}\sum\limits_{h=1}^{T-t}\kappa'^{(h-1)}\right)\Sigma_{\varepsilon 0}^{-1}\right) \\
	-\frac{n+T-1}{2\sigma^2_{\xi 0}} \\
	-\frac{n+T-1}{2}\frac{\partial ln\vert\Sigma_{\varepsilon 0}\vert}{\partial \alpha} \\
	\end{bmatrix}.
	\end{equation*}
	
	\noindent The first term in $\frac{\partial lnL_L^c(\theta_0)}{\partial \theta}$, $\frac{\partial \ln L_{1,L}^c(\theta_0)}{\partial \theta}$, has zero mean and is asymptotically distributed with further assumption of the third and fourth moment of $\xi_{it}$ conditional on $\epsilon_{it}$.

	\begin{proposition}

		\begin{equation*}
		\frac{1}{\sqrt{nT}}\frac{\partial \ln L_{1,L}^c(\theta_0)}{\partial \theta} \xrightarrow{d} N\left(0, \lim\limits_{n,T \rightarrow \infty}\ell_{nT,\theta_0}\right),
		\end{equation*}
		
		\noindent where $\ell_{nT,\theta_0}=-\frac{1}{nT}E\left(\frac{\partial^2 \ln L_L^c(\theta_0)}{\partial \theta \partial \theta'}\right)$ and 	
		
		\begin{align}
			\label{eq8}
			\begin{split}
			&E\left(-\frac{\partial^2 \ln L_L^c(\theta_0)}{\partial \theta \partial \theta'}\right)=\frac{1}{\sigma^2_{\xi 0}}\\
			&
			\begin{bmatrix}
			I_{\lambda \lambda} & * & * & * & * & \textbf{0}_{1 \times J} \\
			ER_L'J_L Q_{1L} & ER_L'J_L R_L & \textbf{0}_{(k_1+2)\times p}& * &\textbf{0}_{(k_1+2)\times 1} & \textbf{0}_{(k_1+2)\times J}\\
			E[\varepsilon_L' J_L Q_{1L}] & \textbf{0}_{p \times (k_1+2)} &  E\epsilon_{L}'J_L\epsilon_L & \textbf{0}_{p \times k_{\phi_2}}& \textbf{0}_{k_{\phi_2}\times 1}& \textbf{0}_{k_{\phi_2}\times J} \\
			+E[\varepsilon_L ' J_L(G_{1L}\varepsilon_L \delta_0)] & & & & & \\
			-\delta_0 \otimes EK_L'J_L Q_{1L} & -\delta_0 \otimes EK_L'J_LR_L &\textbf{0}_{k_{\phi_2}\times p} & I_{\Phi_2\Phi_2}& \textbf{0}_{k_{\phi_2}\times 1} & \textbf{0}_{k_{\phi_2}\times J} \\
			Etr(G_{1L}) & \textbf{0}_{1 \times (k_1+2)} &\textbf{0}_{1 \times p} & \textbf{0}_{1 \times k_{\phi_2}}&\frac{1}{\sigma_{\xi0}^2}\left(\frac{nT}{2}-T-n+1\right) & \textbf{0}_{1 \times J} \\
			\textbf{0}_{J \times 1} & \textbf{0}_{J \times (k_1 +2)}& \textbf{0}_{J \times p} & \textbf{0}_{J \times k_{\phi_2}} & \textbf{0}_{J \times 1} & I_{\alpha\alpha}
			\end{bmatrix}, 
			\end{split}
		\end{align}
		
		\noindent where $J_L=J_T \otimes J_n$, $I_{\lambda\lambda}=E[Q_{1L}'J_LQ_{1L}+\sigma^2_{\xi 0}tr(G^2_{1L}+G'_{1L}J_LG_{1L})], I_{\phi_2\phi_2}=(\sigma_{\xi 0}^2 \Sigma_{\varepsilon 0}^{-1}+\delta_0\delta_0')\otimes EK_L'J_LK_L,$ and $I_{\alpha\alpha}$ is a $J \times J$ matrix with its $(k,j)$ element being $\frac{nT}{2}\sigma^2_{\xi 0}tr\left(\Sigma_{\varepsilon 0}^{-1}\frac{\partial \Sigma_{\varepsilon 0}}{\partial \alpha_k}\Sigma_{\varepsilon 0}^{-1}\frac{\partial \Sigma_{\epsilon 0}}{\partial \alpha_j}\right)$.
	\end{proposition}
	\textbf{Proof.} See Appendix 1. \\
	
	\noindent Furthermore, the bias term $\Delta_L$ is composed of two types of biases such that
	\begin{equation*}
	\Delta_L = (n-1)a_{1,\theta_0} + Ta_{2,\theta_0},
	\end{equation*}
	
	\noindent where $a_{1,\theta_0}$ and $a_{2,\theta_0}$ are of $O(1)$ given as 
	
	\begin{equation*} a_{1,\theta_0}=
	\begin{bmatrix}
	-\frac{1}{n-1}tr(G_{1L}\big(\frac{1}{T}1_T 1_T' \otimes J_n)) \\
	\frac{1}{n-1}[tr(G_{2L}\big(\frac{1}{T}1_T 1_T' \otimes J_n)), tr(G_{3L}\big(\frac{1}{T}1_T 1_T' \otimes J_n)), 0]' \\
	0_{p \times 1} \\
	-vec\left(\frac{1}{T}\left(\sum\limits_{t=1}^{T-1}\sum\limits_{h=1}^{T-t}\kappa_0'^{(h-1)}\right)\Sigma_{\varepsilon 0}^{-1}\right) \\
	-\frac{1}{2\sigma_{\xi_0}^2} \\
	-\frac{1}{2}\frac{\partial ln |\Sigma_{\varepsilon 0}|}{\partial \alpha}
	\end{bmatrix},
	\end{equation*}
	
	\noindent and
	\begin{equation*} a_{2,\theta_0} = 
	\begin{bmatrix}
	-\frac{1}{T}tr(G_{1L}(I_T \otimes \frac{1}{n}1_n 1_n')) \\
	0_{(k_1+2)\times 1} \\
	0_{p \times 1} \\
	0_{p(p+k_2) \times 1} \\
	-\frac{1}{2 \sigma^2_{\xi 0}} \\
	-\frac{1}{2} \frac{\partial ln |\Sigma_{\varepsilon 0}|}{\partial \alpha}
	\end{bmatrix}.
	\end{equation*}
	
	\noindent The bias $a_{1,\theta_0}$ is from individual effects and $a_{2,\theta_0}$ is due to time effects.

	\begin{proposition} (LLN) Under Assumptions 1-6, for any finite integer $m$,
		\begin{align*}
		&\frac{1}{L}\zeta_{1L}'J_LW_{j_1L}W_{L}^m \zeta_{2L}-E\left[\frac{1}{L}\zeta_{1L}'J_LW_{j_1L}W_{L}^m\zeta_{2L}\right]=o_p(1)\\
		&\frac{1}{L}(G_{j_1L}\zeta_{1L})'J_L\zeta_{2L}-E\left[\frac{1}{L}(G_{j_1L}\zeta_{1L})'J_L\zeta_{2L}\right]=o_p(1)\\
		&\frac{1}{L}(G_{j_1L}\zeta_{1L})'J_LG_{j_2L}\zeta_{2L}-E\left[\frac{1}{L}(G_{j_1L}\zeta_{1L})'J_LG_{j_2L}\zeta_{2L}\right]=o_p(1),
		\end{align*}
		where $j_1$ and $j_2$ can be either 1, 2, 3, corresponding to $W_{1L}, W_{2L}, W_{3L}$; $\zeta_{1L}$ and $\zeta_{2L}$ can be either some nonstochastic regressor vectors, $\varepsilon_L, \xi_L$, or $Z_{L,-1}$. 
	\end{proposition}
	\textbf{Proof.} See Appendix 1.
	
	\begin{cor} (ULLN). Under Assumptions 1-7,
		\begin{align*}
		&\sup\limits_{\theta \in \Theta}\left|\frac{1}{L}\zeta_{1L}'(\theta)J_LW_{j_1L}W_{L}^m \zeta_{2L}(\theta)-E\left[\frac{1}{L}\zeta_{1L}'(\theta)J_LW_{j_1L}W_{L}^m\zeta_{2L}(\theta)\right] \right|=o_p(1) \\
		&\sup\limits_{\theta \in \Theta}\left|\frac{1}{L}(G_{j_1L}\zeta_{1L}(\theta))'J_L\zeta_{2L}(\theta)-E\left[\frac{1}{L}(G_{j_1L}\zeta_{1L}(\theta))'J_L\zeta_{2L}(\theta)\right] \right|=o_p(1)\\
		&\sup\limits_{\theta \in \Theta}\left|\frac{1}{L}(G_{j_1L}\zeta_{1L}(\theta))'J_LG_{j_2L}\zeta_{2L}(\theta)-E\left[\frac{1}{L}(G_{j_1L}\zeta_{1L}(\theta))'J_LG_{j_2L}\zeta_{2L}(\theta)\right] \right|=o_p(1).
		\end{align*}
	\end{cor} 
	\textbf{Proof.} See Appendix 1. \\
	
	Now let $\hat{\theta}_{nT}=\underset{\theta \in \Theta}{\text{argmax}} \hspace{1mm}\ln L_L^c(\theta)$ be the MLE. Then the asymptotic distribution of $\hat{\theta}_{nT}$ is  (Qu, Lee, and Yu, 2017)
	\begin{align*}
	&\sqrt{nT}(\hat{\theta}_{nT}-\theta_0)-\sqrt{\frac{n}{T}}b_{1,\theta_0,nT}-\sqrt{\frac{T}{n}}b_{2,\theta_0,nT}+O_p\left(\text{max}\left(\sqrt{\frac{n}{T^3}},\sqrt{\frac{T}{n^3}}\right)\right) \\
	&\hspace{50mm}\xrightarrow{d} N\left(0, \lim\limits_{T \rightarrow \infty} \ell_{nT,\theta_0}^{-1}\right),
	\end{align*}

	\noindent where $b_{1,\theta_0,nT}=\ell_{nT,\theta_0}^{-1}a_{1,\theta_0}$ and $b_{2,\theta_0,nT}=\ell_{nT,\theta_0}^{-1}a_{2,\theta_0}$ are the bias terms with their orders $O(1)$. Note that $\hat{\theta}_{nT}$ has the bias $\frac{1}{T}b_{1,\theta_0,nT}$ and $\frac{1}{n}b_{2,\theta_0,nT}$ with $\frac{n}{T}\rightarrow k$, which implies the bias term is of order $O\left(\text{max}\left(\frac{1}{n},\frac{1}{T}\right)\right).$\footnote{Indeed, a bias term is of order $O\left(\text{max}\left(\frac{1}{n},\frac{1}{T}\right)\right)$ in any cases if $\frac{n}{T}\rightarrow k<\infty, \frac{n}{T}\rightarrow 0,$ or $\frac{n}{T}\rightarrow \infty$.}  A bias-corrected estimator can be defined as
	
	\begin{equation}
		\label{eq9}
		\hat{\theta}_{nT}^1 = \hat{\theta}_{nT}-\frac{\hat{B}_{1,nT}}{T}-\frac{\hat{B}_{2,nT}}{n}, 
	\end{equation}
	
	\noindent where $\hat{B}_{1,nT}=[\ell_{nT,\theta}^{-1} \cdot a_{1,\theta}] \vert  _{\theta=\hat{\theta}_{nT}}$ and $\hat{B}_{2,nT}=[\ell_{nT,\theta}^{-1} \cdot a_{2,\theta}]\vert _{\theta=\hat{\theta}_{nT}}.$ With further assumption, $\hat{\theta}_{nT}^1$ is properly centered.
	
	\begin{assumption}
		$\frac{\partial a_1(\theta)}{\partial \theta}<\infty$ and $\frac{\partial a_2(\theta)}{\partial \theta}<\infty$ in the neighborhood of $\theta_0$.
	\end{assumption}
	
	\begin{proposition}
		Under Assumptions 1 to 8, if $\frac{n}{T^3}\rightarrow 0$ and $\frac{T}{n^3}\rightarrow 0$, then
		
		\begin{equation*}
		\sqrt{nT}(\hat{\theta}^1_{nT}-\theta_0) \xrightarrow{d} N\left(0,\lim\limits_{n,T\rightarrow \infty}\ell^{-1}_{nT,\theta_0}\right).
		\end{equation*}
		
	\end{proposition}
	\textbf{Proof.} See Appendix 1.
	
	\begin{remark}
		The asymptotic distribution of the unbiased score function can be equivalently represented as
		
		\begin{equation*}
		\frac{1}{\sqrt{nT}}\frac{\partial \ln L_L^c(\theta_0)}{\partial \theta}-\frac{\Delta_L}{\sqrt{nT}} \xrightarrow{d} N\left(0,\lim\limits_{n,T\rightarrow \infty}\ell_{nT,\theta_0}\right).
		\end{equation*}
		
		\noindent Since $\Delta_L=(n-1)a_{1,\theta_0}+Ta_{2,\theta_0}$, it is equivalent to
		\begin{equation*}
		\frac{1}{\sqrt{nT}}\frac{\partial \ln L_L^c(\theta_0)}{\partial \theta}-\sqrt{\frac{n}{T}}a_{1,\theta_0}-\sqrt{\frac{T}{n}}a_{2,\theta_0} \xrightarrow{d} N\left(0,\lim\limits_{n,T\rightarrow \infty}\ell_{nT,\theta_0}\right).
		\end{equation*}
		
		\noindent I denote $\sqrt{\frac{n}{T}}a_{1,\theta_0}$ by $\Delta_1$ and $\sqrt{\frac{T}{n}}a_{2,\theta_0}$ by $\Delta_2$ whose orders are $O(1)$, i.e.,
		\begin{equation*}
		\frac{1}{\sqrt{nT}}\frac{\partial \ln L_L^c(\theta_0)}{\partial \theta}-\Delta_1-\Delta_2 \xrightarrow{d} N\left(0,\lim\limits_{n,T\rightarrow \infty}\ell_{nT,\theta_0}\right).
		\end{equation*}
		
		\indent There are three cases: (i) If $\frac{n}{T}\rightarrow k<\infty$, $\Delta_1$ and $\Delta_2$ do not vanish and so $\frac{\partial \ln L_L^c(\theta_0)}{\partial \theta}$ is not centered around zero. (ii) If $\frac{n}{T} \rightarrow 0$, the score function has a degenerating distribution as $\frac{1}{T}\frac{\partial \ln L_L^c(\theta_0)}{\partial \theta}-\sqrt{\frac{n}{T}}\Delta_2 \xrightarrow{p} 0$. (iii) If $\frac{n}{T} \rightarrow \infty$, the score function again has a degenerating distribution as $\frac{1}{n}\frac{\partial \ln L_L^c(\theta_0)}{\partial \theta}-\sqrt{\frac{T}{n}}\Delta_1 \xrightarrow{p} 0.$ The RS test statistic requires a non-degenerating distribution and so I assume $\frac{n}{T}\rightarrow k<\infty.$
	\end{remark}

	\section{The Robust Rao's Score Tests for Endogeneity of Spatial Weights Matrices}
	Consider the following partition of the parameter vector $\theta = (\delta', \eta', \omega')'$, where $\omega=(\beta',\phi_2',\alpha',\sigma_{\xi}^2)'$. In this partition, $(\delta',\eta')'$ represent the parameter vector of interest, while $\omega$ is pure nuisance parameters. From equation $\eqref{eq4}$, I may design a test for endogeneity of $W_{nt}$ by testing if $\delta=0.$ The null hypothesis then can be stated as $H_0: \delta_0 = 0$ versus the alternative hypothesis as $H_1: \delta_0 \neq 0$. Let $I(\theta_0)=\lim\limits_{n,T \rightarrow \infty} \ell_{nT,\theta_0}$ and its estimator $I(\theta)=-\frac{1}{nT}\frac{\partial^2 lnL_L^c(\theta)}{\partial \theta \partial \theta'}.$ Let $L_{\Psi}(\theta)=\frac{1}{nT}\frac{\partial lnL_L^c(\theta)}{\partial \Psi}$ and $L_{\Psi\Psi}(\theta)=\frac{1}{nT}\frac{\partial^2 lnL_L^c(\theta)}{\partial \Psi \partial \Psi'},$ where $\Psi \in\{\delta,\eta,\omega\}.$ I consider the following partition of $I(\theta)$: 
	\begin{equation*} I(\theta_0)=\frac{1}{nT}E\left(-\frac{\partial^2 lnL_L^c(\theta_0)}{\partial \theta \partial \theta'}\right)=
	\begin{bmatrix}
	I_{\delta\delta}(\theta_0) & I_{\delta \eta}(\theta_0) & I_{\delta \omega}(\theta_0) \\
	I_{\eta \delta}(\theta_0) & I_{\eta\eta}(\theta_0) & I_{\eta \omega}(\theta_0) \\
	I_{\omega \delta}(\theta_0) & I_{\omega \eta}(\theta_0) & I_{\omega \omega}(\theta_0)
	\end{bmatrix}
	\end{equation*}
	
	\noindent and the partition of the bias term $\Delta_L(\theta)$:
	
	\begin{equation*}\Delta_L(\theta)=
	\begin{bmatrix}
	\Delta_{L,\delta}(\theta) \\
	\Delta_{L,\eta}(\theta) \\
	\Delta_{L,\omega}(\theta)\\
	\end{bmatrix}
	\end{equation*}
	
	\noindent such that
	
	\begin{equation*}
	\frac{\Delta_L(\theta)}{\sqrt{nT}}=\Delta_{1}(\theta)+\Delta_{2}(\theta),
	\end{equation*}
	
	\noindent as shown in Remark 1. I denote $I\equiv I(\theta_0)$ and let $\tilde{\theta}=(0',0',\tilde{\omega}')'$ be the restricted MLE when the joint null $H_0^{\delta,\eta}: \delta_0=0, \hspace{1mm}\eta_0=0$ holds. I first consider the case of $H_0^{\delta}: \delta_0=0$ when $H_0^{\eta}: \eta_0=0$ holds. Since $L_{\delta}(\theta)$ is not centered around zero as in Proposition 1, I introduce the \textit{bias-corrected score function}. As derived in the Appendix 3 (Proposition 4 proof), the first-order Taylor expansion of $L_\delta(\tilde{\theta})$ gives the following equation:
	
	\begin{equation*}
	\sqrt{nT}L_\delta(\tilde{\theta})=\sqrt{nT}L_\delta(\theta_0)-I_{\delta \omega}I_{\omega \omega}^{-1}\sqrt{nT}L_\omega(\theta_0),
	\end{equation*}
	
	\noindent which implies
	\begin{equation*}
	\sqrt{nT}L_\delta(\tilde{\theta})\xrightarrow{d} N\left((\Delta_{1,\delta}(\theta_0)+\Delta_{2,\delta}(\theta_0))-I_{\delta \omega}I_{\omega\omega}^{-1}(\Delta_{1,\omega}(\theta_0)+\Delta_{2,\omega}(\theta_0)),I_{\delta \cdot \omega}\right)_,
	\end{equation*}
	
	\noindent where $I_{\delta \cdot \omega}:=I_{\delta\delta}-I_{\delta\omega}I_{\omega \omega}^{-1}I_{\omega\delta}$. \textit{The bias-corrected score function} $C_\delta(\tilde{\theta})$ is therefore of the form
	\begin{equation*}
	C_\delta(\tilde{\theta}):=L_\delta(\tilde{\theta})-\frac{1}{\sqrt{nT}}(\Delta_{1,\delta}(\tilde{\theta})+\Delta_{2,\delta}(\tilde{\theta}))+\frac{1}{\sqrt{nT}}I_{\delta\omega}(\tilde{\theta})I_{\omega\omega}^{-1}(\tilde{\theta})(\Delta_{1,\omega}(\tilde{\theta})+\Delta_{2,\omega}(\tilde{\theta})),
	\end{equation*}
	
	\noindent where the asymptotic distribution of $C_\delta(\tilde{\theta})$ is centered around zero. The standard RS test statistic is
	\begin{equation}
		\label{eq10}
		RS_{\delta}(\tilde{\theta})=nT C_{\delta}'(\tilde{\theta})[I_{\delta \cdot \omega}(\tilde{\theta})]^{-1}C_\delta(\tilde{\theta}) \xrightarrow{d} \chi^2_{p}(0). 
	\end{equation}
	
	\noindent Now I investigate the asymptotic distribution of $RS_\delta(\tilde{\theta})$ under the sequences of local alternatives of $H_a^\delta: \delta_0=\zeta/\sqrt{nT}$ and $H_a^\eta: \lambda_0=\nu/\sqrt{nT}$, where $\zeta$ and $\nu$ are bounded constant vectors. From the first-order Taylor expansion of $L_\delta(\tilde{\theta})$, 
	\begin{equation}
		\label{eq11}
		\sqrt{nT}L_\delta(\tilde{\theta})=[I_{p}, -I_{\delta\omega}I_{\omega\omega}^{-1}]
		\times \begin{bmatrix}
		\sqrt{nT}L_\delta(\theta_0) \\
		\sqrt{nT}L_\omega(\theta_0) \\
		\end{bmatrix} -[I_{\delta\omega}I_{\omega\omega}^{-1}I_{\omega\eta}-I_{\delta\eta}]\nu
		-[I_{\delta\omega}I_{\omega\omega}^{-1}I_{\omega\delta}-I_{\delta\delta}]\zeta+o_p(1). 
	\end{equation}
	
	\noindent By Proposition 1,
	\begin{equation}
		\label{eq12}
		\begin{bmatrix} \sqrt{nT}L_{\delta}(\theta_0) \\
		\sqrt{nT}L_{\omega}(\theta_0) \\
		\end{bmatrix} - \begin{bmatrix}
		\Delta_{1,\delta}(\theta_0)+\Delta_{2,\delta}(\theta_0) \\
		\Delta_{1,\omega}(\theta_0)+\Delta_{2,\omega}(\theta_0) \\
		\end{bmatrix} \xrightarrow{d} N\left(0, \begin{bmatrix} 
		I_{\delta\delta} & I_{\delta\omega} \\
		I_{\omega\delta} & I_{\omega\omega}\\
		\end{bmatrix} \right). 
	\end{equation}
	
	\noindent Then, $\eqref{eq11}$ and $\eqref{eq12}$ imply that
	\begin{equation*}
	\sqrt{nT}C_\delta(\tilde{\theta})\xrightarrow{d} N(I_{\delta \cdot \omega}\zeta+I_{\delta\eta \cdot \omega}\nu,I_{\delta \cdot \omega}),
	\end{equation*}
	
	\noindent where $I_{\delta\eta \cdot \omega}=I_{\delta \eta}-I_{\delta \omega}I_{\omega \omega}^{-1}I_{\omega \eta}.$
	That $\sqrt{nT}C_\delta(\tilde{\theta})$ is asymptotically not centered around zero yields that $RS_\delta$ follows $\chi^2$ distribution with non-zero noncentrality parameter. This is a problem to be resolved because the test statistic leads to over-rejection of the null hypothesis  (Davidson and Mackinnon, 1987; Saikkonen, 1989). A robust version of RS test can be constructed by adjusting $C_\delta(\tilde{\theta})$ so that the RS test statistic is centered around zero: see Bera and Yoon (1993); Dogan, Taspinar, and Bera (2018); Bera, Dogan, and Taspinar (2018, 2019); Bera, Dogan, Taspinar, Leiluo (2019); Bera, Bilias, Yoon, Taspinar, and Dogan (2020). The asymptotic behavior of the standard and robust RS test statistics under $H_0^\delta$ and $H_a^\delta$ are provided in the following Proposition.
	
	\begin{proposition} Under the stated assumptions and $\frac{n}{T}\rightarrow k<\infty$, the following results hold.
		\begin{enumerate}
			\item Under $H_a^\delta$ and $H_a^\eta$,
			\begin{equation*}
			RS_\delta(\tilde{\theta}) \xrightarrow{d} \chi^2_{p}(\varphi_1),
			\end{equation*}
			
			\noindent where $\varphi_1 = \zeta'I_{\delta \cdot \omega}\zeta+2\zeta'I_{\delta\eta \cdot \omega}\nu+\nu'I_{\delta\eta \cdot \omega}I_{\delta\cdot\omega}^{-1}I_{\delta\eta \cdot \omega}\nu$ is the non-centrality parameter.
			
			\item Under $H_0^\delta: \delta_0=0$ and irrespective of whether $H_0^\eta$ or $H_a^\eta$ holds, the distribution of the robust test $RS_\delta^*(\tilde{\theta})$ is given by 
			\vspace{2mm}
			\begin{equation*}
			RS_\delta^*(\tilde{\theta})=nT C_\delta^*(\tilde{\theta})[I_{\delta \cdot \omega}(\tilde{\theta})-I_{\delta \eta \cdot \omega}(\tilde{\theta})I_{\eta \cdot \omega}^{-1}(\tilde{\theta})I_{\delta\eta \cdot \omega}'(\tilde{\theta})]^{-1}C_\delta^*(\tilde{\theta}) \xrightarrow{d} \chi^2_{p}(0),
			\end{equation*}
			
			\noindent where $C_\delta^*(\tilde{\theta})=[C_\delta(\tilde{\theta})-I_{\delta\eta \cdot \omega}(\tilde{\theta})I_{\eta \cdot \omega}^{-1}(\tilde{\theta})C_\eta(\tilde{\theta})]$ is the \textit{adjusted} bias-corrected score function.
			
			\item Under $H_a^\delta$ and irrespective of whether $H_0^\eta$ or $H_a^\eta$ holds, 
			\begin{equation*}
			RS_\delta^*(\tilde{\theta}) \xrightarrow{d} \chi^2_{p}(\varphi_2),
			\end{equation*}
			\noindent where $\varphi_2=\zeta'(I_{\delta\cdot\omega}-I_{\delta\eta\cdot\omega}I_{\eta\cdot\omega}^{-1}I_{\delta\eta\cdot\omega}')\zeta$ is the non-centrality parameter.
		\end{enumerate}
	\end{proposition}
	\textbf{Proof.} See Appendix 3. \\
	
	\noindent This result indicates that $RS_\delta^*(\tilde{\theta})$ is a robust test since it fixes the over-rejection rate of the null hypothesis. That is, it gives asymptotically correct size with its asymptotic null distribution being centered chi-square distribution under the sequence of alterantives $H_a^\eta: \eta_0=\nu/\sqrt{nT}$. Also note that under $H_a^\delta$ and $H_0^\eta$, the result shows $RS_\delta^*(\tilde{\theta}) \xrightarrow{d} \chi^2_{p}(\varphi_2)$ and $RS_\delta(\tilde{\theta}) \xrightarrow{d} \chi^2_{p}(\varphi_1)$ with $\varphi_1-\varphi_2 \geq 0$, indicating $RS_\delta^*(\tilde{\theta})$ has less asymptotic power than $RS_\delta(\tilde{\theta})$ when there is no local misspecification, i.e., when $\eta_0=0.$ That is, one pays the premium to have less power if no presence of local misspecification is found. 
	
	A different approach on testing endogeneity of $W$ could be implemented using the Conditional Lagrange Multiplier (CLM) test following the framework of Qu and Lee (2015) and Cheng and Fei Lee (2017). Let $\hat{\theta}=\underset{\theta: \; \delta=0}{\text{argmax}}\ln L_L^c(\theta)$ be the restricted ML estimator under $H_0^\delta$. Then a valid CLM is formulated as
	\begin{equation*}
		LM_{C}(\hat{\theta})=nTC^{'}_\delta(\hat{\theta})[I_{\delta \cdot \Psi}(\hat{\theta})]^{-1}C_\delta(\hat{\theta}),
	\end{equation*}
	\noindent where $\Psi=(\lambda,\phi_1^{'},\phi_2^{'},\alpha',\sigma_\xi^2)'$. Note that $LM(\hat{\theta})$ is asymptotically central chi-squared under $H_0^\delta$ and has the same form of the non-robust $RS_\delta(\theta)$ test at $\hat{\theta}$. Notably, the robust RS test is \textit{computationally efficient} in the sense that it does not require for $\eta=(\lambda,\gamma,\rho)'$ to be estimated,
	while CLM requires the restricted ML estimators obtained under $H_0: \delta_0=0$, i.e., $(\lambda, \phi_1', \phi_2', \alpha', \sigma_\xi^{2'})$ need to be estimated. A comparison over elapsed times between two tests is reported in Section 6. 
	
	\section{The Test Statistics}
	Now I explicitly derive the robust RS test statistics for the employed model (Qu, Lee, and Yu, 2017) using Proposition 4. One may have the following hypotheses: 
	
	1. $H_0^\delta: \delta_0=0$ and $H_0^\eta: \eta_0=0.$ \\
	\indent 2. $H_a^\delta: \delta_0=\zeta/\sqrt{nT}$ and $H_a^\eta: \eta_0=\nu/\sqrt{nT}$. 
	
	\noindent The first joint null hypothesis tests if $W_{nt}$ is endogenous when there is no local misspecification. I then consider the asymptotic distribution of the test statistic under the alternative hypothesis of the endogeneity parameter $\delta$ and local presence of misspecification in $\eta$. 
	
	Recall that the restricted MLE is denoted by $\tilde{\theta}=(0',0',\omega')'$ under the joint null hypothesis, where $\omega=(\beta',\phi_2',\alpha',\sigma_\xi^2)'$. It is highlighted that the robust RS test has computational advantage in the sense that it only requires $\tilde{\theta}$ and one does not need to estimate other parameters under the alternative hypothesis. I firstly consider the asymptotic distribution of the test statistic under the joint null hypothesis $H_0^\delta$ and  $H_0^\eta$. The concentrated log-likelihood function at $\tilde{\theta}$, $\ln L_L^c(\tilde{\theta})$, turns down to
	\begin{align*}
	\ln L_L^c(\tilde{\theta})&=-\frac{nT}{2}\ln2\pi-\frac{nT}{2}\ln\widetilde{\sigma_\xi^2}-\frac{nT}{2}\ln\vert \widetilde{\Sigma_\epsilon}\vert  \\
	&\hspace{4mm}-\frac{1}{2}vec'(Z_L-K_L\widetilde{\Phi_2})(\widetilde{\Sigma_\epsilon}^{-1}\otimes J_L)vec(Z_L-K_L\widetilde{\Phi_2}) \\
	&\hspace{4mm}-\frac{1}{2\widetilde{\sigma_\xi^2}}(Y_L-X_{1L}\tilde{\beta})'J_L(Y_L-X_{1L}\tilde{\beta}),
	\end{align*}
	\noindent which is decomposed into two unrelated components as $\ln L_L^c(\tilde{\theta})=\ln L_L^{C1}(\tilde{\theta})+\ln L_L^{C2}(\tilde{\theta})$, where
	\begin{align*}
	\ln L_L^{C1}(\tilde{\theta})&=-\frac{nT}{2}\ln2\pi-\frac{nT}{2}\ln\widetilde{\sigma_\xi^2}-\frac{1}{2\widetilde{\sigma_\xi^2}}(Y_L-X_{1L}\tilde{\beta})'J_L(Y_L-X_{1L}\tilde{\beta}), \\
	\ln L_L^{C2}(\tilde{\theta})&=-\frac{nT}{2}\ln\vert \widetilde{\Sigma_\epsilon}\vert -\frac{1}{2}vec'(Z_L-K_L\widetilde{\Phi_2})(\widetilde{\Sigma_\epsilon}^{-1}\otimes J_L)vec(Z_L-K_L\widetilde{\Phi_2}).
	\end{align*}
	\noindent Using the bias-corrected ML estimator in \eqref{eq9}, one obtains the restricted ML estimators and the residuals, $\xi_L(\tilde{\theta})=Y_L-X_{1L}\tilde{\beta}$ and $\varepsilon_L(\tilde{\theta})=Z_L-K_L\widetilde{\Phi_2}$. I now provide explicit expressions for $RS_\delta(\tilde{\theta})$ and $RS_\delta^*(\tilde{\theta})$. The test statistic requires the score functions with respect to $\delta$ and $\eta$. The score functions evaluated at $\tilde{\theta}$ are given as
	\begin{align*}
		L_\delta(\tilde{\theta})&=\frac{1}{nT}\frac{\partial \ln L_L^c(\tilde{\theta})}{\partial \delta}=\frac{1}{nT\widetilde{\sigma_\xi^2}}\left(\varepsilon'_L(\tilde{\theta})J_L\xi_L(\tilde{\theta})\right),\\
		L_\eta(\tilde{\theta})&=\frac{1}{nT}\frac{\partial \ln L_L^c(\tilde{\theta})}{\partial \eta}=\frac{1}{nT\widetilde{\sigma_\xi^2}}[\xi'_L(\tilde{\theta})J_LW_{1L}Y_L-tr(W_{1L}), Y_{L,-1}'J_L\xi_L(\tilde{\theta}), \\
		&\hspace{53mm}\text{} (W_{L,-1}Y_{L,-1})'J_L\xi_L(\tilde{\theta})]'.
	\end{align*}

	\noindent From the information matrix in equation (8) as well as the second order score functions using Corollary 1, one can find the consistent estimator for the information matrix as
	\begin{equation*}
	I(\tilde{\theta})=\frac{1}{nT\widetilde{\sigma_\xi^2}}\begin{bmatrix*} 
	J_{\lambda\lambda}(\tilde{\theta}) & * & * & * & * & *\\
	J_{\phi_1\lambda}(\tilde{\theta}) & J_{\phi_1\phi_1}(\tilde{\theta}) & * & * & * & * \\
	J_{\delta \lambda}(\tilde{\theta}) & 0_{p \times (k_1+2)} & J_{\delta\delta}(\tilde{\theta}) & * & * & * \\
	0_{k_{\phi_2} \times 1} & 0_{k_{\phi_2} \times (k_1+2)} & 0_{k_{\phi_2}\times p} & J_{\phi_2 \phi_2}(\tilde{\theta}) & * & * \\
	J_{\sigma_\xi^2\lambda}(\tilde{\theta}) & 0_{1 \times (k_1+2)} & 0_{1 \times p} & 0_{1 \times k_{\phi_2}} & J_{\sigma_\xi^2 \sigma_\xi^2}(\tilde{\theta}) & * \\
	0_{J \times 1} & 0_{J \times (k_1+2)} & 0_{J \times p} & 0_{J \times k_{\phi_2}} & 0_{J \times 1} & J_{\alpha\alpha}(\tilde{\theta}) \end{bmatrix*},
	\end{equation*}
	
	\noindent where
	\begin{align*}
	J_{\lambda\lambda}(\tilde{\theta})&=(W_{1L}Y_L)'J_LW_{1L}Y_L+\widetilde{\sigma_\xi^2}tr(W_{1L}^2), \\
	J_{\phi_1\lambda}(\tilde{\theta})&=R_L'J_LW_{1L}Y_L,\\
	J_{\phi_1\phi_1}(\tilde{\theta})&=R_L'J_LR_L, \\
	J_{\delta\lambda}(\tilde{\theta})&=\varepsilon'_L(\tilde{\theta})J_LW_{1L}Y_L,\\ J_{\delta\delta}(\tilde{\theta})&=\varepsilon_L'(\tilde{\theta})J_L\varepsilon_L'(\tilde{\theta}), \\ J_{\phi_2\phi_2}(\tilde{\theta})&=(\widetilde{\sigma_\xi^2}\widetilde{\Sigma_\varepsilon}^{-1})\otimes (K_L'J_LK_L), \\
	J_{\sigma_\xi^2\lambda}(\tilde{\theta})&=tr(W_{1L}), \\ J_{\sigma_\xi^2\sigma_\xi^2}(\tilde{\theta})&=\frac{1}{\widetilde{\sigma_\xi^2}}\left(\frac{nT}{2}-T-n+1\right), \\
	J_{\alpha\alpha,kj}(\tilde{\theta})&=\frac{nT}{2}\widetilde{\sigma_\xi^2}tr\left(\widetilde{\Sigma_\varepsilon}^{-1}\frac{\partial \Sigma_\varepsilon}{\partial \alpha_k}(\tilde{\theta})\widetilde{\Sigma_\varepsilon}^{-1}\frac{\partial \Sigma_\varepsilon}{\partial \alpha_j}(\tilde{\theta})\right) \hspace{2mm} \text{for} \hspace{1mm} k,j=1,\dots,J.
	\end{align*}	
	\noindent Remark that the estimator for the information matrix at $\tilde{\theta}$, $I(\tilde{\theta})$, forms a block diagonal matrix with respect to $(\phi_2',\alpha)'$. Thus one may regard $\omega=(\beta',\sigma_\xi^2)$. Hence the estimators for the information matrices necessary for computing the test statistic under the joint null $H_0^\delta$ and $H_0^\eta$ are

	\begin{align*}
	I_{\delta\delta}(\tilde{\theta})&=J_{\delta\delta}(\tilde{\theta})/(nT\widetilde{\sigma_\xi^2})=\varepsilon_L'(\tilde{\theta})J_L\varepsilon_L'(\tilde{\theta})/(nT\sigma_\xi^2),\\
	I_{\delta\omega}(\tilde{\theta})&=J_{\delta\omega}(\tilde{\theta})/(nT\widetilde{\sigma_\xi^2})=
	\begin{bmatrix}J_{\delta\beta'}(\tilde{\theta}) & J_{\delta\sigma_\xi^2}(\tilde{\theta})\end{bmatrix}/(nT\widetilde{\sigma_\xi^2})\\
	&\hspace{29.7mm}=\begin{bmatrix} 0_{p \times k_1} & 0 \end{bmatrix}/(nT\widetilde{\sigma_\xi^2})=0_{p \times (k_1+1)},\\
	I_{\delta\eta}(\tilde{\theta})&=J_{\delta\eta}(\tilde{\theta})/(nT\widetilde{\sigma_\xi^2})=\begin{bmatrix}J_{\delta\lambda}(\tilde{\theta}) \quad J_{\delta\gamma}(\tilde{\theta}) \quad J_{\delta\lambda}(\tilde{\theta})\end{bmatrix}/(nT\widetilde{\sigma_\xi^2})\\
	&\hspace{29.5mm}=\begin{bmatrix} \varepsilon'_L(\tilde{\theta})J_LW_{1L}Y_L & 0_{p \times 1} & 0_{p \times 1}\end{bmatrix}/(nT\widetilde{\sigma_\xi^2}),\\ 
	I_{\omega\omega}(\tilde{\theta})&=J_{\omega\omega}(\tilde{\theta})/(nT\widetilde{\sigma_\xi^2})=
	\begin{bmatrix} J_{\beta\beta'}(\tilde{\theta}) & J_{\beta\sigma_\xi^2}(\tilde{\theta}) \\
	J_{\sigma_\xi^2 \beta'}(\tilde{\theta}) & J_{\sigma_\xi^2 \sigma_\xi^2}(\tilde{\theta})\end{bmatrix}/(nT\widetilde{\sigma_\xi^2})\\
	&\hspace{30mm}=\begin{bmatrix} (R_L'J_LR)_{L(3:k_1+2,3:k_1+2)} & 0_{k_1 \times 1} \\
	0_{1 \times k_1} & \frac{1}{\widetilde{\sigma_\xi^2}} \left(\frac{nT}{2}-T-n+1\right) \end{bmatrix}/(nT\widetilde{\sigma_\xi^2}),\\				
	I_{\eta\omega}(\tilde{\theta})&=J_{\eta\omega}(\tilde{\theta})/(nT\widetilde{\sigma_\xi^2})
	=\begin{bmatrix} J_{\lambda\beta'}(\tilde{\theta}) & J_{\lambda\sigma_\xi^2}(\tilde{\theta}) \\
	J_{\gamma\beta'}(\tilde{\theta}) & J_{\gamma\sigma_\xi^2}(\tilde{\theta})\\
	J_{\rho\beta'}(\tilde{\theta}) & J_{\rho\sigma_\xi^2}(\tilde{\theta}) \end{bmatrix}/(nT\widetilde{\sigma_\xi^2})\\
	&\hspace{30mm}=\begin{bmatrix} (R_L'J_LW_{1L}Y_{1L}\tilde{\beta}_{(3:k_1+2,1)})' & tr(W_{1L}) \\
	(R_L'J_LR_L)_{(1,3:k_1+2)} & 0 \\
	(R_L'J_LR_L)_{(2,3:k_1+2)} & 0 \end{bmatrix}/(nT\widetilde{\sigma_\xi^2}),\\
	I_{\eta\eta}(\tilde{\theta})&=J_{\eta\eta}(\tilde{\theta})/(nT\widetilde{\sigma_\xi^2})=
	\begin{bmatrix} J_{\lambda\lambda}(\tilde{\theta}) & * & * \\
	J_{\gamma\lambda}(\tilde{\theta}) & J_{\gamma\gamma}(\tilde{\theta}) & * \\
	J_{\rho\lambda}(\tilde{\theta}) & J_{\rho\gamma}(\tilde{\theta}) & J_{\rho\rho}(\tilde{\theta}) \end{bmatrix}/(nT\widetilde{\sigma_\xi^2})
	\end{align*}
	
	\noindent where $(A)_{(i,j:k)}$ represents the entries in a matrix $A$ located in $i^{th}$ row, $j^{th}$ to $k^{th}$ columns.
	
	The bias terms evaluated at $\tilde{\theta}$ are given as
	\begin{equation*}
	\Delta_L(\tilde{\theta})=(n-1)a_{1,\theta_0}(\tilde{\theta})+Ta_{2,\theta_0}(\tilde{\theta}),
	\end{equation*}
	
	\noindent where
	\begin{equation*} a_{1,\theta_0}(\tilde{\theta})=
	\begin{bmatrix} 
	-\frac{1}{n-1}tr\left(W_{1L}(\tilde{\theta})\frac{1}{T}1_T1_T'\otimes J_n\right) \\
	\frac{1}{n-1}\left(tr[W_{2L}(\tilde{\theta})\left(\frac{1}{T}1_T1_T'\otimes J_n\right))],tr[W_{3L}(\tilde{\theta})\left(\frac{1}{T}1_T1_T'\otimes J_n\right)],0\right)' \\
	0_{p\times 1} \\
	-vec\left(\frac{1}{T}\left(\sum\limits_{t=1}^{T-1}\sum\limits_{h=1}^{T-t}\tilde{\kappa}'^{(h-1)}\right)\widetilde{\Sigma_{\varepsilon}}^{-1}\right) \\
	-\frac{1}{2\widetilde{\sigma_\xi^2}} \\
	-\frac{1}{2}\frac{\partial ln\vert\Sigma_{\varepsilon 0}\vert}{\partial \alpha}\vert_{\tilde{\alpha}} \\
	\end{bmatrix}
	\end{equation*}
	\noindent and
	\begin{equation*} a_{2,\theta_0}=
	\begin{bmatrix}
	-\frac{1}{T}tr\left(W_{1L}(\tilde{\theta})\left(I_T \otimes \frac{1}{n}1_n1_n'\right)\right) \\
	0_{(k_1+2)\times 1} \\
	0_{p\times 1} \\
	0_{p(p+k_2)\times 1} \\
	-\frac{1}{2\widetilde{\sigma_\xi^2}} \\
	-\frac{1}{2}\frac{\partial ln\vert\Sigma_{\varepsilon 0}\vert}{\partial \alpha}\vert_{\tilde{\alpha}} 
	\end{bmatrix},
	\end{equation*}
	
	\noindent with the order of $\theta=(\lambda,\phi_1,\delta,\phi_2,\sigma_\xi^2,\alpha)$. Denoting $\Delta_1=\sqrt{\frac{n}{T}}a_{1,\theta_0}$ and $\Delta_2=\sqrt{\frac{T}{n}}a_{2,\theta_0}$, the bias terms necessary for computing the test statistics under the joint null are
	\begin{align*}
	\Delta_{1,\delta}(\tilde{\theta})&=0_{p \times 1},\\
	\Delta_{2,\delta}(\tilde{\theta})&=0_{p \times 1},\\
	\Delta_{1,\omega}(\tilde{\theta})&=\begin{bmatrix} \Delta_{1,\beta}(\tilde{\theta}) \\ \Delta_{1,\sigma_\xi^2}(\tilde{\theta}) \end{bmatrix}=\begin{bmatrix} 0_{k_1\times 1} \\ \sqrt{\frac{n}{T}}\left(-\frac{1}{2\widetilde{\sigma_\xi^2}}\right)\end{bmatrix},\\
	\Delta_{2,\omega}(\tilde{\theta})&=\begin{bmatrix} \Delta_{2,\beta}(\tilde{\theta}) \\ \Delta_{2,\sigma_\xi^2}(\tilde{\theta})\end{bmatrix}\\
	&=\begin{bmatrix} 0_{k_1 \times 1} \\ \sqrt{\frac{T}{n}}\left(-\frac{1}{2\widetilde{\sigma_\xi^2}}\right))\end{bmatrix},\\
	\Delta_{1,\eta}(\tilde{\theta})&=\begin{bmatrix} \Delta_{1,\lambda}(\tilde{\theta}) \\ \Delta_{1,\gamma}(\tilde{\theta}) \\ \Delta_{1,\rho}(\tilde{\theta}) \end{bmatrix}\\
	&=\frac{1}{\sqrt{nT}}\begin{bmatrix} -tr(W_{1L}\left(\frac{1}{T}1_T1_T' \otimes J_n\right)) \\ tr(W_{2L}\left(\frac{1}{T}1_T1_T' \otimes J_n\right)) \\ tr(W_{3L}\left(\frac{1}{T}1_T1_T' \otimes J_n\right))\end{bmatrix},\\
	\Delta_{2,\eta}(\tilde{\theta})&=\begin{bmatrix} \Delta_{2,\lambda}(\tilde{\theta}) \\ \Delta_{2,\gamma}(\tilde{\theta}) \\ \Delta_{2,\rho}(\tilde{\theta})\end{bmatrix}\\
	&=\frac{1}{\sqrt{nT}}\begin{bmatrix} -tr(W_{1L}\left(I_T \otimes \frac{1}{n}1_n1_n'\right))\\0\\0\end{bmatrix}.
	\end{align*}
	\noindent Given the above quantities, one can obtain the bias-corrected score functions of $C_\delta(\tilde{\theta})$ and $C_\eta(\tilde{\theta})$.\footnote{The explicit forms are provided in Appendix 2.} Then the standard RS test statistic is
	\begin{align}
		\label{eq13}
		\begin{split}
		RS_\delta(\tilde{\theta})&=nTC_\delta'(\tilde{\theta})I^{-1}_{\delta\cdot \omega}(\tilde{\theta})C_\delta(\tilde{\theta}),
		\end{split}
	\end{align} 
	
	\noindent while the robust RS test statistic in Proposition 4 is provided as
	\begin{align*}
	RS_\delta^*(\tilde{\theta})&=nTC_\delta^{*'}(\tilde{\theta})[I_{\delta\cdot \omega}(\tilde{\theta})-K(\tilde{\theta})I'_{\delta\eta \cdot \omega}(\tilde{\theta})]^{-1}C_\delta^*(\tilde{\theta}),
	\end{align*}
	\noindent where $K(\tilde{\theta})=I_{\delta\eta \cdot \omega}(\tilde{\theta})I_{\eta \cdot \omega}^{-1}(\tilde{\theta})$ and $C_\delta^*(\tilde{\theta})=C_\delta(\tilde{\theta})-K(\tilde{\theta})C_\eta(\tilde{\theta})$, respectively. One may observe that two corrections are made in the score function and its variance with $K(\tilde{\theta})$ in order to manage to rectify over-rejection of the null hypothesis due to the local presence of misspecification in another testing parameter. Note that if $K(\tilde{\theta})=0$ or there is no local misspecification, then the robust RS test is simply the standard RS test. 
	
	Now the quantities to compute the CLM test (Qu and Lee, 2015; Cheng and Fei Lee, 2017) are provided. Let $\theta=(\delta',\Psi')'$ where $\Psi=(\lambda,\phi_1^{'},\phi_2^{'},\sigma_\xi^{2},\alpha^{'})'$ and $\hat{\theta}=\underset{\theta:\;\delta=0}{\text{argmax}}\ln L_L^c(\theta)$ be the restricted ML estimator under $H_0^\delta$. Remark that the estimator for the information matrix at $\hat{\theta}$, $I(\hat{\theta})$ forms a block diagonal matrix with respect to $(\phi_2^{'},\alpha')'$, where one thus may regard $\Psi=(\lambda,\phi_1^{'},\sigma_\xi^2)'$. The concentrated log-likelihood under $H_0^\delta$ is then
	\begin{align*}
		\ln L_L^c(\hat{\theta})&=-\frac{nT}{2}\ln 2\pi + \ln \vert S_L(\hat{\eta}) \vert -\frac{nT}{2}\ln \widehat{\sigma_\xi^2} -\frac{nT}{2}\ln \vert \widehat{\Sigma_\varepsilon} \vert \\
						 &\quad -\frac{1}{2}vec'(Z_L-K_L\widehat{\Phi_2)}\cdot(\widehat{\Sigma_\varepsilon}^{-1}\otimes J_L)\cdot vec(Z_L-K_L\widehat{\Phi_2}) \\
						 &\quad -\frac{1}{2\sigma_\xi^2}[S_L(\hat{\eta})Y_L-X_{1L}\hat{\beta}-\ell_0(\hat{\gamma},\hat{\rho})]'\cdot J_L \cdot [S_L(\hat{\eta})Y_L-X_{1L}\hat{\beta}-\ell_0(\hat{\gamma},\hat{\rho})].
	\end{align*}
	The centered score function at $\hat{\theta}$ is
	\begin{equation*}
		C_{\delta}(\hat{\theta})=L_{\delta}(\hat{\theta})-\frac{1}{\sqrt{nT}}(\Delta_{1,\delta}(\hat{\theta})+\Delta_{2,\delta}(\hat{\theta}))+\frac{1}{\sqrt{nT}}I_{\delta\Psi}(\hat{\theta})I_{\Psi\Psi}^{-1}(\hat{\theta})(\Delta_{1,\Psi}(\hat{\theta}),\Delta_{2,\Psi}(\hat{\theta})),
	\end{equation*}
	\noindent with
	\begin{align*}
		&\Delta_{1,\Psi}(\hat{\theta})=\begin{bmatrix}
									  	\Delta_{1,\lambda}(\hat{\theta}) \\ \Delta_{1,\phi_1}(\hat{\theta}) \\ \Delta_{1,\sigma_\xi^2}(\hat{\theta})
									  \end{bmatrix}
								  	=\sqrt{\frac{n}{T}}
								  	  \begin{bmatrix}
									  	  -\frac{1}{n-1}tr\left(\hat{G}_{1L}\left(\frac{1}{T}1_T1_T^{'}\otimes J-n\right)\right) \\
									  	  \frac{1}{n-1}\left[tr\left(\hat{G}_{2L}\left(\frac{1}{T}1_T1_T^{'} \otimes J_n\right)\right),tr\left(\hat{G}_{3L}\left(\frac{1}{T}1_T1_T^{'}\otimes J_n\right)\right),0\right]' \\
									  	  -\frac{1}{2\widehat{\sigma_\xi^2}}
								  	  \end{bmatrix},\\
		&\Delta_{2,\Psi}(\hat{\theta})=\begin{bmatrix}
											\Delta_{2,\lambda}(\hat{\theta})\\
											\Delta_{2,\phi_1}(\hat{\theta})\\
											\Delta_{2,\sigma_\xi^2}(\hat{\theta})
								   \end{bmatrix}
								      =\sqrt{\frac{T}{n}}
								       \begin{bmatrix}
								      		-\frac{1}{T}tr\left(\hat{G}_{1L}\left(I_T\otimes \frac{1}{n}1_n1_n{'}\right)\right)	\\
								      		0_{(k_1+2)\times 1} \\
								      		-\frac{1}{2\widehat{\sigma_\xi^2}}
								       \end{bmatrix}, \\
		&I_{\delta\delta}(\hat{\theta})=J_{\delta\delta}(\hat{\theta})/(nT\widehat{\sigma_\xi^2})=\varepsilon_L^{'}(\hat{\theta})J_L\varepsilon_L^{'}(\hat{\theta})/(nT\widehat{\sigma_\xi^2}),\\					    
		&I_{\delta\Psi}(\hat{\theta})=[J_{\delta\lambda}(\hat{\theta}) \;\; J_{\delta\phi_1}(\hat{\theta}) \;\; J_{\delta\sigma_\xi^2}(\hat{\theta})]/(nT\widehat{\sigma_\xi^2})=[\varepsilon_L^{'}J_L(W_{1L}Y_L)\;\; 0_{p \times (k_1+2)}\;\; 0_{p \times 1}]/(nT\widehat{\sigma_\xi^2}),\\
		&I_{\Psi\Psi}(\hat{\theta})=
			\begin{bmatrix}
				J_{\lambda\lambda}(\hat{\theta}) & J_{\lambda\phi_1}(\hat{\theta}) & J_{\lambda\sigma_\xi^2}(\hat{\theta}) \\
				J_{\phi_1\lambda}(\hat{\theta}) & J_{\phi_1\phi_1}(\hat{\theta}) & J_{\phi_1\sigma_\xi^2}(\hat{\theta}) \\
				J_{\sigma_\xi^2 \lambda}(\hat{\theta}) & J_{\sigma_\xi^2 \phi_1}(\hat{\theta}) & J_{\sigma_\xi^2 \sigma_\xi^2}(\hat{\theta})
			\end{bmatrix}/(nT\widehat{\sigma_\xi^2})\\
		&\quad\quad\quad = \begin{bmatrix}
						(W_{1L}Y_L)^{'}J_L(W_{1L}Y_L)+\widehat{\sigma_\xi^2}tr(\hat{G}_{1L}^{2}) & R_L^{'}J_L(W_{1L}Y_L) & tr(\hat{G}_{1L}) \\
						* & R_L^{'}J_LR_L & 0_{(k_1+2)\times 1} \\
						* & * & \frac{1}{\widehat{\sigma_\xi^2}}\left(\frac{nT}{2}-T-n+1 \right)
					  \end{bmatrix} /(nT\widehat{\sigma_\xi^2}). \\
	\end{align*} 

	\section{Monte Carlo study and Empirical illustration}
	\subsection{Monte Carlo simulation}
	I run a Monte Carlo simulation to explore the finite properties of the test statistic. It is conducted for 1,000 times and the sample size for individuals and time periods are set by four cases where $n$ is relatively large than $T$ or vice versa, followed by (approximately) doubled sample size, respectively: (i) $n=100, T=10$; (ii) $n=196, T=20$; (iii) $n=9, T=100$; and (iv) $n=16, T=200$. The type I error is set as 0.05 and the dimension of $Z$ is set as $p=1$. The parameters for the main and auxiliary equations follow the setup in Qu, Lee, and Yu (2017): $\beta_0=1, \kappa_0=0.2, \Gamma_0=0.3, \alpha_0=1$, respectively. Data are generated by (1) and (2). The magnitude of local misspecification in $\eta=(\lambda,\gamma,\rho)$ increases by 0.05 from 0 to 0.3 and $\delta_0$ is set from 0 to 0.2, which also increases by 0.05.  The initial values of $Y_0, Z_0$ and the deterministic explanatory variables of $X_{1L}, X_{2L}$ are generated from independent standard normal distributions. To generate the two-way fixed effects, data generated from the multivariate normal distribution of nonzero correlation, 0.5 for both fixed effects, with $X_{iL}$ are averaged over time for each spatial unit ($C_{iL0}$) or over spatial unit for each time ($\alpha_{iL0}$), $i=1,2$. The joint pdf of the disturbance terms, $(v_{it},\epsilon_{it})$, follows bivariate normal distribution of $N\left(\begin{pmatrix} 0 \\ 0 \end{pmatrix}, \begin{pmatrix} 1 & \delta_0 \\ \delta_0 & 1 \end{pmatrix}\right).$ In summary, we have
	\begin{align*}
	\theta_0&=(\lambda_0,\gamma_0,\rho_0,\beta_0,\delta_0,\kappa_0,\Gamma_0,\sigma_{\xi 0}^2,\alpha_0)'\\
	&=(\lambda_0,\gamma_0,\rho_0,1,\delta_0,0.2,0.3,1-\delta_0^2,1)'.
	\end{align*}

	I also follow the setup for the spatial weight matrices $W_{nt}$ as in Qu, Lee, and Yu (2017), which is generated by Hadamard product of the physical contiguity and the economic distance: $W_{nt}=W_n^d \circ W_{nt}^e$, i.e. $w_{ij,nt}=w_{ij}^dw_{ij,nt}^e$ and is row normalized afterwards. I explore two kinds of contiguities of Queen \& Rook matrices for $W_n^d$ which characterize different adjacency: Queen allows contiguities over edges \& corners, whereas Rook only allows edges. The economic distance $W_{nt}^e$ is generated by $w_{ij,nt}^e=1/|z_{it}-z_{jt}|$ if $i \neq j$ or zero otherwise. Table 1 summarizes our notations on the test statistics. The notation of A:B:C represents that the parameter starts from A to C, increased by B. I will estimate the size and power of the test statistics under the local misspecification in $\eta_0=(\lambda_0,\gamma_0,\rho_0)$ when $n$ is larger than $T$: $(n,T)=(100,10)$ or the reverse: $(n,T)=(9,100)$, and finally when they get doubled in each case: $(n,T)=(196,20)$ and $(n,T)=(16,200)$. Here, $n$ is set as a square number so that the Queen \& Rook matrices could be constructed accordingly. 
	
		\begin{table}[h!] \tiny
		\caption{Summary of the test statistics}	
		\centering
		\addtolength{\leftskip} {-2cm}
		\addtolength{\rightskip}{-2cm}
		\begin{tabular}{c|ccc|cc}\hline
			\multirow{2}{*}{Null hypothesis} & \multicolumn{3}{c|}{Local misspecification} & \multicolumn{2}{c}{Test statistic}  \\ \cline{2-6}
			& $\lambda_0$ & $\gamma_0$ & $\rho_0$ & Standard RS & Robust RS \\ \hline
			\multirow{4}{*}{$H_0: \delta_0=0$} & 0 & 0 & 0 & \begin{tabular}[c]{@{}c@{}}$RS(\delta_0,\lambda_0)$ with $\lambda_0=0$\\ or $RS(\delta_0,\gamma_0)$ with $\gamma_0=0$\\ or $RS(\delta_0,\rho_0)$ with $\rho_0=0$ \end{tabular}   & \begin{tabular}[c]{@{}c@{}} $RS^*(\delta_0,\lambda_0)$ with $\lambda_0=0$\\ or $RS^*(\delta_0,\gamma_0)$ with $\gamma_0=0$ \\ or $RS^*(\delta,\rho_0)$ with $\rho_0=0$\end{tabular}   \\ 
			& 0.05:0.05:0.3 & 0 & 0 & $RS(\delta_0,\lambda_0)$ & $RS^*(\delta_0,\lambda_0)$  \\ 
			& 0 & 0.05:0.05:0.3 & 0 & $RS(\delta_0,\gamma_0)$ & $RS^*(\delta_0,\gamma_0)$ \\
			& 0 & 0 & 0.05:0.05:0.3 & $RS(\delta_0,\rho_0)$ & $RS^*(\delta_0,\rho_0)$ \\ \hline
			\multirow{4}{*}{$H_A: \delta_0=0.05:0.05:0.2$} & 0 & 0 & 0  & \begin{tabular}[c]{@{}c@{}} $RS(\delta_0=c,\lambda_0)$ with $\lambda_0=0$\\ or $RS(\delta_0=c,\gamma_0)$ with $\gamma_0=0$\\ or $RS(\delta_0=c,\rho_0)$ with $\rho_0=0$ \end{tabular} & \begin{tabular}[c]{@{}c@{}}$RS^*(\delta_0=c,\lambda_0)$ with $\lambda_0=0$\\or $RS^*(\delta_0=c,\gamma_0)$ with $\gamma_0=0$\\  or $RS^*(\delta_0=c,\rho_0)$ with $\rho_0=0$\end{tabular} \\
			& 0.05:0.05:0.3 & 0 & 0 & $RS(\delta_0=c,\lambda_0)$ & $RS^*(\delta_0=c,\lambda_0)$ \\
			& 0 & 0.05:0.05:0.3 & 0 & $RS(\delta_0=c,\gamma_0)$  & $RS^*(\delta_0=c,\gamma_0)$ \\
			& 0 & 0 & 0.05:0.05:0.3 & $RS(\delta_0=c,\rho_0)$ & $RS^*(\delta_0=c,\rho_0)$ \\ \hline
			\multicolumn{1}{l}{} & \multicolumn{1}{l}{} & \multicolumn{1}{l}{} & \multicolumn{1}{l}{} & \multicolumn{1}{l}{} & \multicolumn{1}{l}{}                                 \\
			\multicolumn{1}{l}{}               & \multicolumn{1}{l}{} & \multicolumn{1}{l}{} & \multicolumn{1}{l}{} & \multicolumn{1}{l}{}                                & \multicolumn{1}{l}{}                                 \\
			\multicolumn{1}{l}{}               & \multicolumn{1}{l}{} & \multicolumn{1}{l}{} & \multicolumn{1}{l}{} & \multicolumn{1}{l}{}                                & \multicolumn{1}{l}{}                                
		\end{tabular}
	\end{table}

	The size of the test statistics is summarized in Figure 1 and Table 2. First consider the case when $n$ is larger than $T$. For example, when $(n,T)=(100,10)$, the size of the standard RS generally increases as the magnitude of local misspecification in $\eta_0=(\lambda_0,\gamma_0,\rho_0)$ increases, both in Queen and Rook $W_{nt}^{d'}$s. In particular, the largest increase in size is found in the presence of local misspecification in the contemporaneous spatial dependence ($\lambda_{0}$), whereas those in the spatial time dependence ($\rho_{0}$), and dependence over time ($\gamma_{0}$) slightly increase the size. Results even get clearer when both $n$ and $T$ get doubled. When $(n,T)=(196,20)$, the size explodes under the presence of local misspecification in $\lambda_{0}$. Meanwhile, one may find that the robust RS test stay settled around 0.05 and its performance improves as $n$ and $T$ increase. 
	
	Now consider the other case when $T$ is larger than $n$. For example, consider $(n,T)=(9,100)$. Results show that increase in size is remarkable for the presence of local misspecification in $\lambda_{0}$ for the standard RS test as well as decrease in size for the robust RS test for the presence of local misspecification in $\gamma_{0}$, both in Queen and Rook $W_{nt}^{d'}$s. However, the size of the Robust RS test get settled as $n$ and $T$ get doubled, $(n,T)=(16,200)$, while the standard RS test shows even larger increase in size under the local misspecification of $\lambda_{0}$. This result again supports a nice performance of the robust RS test as $n$ and $T$ increase.

		\begin{landscape}
		\begin{table}[h!]
			\centering
			\caption{Size of the test statistics}
			\scriptsize
			\begin{tabular}{ccccccccccccccc} \hline
				\multicolumn{3}{c}{Local   misspecification}                                     & \multicolumn{6}{c}{$n$ larger than $T$}                                                                                                                            & \multicolumn{6}{c}{$T$ larger than $n$}                                                                                                                            \\ \hline
				\multirow{2}{*}{$\lambda$} & \multirow{2}{*}{$\gamma$} & \multirow{2}{*}{$\rho$} & \multirow{2}{*}{$n$} & \multirow{2}{*}{$T$} & \multicolumn{2}{c}{$W_{nt}^{\text{Queen}}\circ W_{nt}^e$} & \multicolumn{2}{c}{$W_{nt}^{\text{Rook}}\circ W_{nt}^e$} & \multirow{2}{*}{$n$} & \multirow{2}{*}{$T$} & \multicolumn{2}{c}{$W_{nt}^{\text{Queen}}\circ W_{nt}^e$} & \multicolumn{2}{c}{$W_{nt}^{\text{Rook}}\circ W_{nt}^e$} \\ \cline{6-9}\cline{12-15}
				&                           &                         &                      &                      & $RS$                        & $RS^*$                      & $RS$                       & $RS^*$                      &                      &                      & $RS$                        & $RS^*$                      & $RS$                       & $RS^*$                      \\ \hline
				0                          & 0                         & 0                       & 100                  & 10                   & 0.036                       & 0.038                       & 0.036                      & 0.035                       & 9                    & 100                  & 0.050                       & 0.051                       & 0.048                      & 0.048                       \\
				0.05                       & 0                         & 0                       &                      &                      & 0.040                       & 0.039                       & 0.041                      & 0.036                       &                      &                      & 0.053                       & 0.051                       & 0.048                      & 0.048                       \\
				0.1                        & 0                         & 0                       &                      &                      & 0.047                       & 0.039                       & 0.045                      & 0.036                       &                      &                      & 0.058                       & 0.053                       & 0.051                      & 0.050                       \\
				0.15                       & 0                         & 0                       &                      &                      & 0.053                       & 0.037                       & 0.047                      & 0.036                       &                      &                      & 0.062                       & 0.052                       & 0.059                      & 0.049                       \\
				0.2                        & 0                         & 0                       &                      &                      & 0.057                       & 0.036                       & 0.050                      & 0.032                       &                      &                      & 0.066                       & 0.052                       & 0.057                      & 0.048                       \\
				0.25                       & 0                         & 0                       &                      &                      & 0.064                       & 0.033                       & 0.056                      & 0.028                       &                      &                      & 0.067                       & 0.051                       & 0.058                      & 0.045                       \\
				0.3                        & 0                         & 0                       &                      &                      & 0.066                       & 0.029                       & 0.056                      & 0.023                       &                      &                      & 0.072                       & 0.050                       & 0.060                      & 0.044                       \\
				0                          & 0.05                      & 0                       &                      &                      & 0.039                       & 0.039                       & 0.039                      & 0.036                       &                      &                      & 0.048                       & 0.051                       & 0.048                      & 0.048                       \\
				0                          & 0.1                       & 0                       &                      &                      & 0.039                       & 0.038                       & 0.039                      & 0.036                       &                      &                      & 0.050                       & 0.051                       & 0.050                      & 0.047                       \\
				0                          & 0.15                      & 0                       &                      &                      & 0.039                       & 0.039                       & 0.039                      & 0.035                       &                      &                      & 0.049                       & 0.048                       & 0.049                      & 0.045                       \\
				0                          & 0.2                       & 0                       &                      &                      & 0.041                       & 0.037                       & 0.040                      & 0.035                       &                      &                      & 0.051                       & 0.042                       & 0.051                      & 0.044                       \\
				0                          & 0.25                      & 0                       &                      &                      & 0.041                       & 0.037                       & 0.041                      & 0.033                       &                      &                      & 0.055                       & 0.034                       & 0.054                      & 0.038                       \\
				0                          & 0.3                       & 0                       &                      &                      & 0.042                       & 0.034                       & 0.042                      & 0.031                       &                      &                      & 0.052                       & 0.028                       & 0.052                      & 0.028                       \\
				0                          & 0                         & 0.05                    &                      &                      & 0.036                       & 0.038                       & 0.036                      & 0.035                       &                      &                      & 0.049                       & 0.051                       & 0.049                      & 0.048                       \\
				0                          & 0                         & 0.1                     &                      &                      & 0.033                       & 0.038                       & 0.035                      & 0.035                       &                      &                      & 0.048                       & 0.051                       & 0.051                      & 0.048                       \\
				0                          & 0                         & 0.15                    &                      &                      & 0.031                       & 0.037                       & 0.038                      & 0.033                       &                      &                      & 0.049                       & 0.051                       & 0.047                      & 0.048                       \\
				0                          & 0                         & 0.2                     &                      &                      & 0.033                       & 0.036                       & 0.037                      & 0.033                       &                      &                      & 0.048                       & 0.050                       & 0.046                      & 0.046                       \\
				0                          & 0                         & 0.25                    &                      &                      & 0.035                       & 0.034                       & 0.038                      & 0.031                       &                      &                      & 0.046                       & 0.048                       & 0.040                      & 0.045                       \\
				0                          & 0                         & 0.3                     &                      &                      & 0.038                       & 0.034                       & 0.038                      & 0.029                       &                      &                      & 0.044                       & 0.045                       & 0.043                      & 0.043                       \\ \hline
				0                          & 0                         & 0                       & 196                  & 20                   & 0.047                       & 0.048                       & 0.047                      & 0.048                       & 16                   & 200                  & 0.056                       & 0.058                       & 0.056                      & 0.057                       \\
				0.05                       & 0                         & 0                       &                      &                      & 0.052                       & 0.048                       & 0.050                      & 0.048                       &                      &                      & 0.059                       & 0.058                       & 0.058                      & 0.057                       \\
				0.1                        & 0                         & 0                       &                      &                      & 0.055                       & 0.048                       & 0.057                      & 0.046                       &                      &                      & 0.060                       & 0.058                       & 0.059                      & 0.058                       \\
				0.15                       & 0                         & 0                       &                      &                      & 0.056                       & 0.044                       & 0.063                      & 0.042                       &                      &                      & 0.061                       & 0.057                       & 0.061                      & 0.056                       \\
				0.2                        & 0                         & 0                       &                      &                      & 0.065                       & 0.042                       & 0.064                      & 0.042                       &                      &                      & 0.064                       & 0.055                       & 0.062                      & 0.053                       \\
				0.25                       & 0                         & 0                       &                      &                      & 0.074                       & 0.042                       & 0.071                      & 0.039                       &                      &                      & 0.072                       & 0.053                       & 0.064                      & 0.046                       \\
				0.3                        & 0                         & 0                       &                      &                      & 0.082                       & 0.039                       & 0.080                      & 0.035                       &                      &                      & 0.082                       & 0.048                       & 0.066                      & 0.043                       \\
				0                          & 0.05                      & 0                       &                      &                      & 0.048                       & 0.049                       & 0.048                      & 0.048                       &                      &                      & 0.053                       & 0.058                       & 0.053                      & 0.055                       \\
				0                          & 0.1                       & 0                       &                      &                      & 0.049                       & 0.048                       & 0.049                      & 0.047                       &                      &                      & 0.056                       & 0.055                       & 0.056                      & 0.053                       \\
				0                          & 0.15                      & 0                       &                      &                      & 0.049                       & 0.046                       & 0.049                      & 0.046                       &                      &                      & 0.051                       & 0.053                       & 0.051                      & 0.051                       \\
				0                          & 0.2                       & 0                       &                      &                      & 0.048                       & 0.040                       & 0.048                      & 0.043                       &                      &                      & 0.050                       & 0.047                       & 0.050                      & 0.045                       \\
				0                          & 0.25                      & 0                       &                      &                      & 0.041                       & 0.039                       & 0.041                      & 0.038                       &                      &                      & 0.044                       & 0.043                       & 0.044                      & 0.042                       \\
				0                          & 0.3                       & 0                       &                      &                      & 0.042                       & 0.037                       & 0.042                      & 0.037                       &                      &                      & 0.042                       & 0.038                       & 0.042                      & 0.038                       \\
				0                          & 0                         & 0.05                    &                      &                      & 0.046                       & 0.048                       & 0.047                      & 0.048                       &                      &                      & 0.056                       & 0.058                       & 0.057                      & 0.056                       \\
				0                          & 0                         & 0.1                     &                      &                      & 0.049                       & 0.048                       & 0.046                      & 0.047                       &                      &                      & 0.054                       & 0.057                       & 0.057                      & 0.056                       \\
				0                          & 0                         & 0.15                    &                      &                      & 0.045                       & 0.047                       & 0.047                      & 0.046                       &                      &                      & 0.055                       & 0.054                       & 0.058                      & 0.054                       \\
				0                          & 0                         & 0.2                     &                      &                      & 0.044                       & 0.046                       & 0.046                      & 0.045                       &                      &                      & 0.054                       & 0.054                       & 0.057                      & 0.052                       \\
				0                          & 0                         & 0.25                    &                      &                      & 0.045                       & 0.045                       & 0.044                      & 0.042                       &                      &                      & 0.058                       & 0.053                       & 0.056                      & 0.048                       \\
				0                          & 0                         & 0.3                     &                      &                      & 0.045                       & 0.041                       & 0.046                      & 0.040                       &                      &                      & 0.058                       & 0.050                       & 0.058                      & 0.046        \\ \hline              
			\end{tabular}
		\end{table}
	\end{landscape}
	
	\newpage
	\begin{figure}[h!]
		\tiny
		\caption{Size of the test statistics ($n$ larger than $T$)}
		\begin{subfigure}{.5\textwidth}
		\caption{$W_{nt}^{\text{Queen}} \circ W_{nt}^e \;$}
		\centering
		\includegraphics[width=1\textwidth]{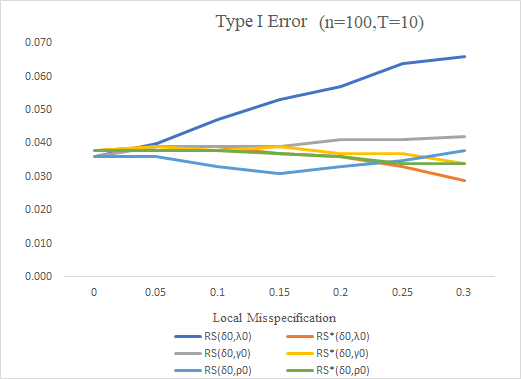}  
		\label{fig:sub-first}
		\end{subfigure}
		\begin{subfigure}{.5\textwidth}
		\centering
		\caption{$W_{nt}^{\text{Rook}} \circ W_{nt}^e$}
		\includegraphics[width=1\textwidth]{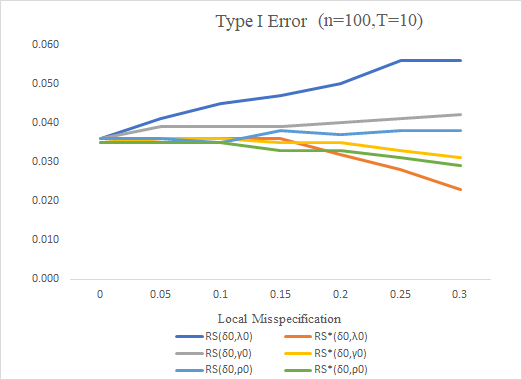}  
		\label{fig:sub-second}
		\end{subfigure}
		\label{fig:fig}
		\medspace
		\begin{subfigure}{.5\textwidth}
		\centering
		\includegraphics[width=1\textwidth]{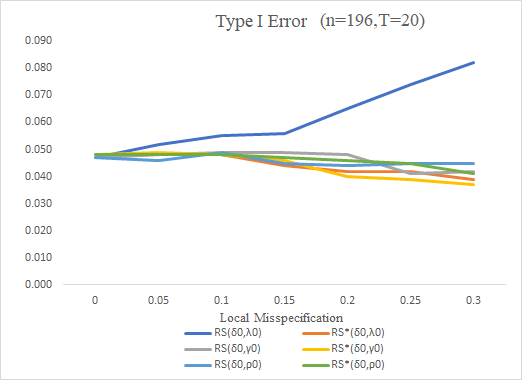}  
		\label{fig:sub-first}
		\end{subfigure}
		\begin{subfigure}{.5\textwidth}
		\centering
		\includegraphics[width=1\textwidth]{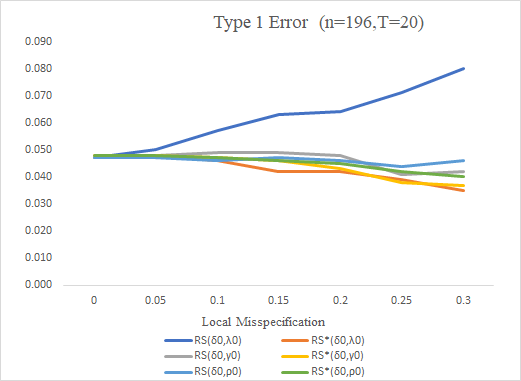}  
		\label{fig:sub-second}
		\end{subfigure}
		\label{fig:fig}
		\begin{subfigure}{.5\textwidth}
			\centering
			\includegraphics[width=1\textwidth]{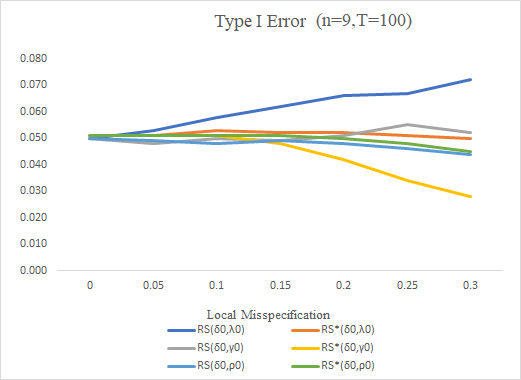}  
			\label{fig:sub-first}
		\end{subfigure}
		\begin{subfigure}{.5\textwidth}
			\centering
			\includegraphics[width=1\textwidth]{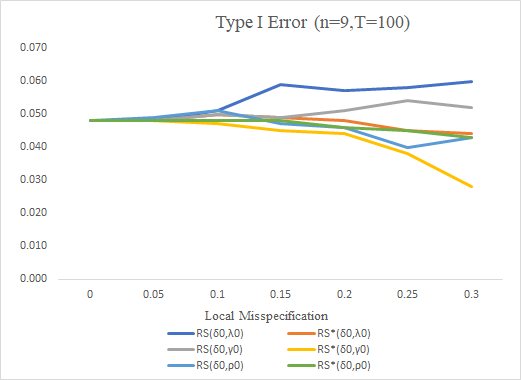}  
			\label{fig:sub-second}
		\end{subfigure}
		\label{fig:fig}
		\medspace
		\begin{subfigure}{.5\textwidth}
			\centering
			\includegraphics[width=1\textwidth]{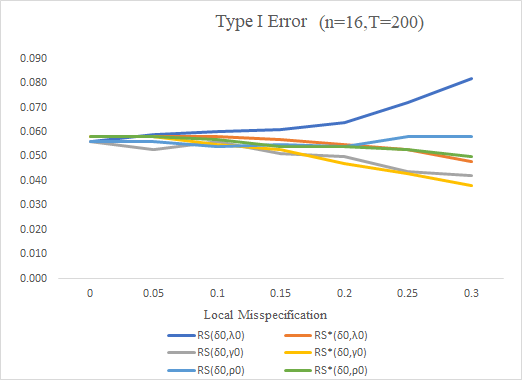}  
			\label{fig:sub-first}
		\end{subfigure}
		\begin{subfigure}{.5\textwidth}
			\centering
			\includegraphics[width=1\textwidth]{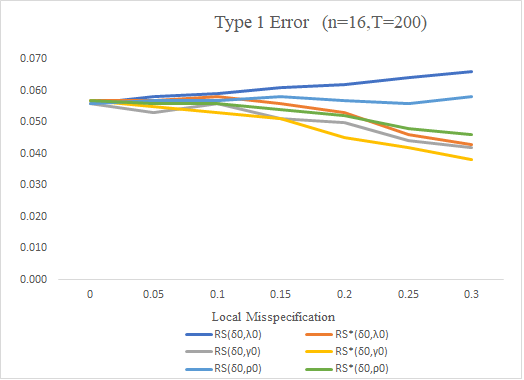}  
			\label{fig:sub-second}
		\end{subfigure}
		\label{fig:fig}
	\end{figure}

	\newpage
	The power of the test statistics is summarized in Figure 2 \& 3 and in Table 4 \& 5. When there is no local misspecification, i.e., $\eta_{0}=(0,0,0)$, the power of the standard RS test is higher than that of the robust RS test, indicating the premium one pays of losing little power when adjusting the standard RS test so that one rectifies the over-rejection of the null hypothesis. For $(n,T)=(100,10)$ or $(n,T)=(9,100)$, the power sharply increases to 1 as $\delta_0$ increases to 0.2. As shown in Figure 2 \& 3 and Table 4, the power of the robust RS is as good as the standard RS without local misspecification for all $\delta_0 \in \{0.05,\;0.1,\;0.15,\;0.2\}$, i.e., one does not lose power in using the robust RS. Similar results are found when $n$ and $T$ get doubled, i.e., $(n,T)=(196,20)$ or $(n,T)=(16,200)$ in Table 5.

		Now the numerical values of the robust RS test- and the conditional LM test statistics are presented in Table 6 \& 7, showing similar results as $n$ and $T$ increase. The robust RS test is expected to have \textit{less} cost in computation since it only requires the ML estimation of the simple fixed-effects model for panel data, whereas the conditional LM test requires that of the SDPD model. As expected, the elapsed time (in seconds) for the Robust RS test is less than that for Conditional LM test (Table 3).
	
	\begin{table}[h!]
		\centering
		\caption{Elasped seconds for Robust RS test versus Conditional LM test}
		\begin{tabular}{cccccc} \hline
			&     & \multicolumn{2}{c}{$W_{nt}^{\text{Queen}}\circ   W_{nt}^e$} & \multicolumn{2}{c}{$W_{nt}^{\text{Rook}}\circ   W_{nt}^e$} \\\cline{3-6}
			n   & T   & $RS^*$                       & $LM_{C}$                     & $RS^*$                      & $LM_{C}$                     \\ \hline
			100 & 10  & 0.511                        & 1.695                        & 0.865                       & 1.192                        \\
			9   & 100 & 1.457                        & 2.067                        & 1.480                       & 1.688                        \\
			196 & 20  & 19.6481                      & 42.0474                      & 17.655                      & 37.0344                      \\
			16  & 200  & 11.1693                      & 25.3624                      & 10.8938                     & 23.0711         \\ \hline            
		\end{tabular}
	\end{table}

		\begin{landscape}
		\begin{table}[h!]
			\centering
			\caption{Power of test statistics ($n$ larger than $T$)}
			\scriptsize
			\begin{tabular}{ccccccccccccccccccccc} \hline
				&                      & \multicolumn{3}{c}{Local misspecification}                                       & \multicolumn{4}{c}{$\delta_0=0.05$}                                                                                  & \multicolumn{4}{c}{$\delta_0=0.1$}                                                                                   & \multicolumn{4}{c}{$\delta_0=0.15$}                                                                                  & \multicolumn{4}{c}{$\delta_0=0.2$}                                                                                   \\ \cline{3-21}
				\multirow{2}{*}{$n$} & \multirow{2}{*}{$T$} & \multirow{2}{*}{$\lambda$} & \multirow{2}{*}{$\gamma$} & \multirow{2}{*}{$\rho$} & \multicolumn{2}{c}{$W_{nt}^{\text{Queen}}\circ W_{nt}^e$} & \multicolumn{2}{c}{$W_{nt}^{\text{Rook}}\circ W_{nt}^e$} & \multicolumn{2}{c}{$W_{nt}^{\text{Queen}}\circ W_{nt}^e$} & \multicolumn{2}{c}{$W_{nt}^{\text{Rook}}\circ W_{nt}^e$} & \multicolumn{2}{c}{$W_{nt}^{\text{Queen}}\circ W_{nt}^e$} & \multicolumn{2}{c}{$W_{nt}^{\text{Rook}}\circ W_{nt}^e$} & \multicolumn{2}{c}{$W_{nt}^{\text{Queen}}\circ W_{nt}^e$} & \multicolumn{2}{c}{$W_{nt}^{\text{Rook}}\circ W_{nt}^e$} \\ \cline{6-21}
				&                      &                            &                           &                         & $RS$                        & $RS^*$                      & $RS$                       & $RS^*$                      & $RS$                        & $RS^*$                      & $RS$                        & $RS^*$                     & $RS$                        & $RS^*$                      & $RS$                       & $RS^*$                      & $RS$                        & $RS^*$                      & $RS$                        & $RS^*$                     \\ \hline
				100                  & 10                   & 0.05                       & 0                         & 0                       & 0.291                       & 0.286                       & 0.292                      & 0.289                       & 0.842                       & 0.833                       & 0.847                       & 0.831                      & 0.997                       & 0.997                       & 0.997                      & 0.998                       & 1.000                       & 1.000                       & 1.000                       & 1.000                      \\
				&                      & 0.1                        & 0                         & 0                       & 0.295                       & 0.285                       & 0.297                      & 0.289                       & 0.851                       & 0.831                       & 0.850                       & 0.829                      & 0.997                       & 0.997                       & 0.997                      & 0.998                       & 1.000                       & 1.000                       & 1.000                       & 1.000                      \\
				&                      & 0.15                       & 0                         & 0                       & 0.295                       & 0.281                       & 0.297                      & 0.281                       & 0.851                       & 0.829                       & 0.850                       & 0.820                      & 0.997                       & 0.997                       & 0.997                      & 0.996                       & 1.000                       & 1.000                       & 1.000                       & 1.000                      \\
				&                      & 0.2                        & 0                         & 0                       & 0.290                       & 0.276                       & 0.284                      & 0.268                       & 0.848                       & 0.826                       & 0.848                       & 0.817                      & 0.997                       & 0.996                       & 0.997                      & 0.996                       & 1.000                       & 1.000                       & 1.000                       & 1.000                      \\
				&                      & 0.25                       & 0                         & 0                       & 0.284                       & 0.268                       & 0.280                      & 0.262                       & 0.843                       & 0.818                       & 0.840                       & 0.808                      & 0.997                       & 0.994                       & 0.997                      & 0.995                       & 1.000                       & 1.000                       & 1.000                       & 1.000                      \\
				&                      & 0.3                        & 0                         & 0                       & 0.284                       & 0.254                       & 0.279                      & 0.252                       & 0.840                       & 0.808                       & 0.833                       & 0.800                      & 0.997                       & 0.993                       & 0.997                      & 0.994                       & 1.000                       & 1.000                       & 1.000                       & 1.000                      \\
				&                      & 0                          & 0.05                      & 0                       & 0.293                       & 0.289                       & 0.293                      & 0.294                       & 0.835                       & 0.833                       & 0.835                       & 0.834                      & 0.997                       & 0.997                       & 0.997                      & 0.998                       & 1.000                       & 1.000                       & 1.000                       & 1.000                      \\
				&                      & 0                          & 0.1                       & 0                       & 0.288                       & 0.286                       & 0.288                      & 0.289                       & 0.832                       & 0.833                       & 0.832                       & 0.831                      & 0.998                       & 0.997                       & 0.998                      & 0.998                       & 1.000                       & 1.000                       & 1.000                       & 1.000                      \\
				&                      & 0                          & 0.15                      & 0                       & 0.288                       & 0.283                       & 0.288                      & 0.283                       & 0.821                       & 0.828                       & 0.821                       & 0.825                      & 0.997                       & 0.997                       & 0.997                      & 0.998                       & 1.000                       & 1.000                       & 1.000                       & 1.000                      \\
				&                      & 0                          & 0.2                       & 0                       & 0.283                       & 0.277                       & 0.283                      & 0.277                       & 0.811                       & 0.822                       & 0.811                       & 0.819                      & 0.996                       & 0.997                       & 0.996                      & 0.997                       & 1.000                       & 1.000                       & 1.000                       & 1.000                      \\
				&                      & 0                          & 0.25                      & 0                       & 0.269                       & 0.272                       & 0.269                      & 0.272                       & 0.790                       & 0.811                       & 0.790                       & 0.809                      & 0.994                       & 0.996                       & 0.994                      & 0.997                       & 1.000                       & 1.000                       & 1.000                       & 1.000                      \\
				&                      & 0                          & 0.3                       & 0                       & 0.263                       & 0.263                       & 0.263                      & 0.266                       & 0.774                       & 0.798                       & 0.774                       & 0.801                      & 0.990                       & 0.995                       & 0.990                      & 0.997                       & 1.000                       & 1.000                       & 1.000                       & 1.000                      \\
				&                      & 0                          & 0                         & 0.05                    & 0.294                       & 0.287                       & 0.292                      & 0.290                       & 0.839                       & 0.833                       & 0.836                       & 0.832                      & 0.997                       & 0.997                       & 0.997                      & 0.998                       & 1.000                       & 1.000                       & 1.000                       & 1.000                      \\
				&                      & 0                          & 0                         & 0.1                     & 0.294                       & 0.285                       & 0.291                      & 0.286                       & 0.838                       & 0.832                       & 0.838                       & 0.828                      & 0.997                       & 0.997                       & 0.997                      & 0.997                       & 1.000                       & 1.000                       & 1.000                       & 1.000                      \\
				&                      & 0                          & 0                         & 0.15                    & 0.294                       & 0.285                       & 0.291                      & 0.284                       & 0.839                       & 0.832                       & 0.832                       & 0.823                      & 0.997                       & 0.997                       & 0.997                      & 0.997                       & 1.000                       & 1.000                       & 1.000                       & 1.000                      \\
				&                      & 0                          & 0                         & 0.2                     & 0.294                       & 0.283                       & 0.293                      & 0.280                       & 0.832                       & 0.828                       & 0.826                       & 0.821                      & 0.997                       & 0.997                       & 0.997                      & 0.997                       & 1.000                       & 1.000                       & 1.000                       & 1.000                      \\
				&                      & 0                          & 0                         & 0.25                    & 0.293                       & 0.276                       & 0.286                      & 0.277                       & 0.825                       & 0.825                       & 0.815                       & 0.815                      & 0.996                       & 0.997                       & 0.995                      & 0.997                       & 1.000                       & 1.000                       & 1.000                       & 1.000                      \\
				&                      & 0                          & 0                         & 0.3                     & 0.285                       & 0.270                       & 0.272                      & 0.268                       & 0.813                       & 0.819                       & 0.801                       & 0.808                      & 0.996                       & 0.997                       & 0.995                      & 0.996                       & 1.000                       & 1.000                       & 1.000                       & 1.000                      \\ \hline
				196                  & 20                   & 0.05                       & 0                         & 0                       & 0.849                       & 0.840                       & 0.846                      & 0.840                       & 1.000                       & 1.000                       & 1.000                       & 1.000                      & 1.000                       & 1.000                       & 1.000                      & 1.000                       & 1.000                       & 1.000                       & 1.000                       & 1.000                      \\
				&                      & 0.1                        & 0                         & 0                       & 0.861                       & 0.839                       & 0.858                      & 0.838                       & 1.000                       & 1.000                       & 1.000                       & 1.000                      & 1.000                       & 1.000                       & 1.000                      & 1.000                       & 1.000                       & 1.000                       & 1.000                       & 1.000                      \\
				&                      & 0.15                       & 0                         & 0                       & 0.867                       & 0.837                       & 0.859                      & 0.834                       & 1.000                       & 1.000                       & 1.000                       & 1.000                      & 1.000                       & 1.000                       & 1.000                      & 1.000                       & 1.000                       & 1.000                       & 1.000                       & 1.000                      \\
				&                      & 0.2                        & 0                         & 0                       & 0.875                       & 0.832                       & 0.858                      & 0.828                       & 1.000                       & 1.000                       & 1.000                       & 1.000                      & 1.000                       & 1.000                       & 1.000                      & 1.000                       & 1.000                       & 1.000                       & 1.000                       & 1.000                      \\
				&                      & 0.25                       & 0                         & 0                       & 0.880                       & 0.825                       & 0.863                      & 0.822                       & 1.000                       & 1.000                       & 1.000                       & 1.000                      & 1.000                       & 1.000                       & 1.000                      & 1.000                       & 1.000                       & 1.000                       & 1.000                       & 1.000                      \\
				&                      & 0.3                        & 0                         & 0                       & 0.878                       & 0.818                       & 0.864                      & 0.810                       & 1.000                       & 1.000                       & 1.000                       & 1.000                      & 1.000                       & 1.000                       & 1.000                      & 1.000                       & 1.000                       & 1.000                       & 1.000                       & 1.000                      \\
				&                      & 0                          & 0.05                      & 0                       & 0.840                       & 0.840                       & 0.840                      & 0.840                       & 1.000                       & 1.000                       & 1.000                       & 1.000                      & 1.000                       & 1.000                       & 1.000                      & 1.000                       & 1.000                       & 1.000                       & 1.000                       & 1.000                      \\
				&                      & 0                          & 0.1                       & 0                       & 0.845                       & 0.840                       & 0.845                      & 0.841                       & 1.000                       & 1.000                       & 1.000                       & 1.000                      & 1.000                       & 1.000                       & 1.000                      & 1.000                       & 1.000                       & 1.000                       & 1.000                       & 1.000                      \\
				&                      & 0                          & 0.15                      & 0                       & 0.838                       & 0.836                       & 0.838                      & 0.835                       & 1.000                       & 1.000                       & 1.000                       & 1.000                      & 1.000                       & 1.000                       & 1.000                      & 1.000                       & 1.000                       & 1.000                       & 1.000                       & 1.000                      \\
				&                      & 0                          & 0.2                       & 0                       & 0.822                       & 0.830                       & 0.822                      & 0.829                       & 1.000                       & 1.000                       & 1.000                       & 1.000                      & 1.000                       & 1.000                       & 1.000                      & 1.000                       & 1.000                       & 1.000                       & 1.000                       & 1.000                      \\
				&                      & 0                          & 0.25                      & 0                       & 0.809                       & 0.826                       & 0.809                      & 0.824                       & 1.000                       & 1.000                       & 1.000                       & 1.000                      & 1.000                       & 1.000                       & 1.000                      & 1.000                       & 1.000                       & 1.000                       & 1.000                       & 1.000                      \\
				&                      & 0                          & 0.3                       & 0                       & 0.787                       & 0.816                       & 0.787                      & 0.818                       & 1.000                       & 1.000                       & 1.000                       & 1.000                      & 1.000                       & 1.000                       & 1.000                      & 1.000                       & 1.000                       & 1.000                       & 1.000                       & 1.000                      \\
				&                      & 0                          & 0                         & 0.05                    & 0.837                       & 0.841                       & 0.841                      & 0.840                       & 1.000                       & 1.000                       & 1.000                       & 1.000                      & 1.000                       & 1.000                       & 1.000                      & 1.000                       & 1.000                       & 1.000                       & 1.000                       & 1.000                      \\
				&                      & 0                          & 0                         & 0.1                     & 0.837                       & 0.840                       & 0.841                      & 0.840                       & 1.000                       & 1.000                       & 1.000                       & 1.000                      & 1.000                       & 1.000                       & 1.000                      & 1.000                       & 1.000                       & 1.000                       & 1.000                       & 1.000                      \\
				&                      & 0                          & 0                         & 0.15                    & 0.837                       & 0.838                       & 0.839                      & 0.839                       & 1.000                       & 1.000                       & 1.000                       & 1.000                      & 1.000                       & 1.000                       & 1.000                      & 1.000                       & 1.000                       & 1.000                       & 1.000                       & 1.000                      \\
				&                      & 0                          & 0                         & 0.2                     & 0.837                       & 0.837                       & 0.832                      & 0.835                       & 1.000                       & 1.000                       & 1.000                       & 1.000                      & 1.000                       & 1.000                       & 1.000                      & 1.000                       & 1.000                       & 1.000                       & 1.000                       & 1.000                      \\
				&                      & 0                          & 0                         & 0.25                    & 0.833                       & 0.833                       & 0.829                      & 0.831                       & 1.000                       & 1.000                       & 1.000                       & 1.000                      & 1.000                       & 1.000                       & 1.000                      & 1.000                       & 1.000                       & 1.000                       & 1.000                       & 1.000                      \\
				&                      & 0                          & 0                         & 0.3                     & 0.830                       & 0.825                       & 0.825                      & 0.824                       & 1.000                       & 1.000                       & 1.000                       & 1.000                      & 1.000                       & 1.000                       & 1.000                      & 1.000                       & 1.000                       & 1.000                       & 1.000                       & 1.000       \\ \hline              
			\end{tabular}
		\end{table}
	\end{landscape}
	
	\begin{landscape}
		\begin{table}[h!]
			\centering
			\caption{Power of test statistics ($T$ larger than $n$)}
			\scriptsize
			\begin{tabular}{ccccccccccccccccccccc} \hline
				&                      & \multicolumn{3}{c}{Local misspecification}                                       & \multicolumn{4}{c}{$\delta_0=0.05$}                                                                                  & \multicolumn{4}{c}{$\delta_0=0.1$}                                                                                   & \multicolumn{4}{c}{$\delta_0=0.15$}                                                                                  & \multicolumn{4}{c}{$\delta_0=0.2$}                                                                                   \\ \cline{3-21}
				\multirow{2}{*}{$n$} & \multirow{2}{*}{$T$} & \multirow{2}{*}{$\lambda$} & \multirow{2}{*}{$\gamma$} & \multirow{2}{*}{$\rho$} & \multicolumn{2}{c}{$W_{nt}^{\text{Queen}}\circ W_{nt}^e$} & \multicolumn{2}{c}{$W_{nt}^{\text{Rook}}\circ W_{nt}^e$} & \multicolumn{2}{c}{$W_{nt}^{\text{Queen}}\circ W_{nt}^e$} & \multicolumn{2}{c}{$W_{nt}^{\text{Rook}}\circ W_{nt}^e$} & \multicolumn{2}{c}{$W_{nt}^{\text{Queen}}\circ W_{nt}^e$} & \multicolumn{2}{c}{$W_{nt}^{\text{Rook}}\circ W_{nt}^e$} & \multicolumn{2}{c}{$W_{nt}^{\text{Queen}}\circ W_{nt}^e$} & \multicolumn{2}{c}{$W_{nt}^{\text{Rook}}\circ W_{nt}^e$} \\ \cline{6-21}
				&                      &                            &                           &                         & $RS$                        & $RS^*$                      & $RS$                       & $RS^*$                      & $RS$                        & $RS^*$                      & $RS$                        & $RS^*$                     & $RS$                        & $RS^*$                      & $RS$                       & $RS^*$                      & $RS$                        & $RS^*$                      & $RS$                        & $RS^*$                     \\ \hline
				9                    & 100                  & 0.05                       & 0                         & 0                       & 0.312                       & 0.301                       & 0.308                      & 0.300                       & 0.811                       & 0.805                       & 0.805                       & 0.806                      & 0.992                       & 0.990                       & 0.991                      & 0.990                       & 1.000                       & 1.000                       & 1.000                       & 1.000                      \\
				&                      & 0.1                        & 0                         & 0                       & 0.325                       & 0.306                       & 0.315                      & 0.302                       & 0.821                       & 0.809                       & 0.809                       & 0.808                      & 0.993                       & 0.991                       & 0.992                      & 0.990                       & 1.000                       & 1.000                       & 1.000                       & 1.000                      \\
				&                      & 0.15                       & 0                         & 0                       & 0.339                       & 0.309                       & 0.315                      & 0.301                       & 0.824                       & 0.811                       & 0.814                       & 0.806                      & 0.993                       & 0.992                       & 0.993                      & 0.990                       & 1.000                       & 1.000                       & 1.000                       & 1.000                      \\
				&                      & 0.2                        & 0                         & 0                       & 0.345                       & 0.309                       & 0.320                      & 0.294                       & 0.832                       & 0.812                       & 0.814                       & 0.802                      & 0.994                       & 0.992                       & 0.993                      & 0.990                       & 1.000                       & 1.000                       & 1.000                       & 1.000                      \\
				&                      & 0.25                       & 0                         & 0                       & 0.351                       & 0.306                       & 0.326                      & 0.288                       & 0.834                       & 0.810                       & 0.812                       & 0.794                      & 0.994                       & 0.992                       & 0.993                      & 0.990                       & 1.000                       & 1.000                       & 1.000                       & 1.000                      \\
				&                      & 0.3                        & 0                         & 0                       & 0.362                       & 0.303                       & 0.324                      & 0.279                       & 0.836                       & 0.807                       & 0.803                       & 0.786                      & 0.993                       & 0.992                       & 0.992                      & 0.988                       & 1.000                       & 1.000                       & 1.000                       & 1.000                      \\
				&                      & 0                          & 0.05                      & 0                       & 0.297                       & 0.290                       & 0.295                      & 0.289                       & 0.799                       & 0.798                       & 0.800                       & 0.801                      & 0.990                       & 0.989                       & 0.990                      & 0.990                       & 1.000                       & 1.000                       & 1.000                       & 1.000                      \\
				&                      & 0                          & 0.1                       & 0                       & 0.288                       & 0.287                       & 0.288                      & 0.284                       & 0.806                       & 0.794                       & 0.804                       & 0.799                      & 0.987                       & 0.989                       & 0.987                      & 0.989                       & 1.000                       & 1.000                       & 1.000                       & 1.000                      \\
				&                      & 0                          & 0.15                      & 0                       & 0.279                       & 0.277                       & 0.280                      & 0.277                       & 0.798                       & 0.790                       & 0.798                       & 0.791                      & 0.983                       & 0.989                       & 0.983                      & 0.988                       & 1.000                       & 1.000                       & 1.000                       & 1.000                      \\
				&                      & 0                          & 0.2                       & 0                       & 0.272                       & 0.257                       & 0.271                      & 0.266                       & 0.785                       & 0.781                       & 0.785                       & 0.779                      & 0.981                       & 0.984                       & 0.981                      & 0.986                       & 1.000                       & 1.000                       & 1.000                       & 1.000                      \\
				&                      & 0                          & 0.25                      & 0                       & 0.259                       & 0.241                       & 0.259                      & 0.244                       & 0.756                       & 0.767                       & 0.756                       & 0.768                      & 0.974                       & 0.983                       & 0.974                      & 0.985                       & 1.000                       & 1.000                       & 1.000                       & 1.000                      \\
				&                      & 0                          & 0.3                       & 0                       & 0.242                       & 0.218                       & 0.242                      & 0.223                       & 0.736                       & 0.745                       & 0.737                       & 0.750                      & 0.967                       & 0.976                       & 0.968                      & 0.979                       & 1.000                       & 1.000                       & 1.000                       & 1.000                      \\
				&                      & 0                          & 0                         & 0.05                    & 0.299                       & 0.289                       & 0.300                      & 0.289                       & 0.796                       & 0.801                       & 0.798                       & 0.803                      & 0.991                       & 0.989                       & 0.990                      & 0.990                       & 1.000                       & 1.000                       & 1.000                       & 1.000                      \\
				&                      & 0                          & 0                         & 0.1                     & 0.296                       & 0.291                       & 0.294                      & 0.289                       & 0.793                       & 0.798                       & 0.790                       & 0.802                      & 0.991                       & 0.989                       & 0.990                      & 0.989                       & 1.000                       & 1.000                       & 1.000                       & 1.000                      \\
				&                      & 0                          & 0                         & 0.15                    & 0.289                       & 0.288                       & 0.287                      & 0.284                       & 0.793                       & 0.795                       & 0.787                       & 0.797                      & 0.992                       & 0.989                       & 0.990                      & 0.989                       & 1.000                       & 1.000                       & 1.000                       & 1.000                      \\
				&                      & 0                          & 0                         & 0.2                     & 0.277                       & 0.283                       & 0.284                      & 0.279                       & 0.788                       & 0.792                       & 0.776                       & 0.791                      & 0.992                       & 0.988                       & 0.989                      & 0.988                       & 1.000                       & 1.000                       & 1.000                       & 1.000                      \\
				&                      & 0                          & 0                         & 0.25                    & 0.273                       & 0.278                       & 0.276                      & 0.272                       & 0.786                       & 0.788                       & 0.767                       & 0.784                      & 0.989                       & 0.988                       & 0.986                      & 0.987                       & 1.000                       & 1.000                       & 1.000                       & 1.000                      \\
				&                      & 0                          & 0                         & 0.3                     & 0.266                       & 0.273                       & 0.264                      & 0.258                       & 0.779                       & 0.783                       & 0.756                       & 0.778                      & 0.987                       & 0.987                       & 0.984                      & 0.985                       & 1.000                       & 1.000                       & 1.000                       & 1.000                      \\
				16                   & 200                  & 0.05                       & 0                         & 0                       & 0.786                       & 0.777                       & 0.785                      & 0.777                       & 1.000                       & 1.000                       & 1.000                       & 1.000                      & 1.000                       & 1.000                       & 1.000                      & 1.000                       & 1.000                       & 1.000                       & 1.000                       & 1.000                      \\ \hline
				&                      & 0.1                        & 0                         & 0                       & 0.792                       & 0.777                       & 0.790                      & 0.777                       & 1.000                       & 1.000                       & 1.000                       & 1.000                      & 1.000                       & 1.000                       & 1.000                      & 1.000                       & 1.000                       & 1.000                       & 1.000                       & 1.000                      \\
				&                      & 0.15                       & 0                         & 0                       & 0.795                       & 0.777                       & 0.794                      & 0.776                       & 1.000                       & 1.000                       & 1.000                       & 1.000                      & 1.000                       & 1.000                       & 1.000                      & 1.000                       & 1.000                       & 1.000                       & 1.000                       & 1.000                      \\
				&                      & 0.2                        & 0                         & 0                       & 0.798                       & 0.775                       & 0.792                      & 0.770                       & 1.000                       & 1.000                       & 1.000                       & 1.000                      & 1.000                       & 1.000                       & 1.000                      & 1.000                       & 1.000                       & 1.000                       & 1.000                       & 1.000                      \\
				&                      & 0.25                       & 0                         & 0                       & 0.796                       & 0.773                       & 0.791                      & 0.764                       & 1.000                       & 1.000                       & 1.000                       & 1.000                      & 1.000                       & 1.000                       & 1.000                      & 1.000                       & 1.000                       & 1.000                       & 1.000                       & 1.000                      \\
				&                      & 0.3                        & 0                         & 0                       & 0.800                       & 0.766                       & 0.788                      & 0.752                       & 1.000                       & 1.000                       & 1.000                       & 1.000                      & 1.000                       & 1.000                       & 1.000                      & 1.000                       & 1.000                       & 1.000                       & 1.000                       & 1.000                      \\
				&                      & 0                          & 0.05                      & 0                       & 0.774                       & 0.777                       & 0.775                      & 0.777                       & 1.000                       & 1.000                       & 1.000                       & 1.000                      & 1.000                       & 1.000                       & 1.000                      & 1.000                       & 1.000                       & 1.000                       & 1.000                       & 1.000                      \\
				&                      & 0                          & 0.1                       & 0                       & 0.760                       & 0.773                       & 0.760                      & 0.774                       & 1.000                       & 1.000                       & 1.000                       & 1.000                      & 1.000                       & 1.000                       & 1.000                      & 1.000                       & 1.000                       & 1.000                       & 1.000                       & 1.000                      \\
				&                      & 0                          & 0.15                      & 0                       & 0.752                       & 0.767                       & 0.752                      & 0.769                       & 1.000                       & 1.000                       & 1.000                       & 1.000                      & 1.000                       & 1.000                       & 1.000                      & 1.000                       & 1.000                       & 1.000                       & 1.000                       & 1.000                      \\
				&                      & 0                          & 0.2                       & 0                       & 0.739                       & 0.757                       & 0.739                      & 0.758                       & 1.000                       & 1.000                       & 1.000                       & 1.000                      & 1.000                       & 1.000                       & 1.000                      & 1.000                       & 1.000                       & 1.000                       & 1.000                       & 1.000                      \\
				&                      & 0                          & 0.25                      & 0                       & 0.715                       & 0.740                       & 0.715                      & 0.742                       & 1.000                       & 1.000                       & 1.000                       & 1.000                      & 1.000                       & 1.000                       & 1.000                      & 1.000                       & 1.000                       & 1.000                       & 1.000                       & 1.000                      \\
				&                      & 0                          & 0.3                       & 0                       & 0.699                       & 0.725                       & 0.699                      & 0.720                       & 1.000                       & 1.000                       & 1.000                       & 1.000                      & 1.000                       & 1.000                       & 1.000                      & 1.000                       & 1.000                       & 1.000                       & 1.000                       & 1.000                      \\
				&                      & 0                          & 0                         & 0.05                    & 0.777                       & 0.777                       & 0.773                      & 0.776                       & 1.000                       & 1.000                       & 1.000                       & 1.000                      & 1.000                       & 1.000                       & 1.000                      & 1.000                       & 1.000                       & 1.000                       & 1.000                       & 1.000                      \\
				&                      & 0                          & 0                         & 0.1                     & 0.778                       & 0.775                       & 0.782                      & 0.775                       & 1.000                       & 1.000                       & 1.000                       & 1.000                      & 1.000                       & 1.000                       & 1.000                      & 1.000                       & 1.000                       & 1.000                       & 1.000                       & 1.000                      \\
				&                      & 0                          & 0                         & 0.15                    & 0.773                       & 0.770                       & 0.773                      & 0.772                       & 1.000                       & 1.000                       & 1.000                       & 1.000                      & 1.000                       & 1.000                       & 1.000                      & 1.000                       & 1.000                       & 1.000                       & 1.000                       & 1.000                      \\
				&                      & 0                          & 0                         & 0.2                     & 0.769                       & 0.770                       & 0.769                      & 0.766                       & 1.000                       & 1.000                       & 1.000                       & 1.000                      & 1.000                       & 1.000                       & 1.000                      & 1.000                       & 1.000                       & 1.000                       & 1.000                       & 1.000                      \\
				&                      & 0                          & 0                         & 0.25                    & 0.768                       & 0.766                       & 0.767                      & 0.762                       & 1.000                       & 1.000                       & 1.000                       & 1.000                      & 1.000                       & 1.000                       & 1.000                      & 1.000                       & 1.000                       & 1.000                       & 1.000                       & 1.000                      \\
				&                      & 0                          & 0                         & 0.3                     & 0.764                       & 0.761                       & 0.756                      & 0.756                       & 1.000                       & 1.000                       & 1.000                       & 1.000                      & 1.000                       & 1.000                       & 1.000                      & 1.000                       & 1.000                       & 1.000                       & 1.000                       & 1.000   \\ \hline                  
			\end{tabular}
		\end{table}
	\end{landscape}
	
		\begin{figure}[h!]
		\caption{\small Power under local misspecification ($n$ larger than $T$, $(n,T)=(100,10)$)}
		\begin{subfigure}{.5\textwidth}
			\caption{$Queen\; W_{nt}^d \circ W_{nt}^e$}
			\centering
			\includegraphics[width=\textwidth, height=0.9\textheight]{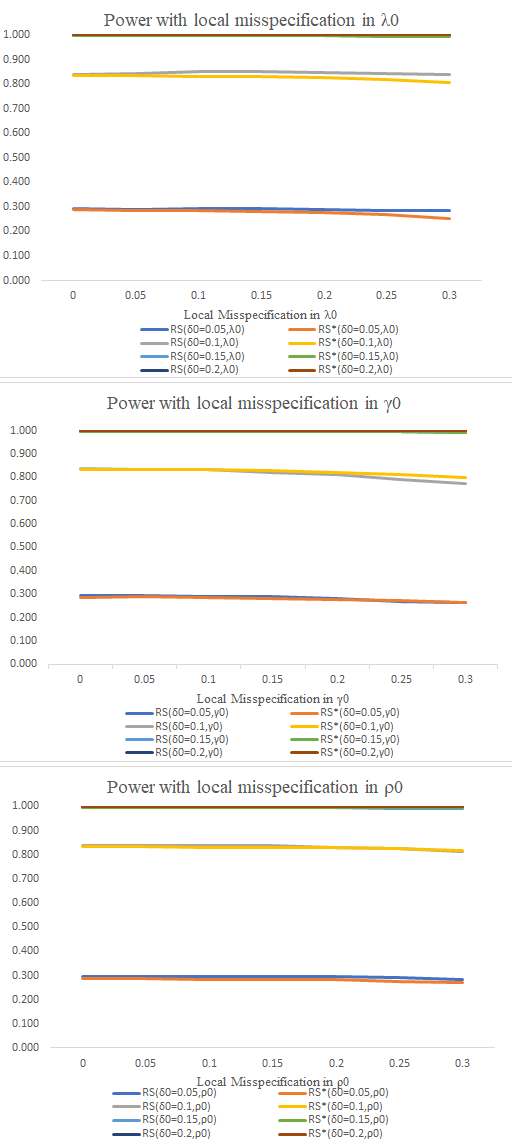}  
			\label{fig:sub-first}
		\end{subfigure}
		\hfill
		\begin{subfigure}{.5\textwidth}
			\caption{$Rook\; W_{nt}^d \circ W_{nt}^e$}
			\centering
			\includegraphics[width=\textwidth, height=0.9\textheight]{Q_power_n100_T10}  
			\label{fig:sub-second}
		\end{subfigure}
		\label{fig:fig}
	\end{figure}
	
	\begin{figure}[h!]
		\caption{\small Power under local misspecification ($T$ larger than $n$, $(n,T)=(9,100)$)}
		\begin{subfigure}{.5\textwidth}
			\caption{$Queen\; W_{nt}^d \circ W_{nt}^e$}
			\centering
			\includegraphics[width=\textwidth, height=0.9\textheight]{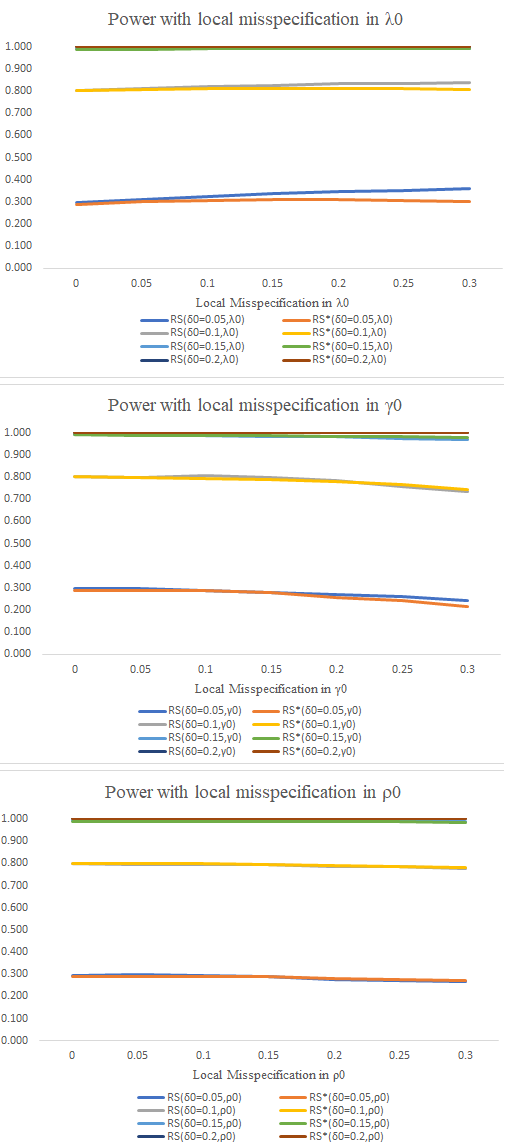}  
			\label{fig:sub-first}
		\end{subfigure}
		\hfill
		\begin{subfigure}{.5\textwidth}
			\caption{$Rook\; W_{nt}^d \circ W_{nt}^e$}
			\centering
			\includegraphics[width=\textwidth, height=0.9\textheight]{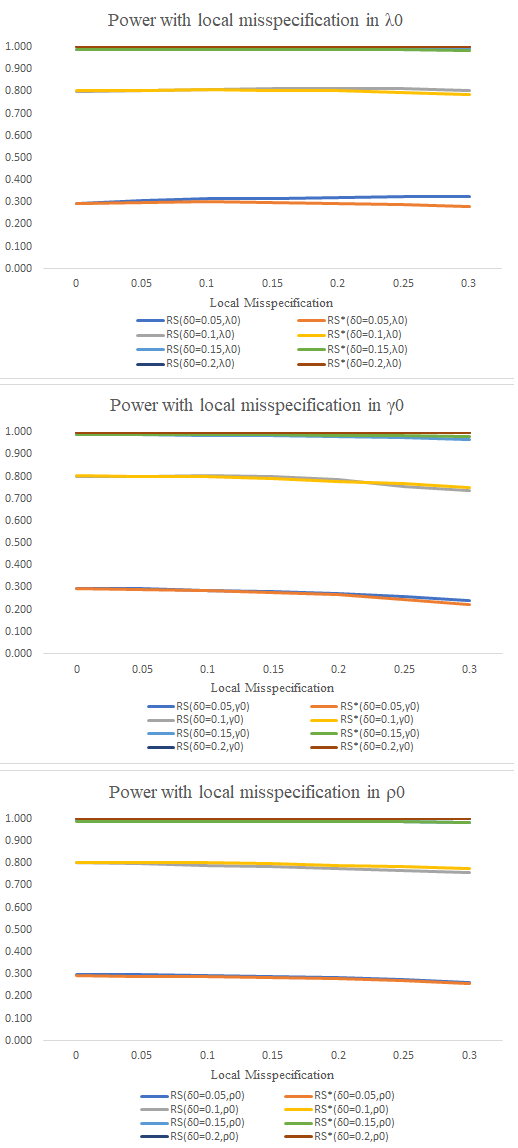}  
			\label{fig:sub-second}
		\end{subfigure}
		\label{fig:fig}
	\end{figure}
	
		\begin{landscape}
		\begin{table}[h!]
			\centering
			\caption{Robust RS test statistic \& Conditional LM test statistic ($n$ larger than $T$)}
			\scriptsize
			\begin{tabular}{ccccccccccccccccc} \hline
				\multicolumn{1}{l}{} & \multicolumn{1}{l}{} & \multicolumn{3}{c}{Local misspecification}                                       & \multicolumn{4}{c}{$\delta_0=0.00$}                                                                                  & \multicolumn{4}{c}{$\delta_0=0.05$}                                                                                  & \multicolumn{4}{c}{$\delta_0=0.1$}    \\ \cline{3-17}
				\multirow{2}{*}{$n$} & \multirow{2}{*}{$T$} & \multirow{2}{*}{$\lambda$} & \multirow{2}{*}{$\gamma$} & \multirow{2}{*}{$\rho$} & \multicolumn{2}{c}{$W_{nt}^{\text{Queen}}\circ W_{nt}^e$} & \multicolumn{2}{c}{$W_{nt}^{\text{Rook}}\circ W_{nt}^e$} & \multicolumn{2}{c}{$W_{nt}^{\text{Queen}}\circ W_{nt}^e$} & \multicolumn{2}{c}{$W_{nt}^{\text{Rook}}\circ W_{nt}^e$} & \multicolumn{2}{c}{$W_{nt}^{\text{Queen}}\circ W_{nt}^e$} & \multicolumn{2}{c}{$W_{nt}^{\text{Rook}}\circ W_{nt}^e$} \\ \cline{6-17} 
				&                      &                            &                           &                         & $RS^*$                        & $LM_{C}$                      & $RS^*$                       & $LM_{C}$                      & $RS^*$                        & $LM_{C}$                      & $RS^*$                        & $LM_{C}$                     & $RS^*$                         & $LM_{C}$                     & $RS^*$                        & $LM_{C}$                     \\ \hline
				100                  & 10                   & 0                          & 0                         & 0                       & 0.369                       & 0.488                       & 0.339                      & 0.459                       & 0.867                       & 0.706                       & 0.900                       & 0.730                      & 6.077                        & 5.637                      & 6.184                       & 5.728                      \\
				&                      & 0.05                       & 0                         & 0                       & 0.370                       & 0.484                       & 0.338                      & 0.460                       & 0.871                       & 0.707                       & 0.897                       & 0.727                      & 6.075                        & 5.634                      & 6.155                       & 5.722                      \\
				&                      & 0.1                        & 0                         & 0                       & 0.369                       & 0.490                       & 0.334                      & 0.462                       & 0.868                       & 0.708                       & 0.886                       & 0.723                      & 6.030                        & 5.631                      & 6.071                       & 5.713                      \\
				&                      & 0.15                       & 0                         & 0                       & 0.366                       & 0.492                       & 0.327                      & 0.465                       & 0.859                       & 0.709                       & 0.868                       & 0.719                      & 5.943                        & 5.626                      & 5.934                       & 5.703                      \\
				&                      & 0.2                        & 0                         & 0                       & 0.361                       & 0.493                       & 0.319                      & 0.467                       & 0.845                       & 0.710                       & 0.842                       & 0.714                      & 5.812                        & 5.618                      & 5.746                       & 5.690                      \\
				&                      & 0.25                       & 0                         & 0                       & 0.353                       & 0.494                       & 0.309                      & 0.469                       & 0.824                       & 0.711                       & 0.809                       & 0.710                      & 5.639                        & 5.610                      & 5.509                       & 5.675                      \\
				&                      & 0.3                        & 0                         & 0                       & 0.343                       & 0.494                       & 0.297                      & 0.472                       & 0.798                       & 0.711                       & 0.769                       & 0.704                      & 5.424                        & 5.600                      & 5.225                       & 5.658                      \\
				&                      & 0                          & 0.05                      & 0                       & 0.381                       & 0.514                       & 0.352                      & 0.487                       & 0.856                       & 0.675                       & 0.884                       & 0.695                      & 6.062                        & 5.545                      & 6.158                       & 5.624                      \\
				&                      & 0                          & 0.1                       & 0                       & 0.392                       & 0.543                       & 0.364                      & 0.516                       & 0.835                       & 0.645                       & 0.858                       & 0.659                      & 5.985                        & 5.451                      & 6.071                       & 5.518                      \\
				&                      & 0                          & 0.15                      & 0                       & 0.402                       & 0.574                       & 0.375                      & 0.547                       & 0.804                       & 0.613                       & 0.822                       & 0.622                      & 5.849                        & 5.353                      & 5.924                       & 5.408                      \\
				&                      & 0                          & 0.2                       & 0                       & 0.410                       & 0.621                       & 0.384                      & 0.581                       & 0.764                       & 0.580                       & 0.777                       & 0.585                      & 5.657                        & 5.250                      & 5.721                       & 5.292                      \\
				&                      & 0                          & 0.25                      & 0                       & 0.416                       & 0.658                       & 0.392                      & 0.617                       & 0.716                       & 0.547                       & 0.725                       & 0.546                      & 5.416                        & 5.140                      & 5.470                       & 5.170                      \\
				&                      & 0                          & 0.3                       & 0                       & 0.422                       & 0.682                       & 0.399                      & 0.657                       & 0.662                       & 0.469                       & 0.668                       & 0.507                      & 5.132                        & 5.022                      & 5.177                       & 5.041                      \\
				&                      & 0                          & 0                         & 0.05                    & 0.374                       & 0.497                       & 0.347                      & 0.474                       & 0.855                       & 0.693                       & 0.882                       & 0.712                      & 6.038                        & 5.609                      & 6.114                       & 5.685                      \\
				&                      & 0                          & 0                         & 0.1                     & 0.378                       & 0.505                       & 0.354                      & 0.489                       & 0.841                       & 0.682                       & 0.860                       & 0.696                      & 5.982                        & 5.585                      & 6.017                       & 5.645                      \\
				&                      & 0                          & 0                         & 0.15                    & 0.381                       & 0.512                       & 0.360                      & 0.502                       & 0.825                       & 0.672                       & 0.835                       & 0.681                      & 5.908                        & 5.567                      & 5.893                       & 5.609                      \\
				&                      & 0                          & 0                         & 0.2                     & 0.381                       & 0.526                       & 0.364                      & 0.515                       & 0.808                       & 0.663                       & 0.807                       & 0.667                      & 5.818                        & 5.553                      & 5.746                       & 5.575                      \\
				&                      & 0                          & 0                         & 0.25                    & 0.381                       & 0.531                       & 0.367                      & 0.528                       & 0.789                       & 0.657                       & 0.775                       & 0.655                      & 5.713                        & 5.492                      & 5.575                       & 5.545                      \\
				&                      & 0                          & 0                         & 0.3                     & 0.379                       & 0.535                       & 0.369                      & 0.539                       & 0.769                       & 0.652                       & 0.742                       & 0.644                      & 5.594                        & 5.539                      & 5.385                       & 5.521                      \\ \hline
				196                  & 20                   & 0                          & 0                         & 0                       & 0.127                       & 0.140                       & 0.121                      & 0.133                       & 11.656                      & 11.790                      & 11.463                      & 11.713                     & 41.747                       & 41.352                     & 41.662                      & 41.939                     \\
				&                      & 0.05                       & 0                         & 0                       & 0.128                       & 0.140                       & 0.121                      & 0.136                       & 11.674                      & 11.793                      & 11.589                      & 11.709                     & 41.780                       & 42.018                     & 41.690                      & 42.314                     \\
				&                      & 0.1                        & 0                         & 0                       & 0.126                       & 0.140                       & 0.119                      & 0.132                       & 11.611                      & 11.794                      & 11.497                      & 11.702                     & 41.547                       & 42.025                     & 41.363                      & 41.911                     \\
				&                      & 0.15                       & 0                         & 0                       & 0.123                       & 0.140                       & 0.115                      & 0.131                       & 11.468                      & 11.792                      & 11.302                      & 11.692                     & 41.043                       & 42.023                     & 40.687                      & 41.887                     \\
				&                      & 0.2                        & 0                         & 0                       & 0.118                       & 0.140                       & 0.109                      & 0.131                       & 11.242                      & 11.788                      & 11.006                      & 11.680                     & 40.262                       & 42.013                     & 39.673                      & 41.854                     \\
				&                      & 0.25                       & 0                         & 0                       & 0.112                       & 0.140                       & 0.101                      & 0.130                       & 10.934                      & 11.781                      & 10.616                      & 11.67                      & 39.203                       & 41.998                     & 38.335                      & 41.814                     \\
				&                      & 0.3                        & 0                         & 0                       & 0.105                       & 0.140                       & 0.092                      & 0.129                       & 10.544                      & 11.773                      & 10.136                      & 11.649                     & 37.865                       & 41.971                     & 36.694                      & 41.768                     \\
				&                      & 0                          & 0.05                      & 0                       & 0.129                       & 0.143                       & 0.123                      & 0.137                       & 11.675                      & 11.811                      & 11.601                      & 11.739                     & 41.797                       & 41.389                     & 41.718                      & 41.981                     \\
				&                      & 0                          & 0.1                       & 0                       & 0.131                       & 0.147                       & 0.124                      & 0.140                       & 11.594                      & 11.834                      & 11.525                      & 11.768                     & 41.479                       & 41.430                     & 41.405                      & 42.025                     \\
				&                      & 0                          & 0.15                      & 0                       & 0.132                       & 0.151                       & 0.125                      & 0.144                       & 11.415                      & 11.861                      & 11.353                      & 11.798                     & 40.802                       & 42.123                     & 40.733                      & 42.074                     \\
				&                      & 0                          & 0.2                       & 0                       & 0.132                       & 0.143                       & 0.126                      & 0.149                       & 11.143                      & 11.889                      & 11.088                      & 11.831                     & 39.787                       & 41.522                     & 39.724                      & 42.123                     \\
				&                      & 0                          & 0.25                      & 0                       & 0.132                       & 0.148                       & 0.126                      & 0.154                       & 10.786                      & 11.919                      & 10.739                      & 11.865                     & 38.462                       & 42.221                     & 38.405                      & 42.173                     \\
				&                      & 0                          & 0.3                       & 0                       & 0.131                       & 0.152                       & 0.126                      & 0.159                       & 10.353                      & 11.949                      & 10.315                      & 11.902                     & 36.861                       & 41.620                     & 36.811                      & 42.225                     \\
				&                      & 0                          & 0                         & 0.05                    & 0.125                       & 0.137                       & 0.120                      & 0.132                       & 11.607                      & 11.748                      & 11.544                      & 11.700                     & 41.646                       & 41.297                     & 41.553                      & 41.914                     \\
				&                      & 0                          & 0                         & 0.1                     & 0.122                       & 0.134                       & 0.118                      & 0.131                       & 11.522                      & 11.704                      & 11.463                      & 11.686                     & 41.419                       & 41.240                     & 41.262                      & 41.886                     \\
				&                      & 0                          & 0                         & 0.15                    & 0.119                       & 0.130                       & 0.116                      & 0.130                       & 11.401                      & 11.658                      & 11.331                      & 11.671                     & 41.067                       & 41.179                     & 40.792                      & 41.856                     \\
				&                      & 0                          & 0                         & 0.2                     & 0.115                       & 0.127                       & 0.114                      & 0.129                       & 11.246                      & 11.610                      & 11.151                      & 11.655                     & 40.594                       & 41.116                     & 40.147                      & 41.825                     \\
				&                      & 0                          & 0                         & 0.25                    & 0.112                       & 0.124                       & 0.111                      & 0.128                       & 11.057                      & 11.561                      & 10.924                      & 11.638                     & 40.005                       & 41.050                     & 39.333                      & 41.790                     \\
				&                      & 0                          & 0                         & 0.3                     & 0.108                       & 0.120                       & 0.108                      & 0.127                       & 10.837                      & 11.510                      & 10.652                      & 11.621                     & 39.302                       & 40.984                     & 38.357                      & 41.754    \\ \hline                
			\end{tabular}
		\end{table}
	\end{landscape}
	
	\begin{landscape}
		\begin{table}[h!]
			\centering
			\caption{Robust RS test statistic \& Conditional LM test statistic ($T$ larger than $n$)}
			\scriptsize
			\begin{tabular}{ccccccccccccccccc} \hline
				\multicolumn{1}{l}{} & \multicolumn{1}{l}{} & \multicolumn{3}{c}{Local misspecification}                                       & \multicolumn{4}{c}{$\delta_0=0.00$}                                                                                  & \multicolumn{4}{c}{$\delta_0=0.05$}                                                                                  & \multicolumn{4}{c}{$\delta_0=0.1$}                                                                                   \\ \cline{3-17}
				\multirow{2}{*}{$n$} & \multirow{2}{*}{$T$} & \multirow{2}{*}{$\lambda$} & \multirow{2}{*}{$\gamma$} & \multirow{2}{*}{$\rho$} & \multicolumn{2}{c}{$W_{nt}^{\text{Queen}}\circ W_{nt}^e$} & \multicolumn{2}{c}{$W_{nt}^{\text{Rook}}\circ W_{nt}^e$} & \multicolumn{2}{c}{$W_{nt}^{\text{Queen}}\circ W_{nt}^e$} & \multicolumn{2}{c}{$W_{nt}^{\text{Rook}}\circ W_{nt}^e$} & \multicolumn{2}{c}{$W_{nt}^{\text{Queen}}\circ W_{nt}^e$} & \multicolumn{2}{c}{$W_{nt}^{\text{Rook}}\circ W_{nt}^e$} \\ \cline{6-17}
				&                      &                            &                           &                         & $RS^*$                        & $LM_{C}$                      & $RS^*$                       & $LM_{C}$                      & $RS^*$                        & $LM_{C}$                      & $RS^*$                        & $LM_{C}$                     & $RS^*$                         & $LM_{C}$                     & $RS^*$                        & $LM_{C}$                     \\ \hline
				9                    & 100                  & 0                          & 0                         & 0                       & 0.510                       & 0.543                       & 0.441                      & 0.463                       & 4.570                       & 4.800                       & 4.301                      & 4.501                       & 12.470                       & 12.961                     & 12.089                      & 12.616                     \\
				&                      & 0.05                       & 0                         & 0                       & 0.533                       & 0.549                       & 0.450                      & 0.459                       & 4.698                       & 4.823                       & 4.384                      & 4.487                       & 12.736                       & 13.009                     & 12.300                      & 12.581                     \\
				&                      & 0.1                        & 0                         & 0                       & 0.553                       & 0.554                       & 0.454                      & 0.455                       & 4.801                       & 4.841                       & 4.431                      & 4.421                       & 12.936                       & 13.042                     & 12.416                      & 12.533                     \\
				&                      & 0.15                       & 0                         & 0                       & 0.569                       & 0.546                       & 0.453                      & 0.450                       & 4.875                       & 4.825                       & 4.443                      & 4.466                       & 13.065                       & 13.200                     & 12.437                      & 12.470                     \\
				&                      & 0.2                        & 0                         & 0                       & 0.581                       & 0.598                       & 0.447                      & 0.445                       & 4.920                       & 5.105                       & 4.419                      & 4.417                       & 13.118                       & 13.060                     & 12.360                      & 12.396                     \\
				&                      & 0.25                       & 0                         & 0                       & 0.589                       & 0.565                       & 0.437                      & 0.438                       & 4.933                       & 4.864                       & 4.359                      & 4.383                       & 13.091                       & 13.044                     & 12.186                      & 12.306                     \\
				&                      & 0.3                        & 0                         & 0                       & 0.592                       & 0.567                       & 0.421                      & 0.431                       & 4.913                       & 4.860                       & 4.263                      & 4.344                       & 12.981                       & 13.009                     & 11.916                      & 12.202                     \\
				&                      & 0                          & 0.05                      & 0                       & 0.504                       & 0.544                       & 0.436                      & 0.462                       & 4.531                       & 4.808                       & 4.259                      & 4.496                       & 12.356                       & 12.971                     & 11.974                      & 12.601                     \\
				&                      & 0                          & 0.1                       & 0                       & 0.494                       & 0.547                       & 0.428                      & 0.461                       & 4.451                       & 4.821                       & 4.182                      & 4.494                       & 12.130                       & 12.990                     & 11.758                      & 12.594                     \\
				&                      & 0                          & 0.15                      & 0                       & 0.480                       & 0.550                       & 0.418                      & 0.462                       & 4.331                       & 4.837                       & 4.073                      & 4.495                       & 11.798                       & 13.018                     & 11.445                      & 12.591                     \\
				&                      & 0                          & 0.2                       & 0                       & 0.463                       & 0.555                       & 0.405                      & 0.463                       & 4.174                       & 4.857                       & 3.935                      & 4.500                       & 11.367                       & 13.056                     & 11.045                      & 12.595                     \\
				&                      & 0                          & 0.25                      & 0                       & 0.441                       & 0.560                       & 0.391                      & 0.465                       & 3.983                       & 4.882                       & 3.769                      & 4.508                       & 10.846                       & 13.102                     & 10.566                      & 12.605                     \\
				&                      & 0                          & 0.3                       & 0                       & 0.417                       & 0.567                       & 0.375                      & 0.468                       & 3.763                       & 4.909                       & 3.580                      & 4.521                       & 10.249                       & 13.156                     & 10.018                      & 12.625                     \\
				&                      & 0                          & 0                         & 0.05                    & 0.488                       & 0.516                       & 0.438                      & 0.455                       & 4.506                       & 4.721                       & 4.300                      & 4.470                       & 12.370                       & 12.835                     & 12.111                      & 12.562                     \\
				&                      & 0                          & 0                         & 0.1                     & 0.463                       & 0.489                       & 0.433                      & 0.446                       & 4.424                       & 4.636                       & 4.280                      & 4.440                       & 12.228                       & 12.699                     & 12.078                      & 12.507                     \\
				&                      & 0                          & 0                         & 0.15                    & 0.437                       & 0.461                       & 0.426                      & 0.439                       & 4.325                       & 4.547                       & 4.240                      & 4.410                       & 12.046                       & 12.554                     & 11.991                      & 12.451                     \\
				&                      & 0                          & 0                         & 0.2                     & 0.410                       & 0.432                       & 0.418                      & 0.431                       & 4.211                       & 4.452                       & 4.183                      & 4.381                       & 11.825                       & 12.400                     & 11.850                      & 12.395                     \\
				&                      & 0                          & 0                         & 0.25                    & 0.381                       & 0.403                       & 0.410                      & 0.425                       & 4.083                       & 4.352                       & 4.108                      & 4.353                       & 11.567                       & 12.236                     & 11.654                      & 12.338                     \\
				&                      & 0                          & 0                         & 0.3                     & 0.352                       & 0.373                       & 0.400                      & 0.419                       & 3.942                       & 4.248                       & 4.015                      & 4.325                       & 11.274                       & 12.062                     & 11.405                      & 12.283                     \\ \hline
				16                   & 200                  & 0                          & 0                         & 0                       & 0.024                       & 0.026                       & 0.023                      & 0.024                       & 6.484                       & 6.538                       & 6.541                      & 6.578                       & 27.577                       & 27.860                     & 27.729                      & 27.894                     \\
				&                      & 0.05                       & 0                         & 0                       & 0.024                       & 0.022                       & 0.023                      & 0.025                       & 6.520                       & 6.503                       & 6.558                      & 6.543                       & 27.743                       & 27.791                     & 27.796                      & 27.809                     \\
				&                      & 0.1                        & 0                         & 0                       & 0.025                       & 0.027                       & 0.023                      & 0.024                       & 6.504                       & 6.487                       & 6.502                      & 6.517                       & 27.709                       & 27.704                     & 27.595                      & 27.674                     \\
				&                      & 0.15                       & 0                         & 0                       & 0.026                       & 0.028                       & 0.025                      & 0.025                       & 6.435                       & 6.454                       & 6.376                      & 6.480                       & 27.469                       & 27.599                     & 27.130                      & 27.538                     \\
				&                      & 0.2                        & 0                         & 0                       & 0.027                       & 0.029                       & 0.028                      & 0.026                       & 6.312                       & 6.416                       & 6.180                      & 6.438                       & 27.020                       & 27.476                     & 26.407                      & 27.385                     \\
				&                      & 0.25                       & 0                         & 0                       & 0.030                       & 0.030                       & 0.032                      & 0.026                       & 6.134                       & 6.373                       & 5.919                      & 6.391                       & 26.360                       & 27.334                     & 25.440                      & 27.214                     \\
				&                      & 0.3                        & 0                         & 0                       & 0.033                       & 0.031                       & 0.038                      & 0.027                       & 5.902                       & 6.325                       & 5.597                      & 6.339                       & 25.491                       & 27.174                     & 24.245                      & 27.026                     \\
				&                      & 0                          & 0.05                      & 0                       & 0.023                       & 0.025                       & 0.021                      & 0.023                       & 6.470                       & 6.544                       & 6.531                      & 6.588                       & 27.479                       & 27.865                     & 27.644                      & 27.913                     \\
				&                      & 0                          & 0.1                       & 0                       & 0.021                       & 0.025                       & 0.020                      & 0.023                       & 6.390                       & 6.548                       & 6.454                      & 6.596                       & 27.091                       & 27.866                     & 27.273                      & 27.928                     \\
				&                      & 0                          & 0.15                      & 0                       & 0.019                       & 0.024                       & 0.018                      & 0.022                       & 6.244                       & 6.551                       & 6.313                      & 6.603                       & 26.425                       & 27.863                     & 26.628                      & 27.939                     \\
				&                      & 0                          & 0.2                       & 0                       & 0.017                       & 0.024                       & 0.016                      & 0.022                       & 6.037                       & 6.553                       & 6.110                      & 6.608                       & 25.500                       & 27.854                     & 25.728                      & 27.945                     \\
				&                      & 0                          & 0.25                      & 0                       & 0.015                       & 0.023                       & 0.015                      & 0.022                       & 5.775                       & 6.552                       & 5.853                      & 6.611                       & 24.345                       & 27.840                     & 24.600                      & 27.946                     \\
				&                      & 0                          & 0.3                       & 0                       & 0.013                       & 0.023                       & 0.013                      & 0.014                       & 5.466                       & 6.550                       & 5.549                      & 6.612                       & 22.992                       & 27.816                     & 23.275                      & 27.941                     \\
				&                      & 0                          & 0                         & 0.05                    & 0.023                       & 0.026                       & 0.022                      & 0.025                       & 6.484                       & 6.535                       & 6.521                      & 6.560                       & 27.567                       & 27.846                     & 27.653                      & 27.862                     \\
				&                      & 0                          & 0                         & 0.1                     & 0.022                       & 0.026                       & 0.022                      & 0.026                       & 6.462                       & 6.531                       & 6.474                      & 6.544                       & 27.461                       & 27.829                     & 27.455                      & 27.829                     \\
				&                      & 0                          & 0                         & 0.15                    & 0.021                       & 0.026                       & 0.021                      & 0.027                       & 6.420                       & 6.527                       & 6.400                      & 6.529                       & 27.261                       & 27.808                     & 27.137                      & 27.795                     \\
				&                      & 0                          & 0                         & 0.2                     & 0.020                       & 0.026                       & 0.020                      & 0.027                       & 6.356                       & 6.522                       & 6.302                      & 6.515                       & 26.968                       & 27.782                     & 26.705                      & 27.760                     \\
				&                      & 0                          & 0                         & 0.25                    & 0.019                       & 0.026                       & 0.019                      & 0.028                       & 6.272                       & 6.516                       & 6.180                      & 6.503                       & 26.584                       & 27.752                     & 26.163                      & 27.725                     \\
				&                      & 0                          & 0                         & 0.3                     & 0.017                       & 0.026                       & 0.017                      & 0.028                       & 6.168                       & 6.510                       & 6.035                      & 6.493                       & 26.113                       & 27.718                     & 25.516                      & 27.691  \\ \hline                   
			\end{tabular}
		\end{table}
	\end{landscape}

	\newpage
	\subsection{Empirical illustration}
	Using the Penn World Tables (PWT version 6.1), I introduce how to test the endogeneity of the spatial weights matrices ($W$) before regular estimation. As in Ertur and Koch (2007), I measure the variables in Solow-Swan growth model of the logarithm of savings (\textit{lns}), the logarithm of the growth of the working-age population (ages 15 to 64) summed up with the growth rate and the depreciation rate in capital (\textit{ln(n+g+$\delta$)}) in year 1960-1995. I suppose that $g+\delta=0.05$ as in MRW (1992) and Romer (1989). As a general sense, I consider SDPD growth model to capture the contemporaneous dependence over space, dependence over time, or spatial time dependence.
	
	Now I suspect that $W_{nt}$ is a function of the share of gross consumption in GDP (\textit{kc}) and real Gross Domestic Income (GDI) for terms of trade changes (\textit{rgdptt}). As mentioned in Qu, Lee, and Yu (2017), identification is not an issue here and thus one is allowed to have $X_{1nt}$ and $X_{2nt}$ share the common variables. The equations are therefore
	\begin{align*}
	\ln y_{nt}&=\lambda W_{nt}\ln y_{nt}+\rho W_{n,t-1}\ln y_{n,t-1}+\gamma y_{n,t-1}\\
	&\hspace{5mm}+X_{1nt}\beta+c_{n1}+\alpha_{t1}1_n+V_{nt},\\
	(W_{nt})_{ij}&=w_{ij,nt}=h(z_{i,nt},z_{j,nt}),\\
	Z_{nt}&=Z_{n,t-1}\kappa+X_{2nt}\Gamma+c_{n2}+1_n\alpha'_{t2}+\epsilon_{nt},
	\end{align*}
	\noindent where $t=1,\dots,T$, $y_{nt}$ is the real income per worker at time $t$, $X_{1nt}=(lns_{nt},ln(n_{nt}+g+\delta),W_{nt}lns_{nt},W_{nt}ln(n_{nt}+g+\delta))'$, $\beta=(\beta_1,\beta_2,\beta_3,\beta_4)'$, and $Z_{nt}\in \{kc_{nt},rgdptt_{nt},(kc_{nt},rgdptt_{nk})\}$,  $\kappa$ is the associated parameter, $X_{2nt}=(lns_{nt},ln(n_{nt}+g+\delta))'$, and $\Gamma=(\Gamma_1,\Gamma_2)$. The equations above can be augmented as
	\begin{align*}
	lny_{nt}&=\lambda W_{nt}lny_{nt}(1995)+\rho W_{n,t-1}lny_{n,t-1}\\
	&\hspace{5mm}+\gamma y_{n,t-1}+X_{1nt}\beta+c_{n1}+\alpha_{t1}1_n\\
	&\hspace{5mm}+(Z_{nt}-Z_{n,t-1}\kappa-X_{2nt}\Gamma-c_{n2}-1_n\alpha'_{t2})\delta+\xi_{nt},
	\end{align*}
	\noindent where $\xi_{nt}\sim N(0,\sigma_\xi^2I_{n})$. Recall that $(Z_{nt}-Z_{n,t-1}\kappa-X_{2nt}\Gamma-c_{n2}-1_n\alpha'_{t2})$ is our control variables for the endogenous $W_{nt}$. 
	
	Now I use the test statistic, the Robust RS test, to determine if $W_{nt}$ is exogenous, i.e., $H_0^\delta: \delta_0=0$. The results are presented in Table 8. One may find that Robust RS test statistic is far smaller than Standard RS, indicating the standard RS generally leads to over-rejection of the null hypothesis. It also implies the potential problem of $W$ being endogenous when economic distances are effective in constructing $W$, urging the importance of testing the endogeneity of $W$ as a basic work for analysis. \\
	
	\begin{table}[hbt!]
		\centering
		\small
		\caption{Empirical results (Penn World Table version 6.1; year 1960-1995)}
		\begin{tabular}{cccc} \hline
			$Z_{nt}$              & Standard RS                                                              & Robust RS                                                         & Critical values ($\chi^2_{p}$)\\ \hline
			$kc_{nt}$              & 23.631       & 11.192 & 3.8415 \\
			$rgdptt_{nt}$        & 77.562     & 56.914  & 3.8415 \\
			$kc_{nt},rgdptt_{nt}$       & 897.266       & 57.603   & 5.9915 \\ \hline
		\end{tabular}
	\end{table}
	
	\section{Conclusion}
	Even though conventional uses for the spatial weights matrices ($W$) have been found in the predetermined geography, one may allow $W$ to include economic distances, following the accumulating evidence in economics literature. However, this may lead to the violation of the exogenous assumption for the ordinary spatial autoregressive (SAR) estimators as well as for the related test statistics. For this purpose, I propose Robust Rao's Score (RS) test to determine endogeneity of spatial weights matrices ($W$) in spatial dynamic panel data (SDPD) models. 

	The robust Rao's Score (RS) test is \textit{robust} in the sense that it is asymptotically central chi-squared under the null regardless of the presence of local misspecifications in the contemporaneous dependence over space, dependence over time, and spatial time dependence, keeping the type I error fixed. It is also \textit{computationally efficient} in the sense that it only requires the restricted ML estimators under the null where the parameters above are assumed to be zero, reducing the spatial dynamic panel data models to the simple fixed-effects model.
	
	A Monte Carlo simulation supports the analytics and shows nice finite sample properties as $n$ and $T$ increase. Subsequently, an empirical illustration using Penn World Table version 6.1 shows how large the robust RS test adjusts toward the local misspecifications in parameters compared to the standard RS test. Also, it reaffirms the importance of testing the endogeneity of $W$ as a basic work for analysis due to the potential problem of $W$ being endogenous when economic distances are effective in constructing $W$.
	
	\newpage
	\section*{Reference} \scriptsize
	\begin{indentblock}
		Anselin, Luc. (1988). \textit{Spatial Econometrics: Methods and Models}. Springer, New York.
	\end{indentblock}

	\begin{indentblock}
		Anselin, Luc, et al. (1996). Simple diagnostic tests for spatial dependence. \textit{Regional Science and Urban
		Economics}. 26 (1), 77–104.
	\end{indentblock}

	\begin{indentblock}
		 Anselin, Luc. (2001). Rao's score test in spatial econometrics. \textit{Journal of statistical planning and inference}, 97(1): 113-139.
	\end{indentblock}

	\begin{indentblock}
		Baltagi, Badi H., Li, Dong. (2001). LM tests for functional form and spatial error
		correlation. \textit{International Regional Science Review}. 24(2), 194–225.
	\end{indentblock}

	\begin{indentblock}
		Baltagi, Badi H., Song, Seuck Heun, Koh, Won. (2003). Testing panel data regression
		models with spatial error correlation. \textit{Journal of Econometrics}, 117 (1), 123–150.
	\end{indentblock}

	\begin{indentblock}
		Baltagi, Badi H., Song, Seuck Heun., Jung, Byoung Cheol., Koh, Won. (2007). Testing for serial correlation, spatial autocorrelation and
		random effects using panel data. \textit{Journal of Econometrics}, 140 (1), 5–51.
	\end{indentblock}
	
	\begin{indentblock}
		Baltagi, Badi H., Liu, Long. (2008). Testing for random effects and spatial lag dependence
		in panel data models. \textit{Statistics \& Probability Letters}, 78 (18), 3304–3306.
	\end{indentblock}

	\begin{indentblock}
		Baltagi, Badi H., Song, Seuck Heun, Kwon, Jae Hyeok. (2009). Testing for heteroskedasticity and spatial correlation in a random effects panel data model. \textit{Computational Statistics \& Data Analysis}, 53 (8), 2897–2922.
	\end{indentblock}
	
	\begin{indentblock}
		Baltagi, Badi H., Yang, Zhenlin. (2013). Standardized LM tests for spatial error
		dependence in linear or panel regressions. \textit{The Economic Journal}, 16 (1), 103–134.
	\end{indentblock}
	
	\begin{indentblock}
	Baxter, M., Kouparitsas, M. (2005). Determinants of business cycle comovement: a robust analysis. \textit{Journal of Monetary Economics}, 52: 113–157.
	\end{indentblock}

	\begin{indentblock}
	 	Bera, Anil K., Yoon, Mann J. (1993). Specification testing with locally misspecified alternatives. \textit{Economic Theory}, 9(4). 
	\end{indentblock}

	\begin{indentblock}
	 	Bera, Anil K., Doğan, Osman., Taşpınar, Süleyman. (2018). Simple tests for endogeneity of spatial weights matrices. \textit{Regional Science and Urban Economics}, 69: 130-142. 
	\end{indentblock}

	\begin{indentblock}
	 	Bera, Anil K., Doğan, Osman., Taşpınar, Süleyman. (2019).  Testing Spatial Dependence in Spatial Models with Endogenous Weights Matrices. \textit{Journal of Econometric Methods}, 8(1): 1-33.
	\end{indentblock}

	\begin{indentblock}
		 Bera, Anil K., Doğan, Osman., Taşpınar, Süleyman. Leiluo, Yufan. (2019). Robust LM Tests for Spatial Dynamic Panel Data Models. \textit{Regional Science and Urban Economics}, 76: 47-66.
	\end{indentblock}

	\begin{indentblock}
	 	Bera, Anil K., Bilias, Yannis, Yoon, Mann J., Taşpınar, Süleyman, Doğan, Osman. (2020).  Adjustments of Rao's Score Test for Distributional and Local Parametric Misspecifications. \textit{Journal of Econometric Methods}, 9(1): 1-29. 
	\end{indentblock}

	\begin{indentblock}
	 	Cheng, Wei., and Lung-fei Lee. (2017). Testing endogeneity of spatial and social networks. \textit{Regional Science and Urban Economics}, 64: 81-97.
	\end{indentblock}

	\begin{indentblock}
	 	Cliff, Andrew., Ord, Keith (1972). Testing for spatial autocorrelation among regression residuals. \textit{Geographical analysis}, 4(3): 267-284.
	\end{indentblock}

	\begin{indentblock}
	 	Conley, Timothy G., Ligon, Ethan. (2002). Economic distance and cross-country spillovers. \textit{Journal of Economic Growth}, 7(2): 157-187.
	\end{indentblock}

	\begin{indentblock}
	 	Conley, Timothy G., Topa, Giorgio. (2002). Socio-economic distance and spatial patterns in unemployment. \textit{Journal of Applied Econometrics}, 17(4): 303-327. 
	\end{indentblock}

	\begin{indentblock}
	 	Davidson, Russell. and Mackinnon, James G. (1987). Implicit alternatives and the local power of test statistics, \textit{Econometrica}, 55 (6). 
	\end{indentblock}

	\begin{indentblock}
		Debarsy, Nicolas, Ertur, Cem. (2010). Testing for spatial autocorrelation in a fixed effects
		panel data model. \textit{Regional Science and Urban Economics}, 40 (6), 453–470.
	\end{indentblock}

	\begin{indentblock}
	 	De Long, J. Bradford., Summers, Lawrence H. (2018). Equipment Investment and Economic Growth, \textit{The Quarterly Journal of Economics}, 106(2): 445–502.
	\end{indentblock}

	\begin{indentblock}
	 	Ditzen, Jan. (2018). Cross-country convergence in a general Lotka–Volterra model. \textit{Spatial Economic Analysis}, 13:2, 191-211. 
	\end{indentblock}

	\begin{indentblock}
	 	Doğan, Osman., Taspinar, Süleyman. (2013). GMM estimation of spatial autoregressive models with moving average disturbances. \textit{Regional science and urban economics}, 43(6): 903-926.
	\end{indentblock}

	\begin{indentblock}
	 	Doğan, Osman., Taşpınar, Süleyman., Bera, Anil K. (2018). Simple tests for social interaction models with network structures. \textit{Spatial econometrics}, 13(2), 212-246.
	\end{indentblock}
	
	\begin{indentblock}
	 	Elhorst, J.Paul. (2010). Applied spatial econometrics: raising the bar. \textit{Spatial economic analysis}, 5(1): 9-28.
	\end{indentblock}

	\begin{indentblock}
	 	Elhorst, J. Paul. (2014). \textit{Spatial Econometrics: from Cross-sectional Data to Spatial Panels}. Springer Briefs in Regional Science. Springer Berlin Heidelberg, New York.
	\end{indentblock}

	\begin{indentblock}
	 	Ertur, Cem. and Koch, Wilfried. (2007). Growth, technological interdependence and spatial externalities: theory and evidence. \textit{Journal of Applied Econometrics}, 22(6), 1033-1062. 
	\end{indentblock}

	\begin{indentblock}
	 	Ertur, Cem. and Koch, Wilfried. (2011). A contribution to the theory and empirics of Schumpeterian growth with worldwide interactions. \textit{Journal of Economic Growth}, 16(3), 215-255. doi: 10.1007/s10887-011-9067-0.  
	\end{indentblock}

	\begin{indentblock}
	 	Frankel, J., Rose, A. (1998). The endogeneity of the optimum currency area criteria. \textit{The Economic Journal}, 108: 1009–1025.
	\end{indentblock}

	\begin{indentblock}
	 	Ho, Chun-Yu., Wang, Wei., Yu, Jihai. (2013). Growth spillover through trade: A spatial dynamic panel data approach. \textit{Economics Letters}, 120(3):450–453.
	\end{indentblock}

	\begin{indentblock}
	 	Jenish, Nazgul., Prucha, Ingmar R. (2009). Central limit theorems and uniform laws of large numbers for arrays of random fields. \textit{Journal of Econometrics}, 150(1): 86-98. 
	\end{indentblock}

	\begin{indentblock}
	 	Jenish, Nazgul., Prucha, Ingmar R. (2012). On spatial processes and asymptotic inference under near-epoch dependence. \textit{Journal of Econometrics}, 170(1): 178-190.
	\end{indentblock}

	\begin{indentblock}
	 	Kelejian, Harry H., Robinson, Dennis P. (1992). Spatial autocorrelation: a new computationally simple test with an application to per-capita county police
		expenditures. \textit{Regional Science and Urban Economics}, 22(3), 317–331 Special Issue Space and
		Applied Econometrics.
	\end{indentblock}

	\begin{indentblock}
	 	Kelejian, Harry H., Prucha, Ingmar R. (2010). Specification and estimation of spatial autoregressive models with autoregressive and heteroskedastic disturbances. \textit{Journal of econometrics}, 157: 53-67.
	\end{indentblock}

	\begin{indentblock}
	 	Keller, Wolfgang. (2002). Geographic Localization of International Technology Diffusion. \textit{American Economic Review}, 92(1): 120-142. 
	\end{indentblock}
	
	\begin{indentblock}
	 	Lee, Lung-fei., Yu, Jihai. (2010). A spatial dynamic panel data model with both time and individual fixed effects. \textit{Econometric Theory}, 26(2): 564-597.
	\end{indentblock}

	\begin{indentblock}
	 	Lee, Lung-fei., Yu, Jihai. (2012). QML Estimation of Spatial Dynamic Panel Data Models with Time Varying Spatial Weights Matrices. \textit{Spatial Economic Analysis}, 7(1): 31-74.
	\end{indentblock}
	
	\begin{indentblock}
	 	Mankiw, NG., Romer, D., Weil, DN. (1992). A contribution to the empirics of economic growth. \textit{Quarterly Journal of Economics}, 107: 407-437.
	\end{indentblock}

	\begin{indentblock}
	 	Moran, PAP. (1950). A test for the serial independence of residuals. \textit{Biometrika}, 37(1/2): 178-181.
	\end{indentblock}
	
	\begin{indentblock}
	 	Parent, Olivier., LeSage, James P. (2008). Using the variance structure of the conditional autoregressive spatial specification to model knowledge spillovers. \textit{Journal of applied Econometrics}, 23(2): 235-256.
	\end{indentblock}

	\begin{indentblock}
	 	Pinkse, Joris., Slade, Margaret E. (2010). The future of spatial econometrics. \textit{Journal of Regional Science}, 50(1): 103-117.
	\end{indentblock}

	\begin{indentblock}
	 	Qu, Xi., and Lee, Lung-fei. (2015). Estimating a spatial autoregressive model with an endogenous spatial weight matrix. \textit{Journal of Econometrics}, Elsevier, vol. 184(2), pages 209-232.
	\end{indentblock}

	\begin{indentblock}
	 	Qu, Xi., Lee, Lung-fei., Yu, Jihai. (2017). QML estimation of spatial dynamic panel data models with endogenous time varying spatial weights matrices. \textit{Journal of Econometrics}, 197(2): 173-201.
	\end{indentblock}
	
	\begin{indentblock}
		 Saikkonen, Pentti. (1989). Asymptotic relative efficiency of the classical test statistics under misspecification. \textit{Journal of Economics}, 42(3), 351-369.
	\end{indentblock}

	\begin{indentblock}
	 	Skevas, T., Skevas, I., Cabrera, V. (2021). Farm Profitability as a Driver of Spatial Spillovers: The Case of Somatic Cell Counts on Wisconsin Dairies. \textit{Agricultural and Resource Economics Review}, 50(1): 187-200.
	\end{indentblock}
	
	\begin{indentblock}
		Taşpınar, Suleyman., Doğan, Osman, Bera, Anil K. (2017). GMM gradient tests for spatial dynamic panel data models. \textit{Regional Science and Urban Economics}, 65, 65-88.
	\end{indentblock}

	\begin{indentblock}
	Yang, Zhenlin. (2010). A robust LM test for spatial error components. \textit{Regional Science and Urban Economics}, 40 (5), 299–310 Advances In Spatial Econometrics.
	\end{indentblock}
	
	\begin{indentblock}
		Yang, Zhenlin. (2021). Joint Tests for Dynamic and Spatial Effects in Short Panel Data
		Models with Fixed Effects. \textit{Empirical Economics}, 60(1), 51-92.
	\end{indentblock}

	\begin{indentblock}
	 	Yu, Jihai., Jong, Robert de., Lee, Lung Fei. (2008). Quasi-maximum likelihood estimators for spatial dynamic panel data with fixed effects when both n and T are large. \textit{Journal of Econometrics}, 146(1): 118-134.
	\end{indentblock}

	\newpage
	\section*{Appendix}
	\captionsetup{labelformat=AppendixTables}
	\setcounter{table}{0}
	
	\subsection*{1. Proposition 1 \& 2 \& 3 proof}
	They are provided in Qu, Lee, and Yu (2017) and hence omitted here.
		
	\subsection*{2. The explicit forms of the bias-corrected score functions}	
		\begin{align*}
			C_\delta(\tilde{\theta})&=L_\delta(\tilde{\theta})-\frac{1}{\sqrt{nT}}(\underbrace{\Delta_{1,\delta}(\tilde{\theta})}_\text{$=0$}+\underbrace{\Delta_{2,\delta}(\tilde{\theta}))}_\text{$=0$}+\frac{1}{\sqrt{nT}}\underbrace{I_{\delta\omega}(\tilde{\theta})}_\text{$=0$}I_{\omega\omega}^{-1}(\tilde{\theta})(\Delta_{1, \omega}(\tilde{\theta})+\Delta_{2,\omega}(\tilde{\theta})) \\
			&=L_\delta(\tilde{\theta})=\frac{1}{nT\widetilde{\sigma_\xi^2}}\left(\varepsilon'_L(\tilde{\theta})J_L\xi_L(\tilde{\theta})\right), \\
			C_\eta(\tilde{\theta})&=L_\eta(\tilde{\theta})-\frac{1}{\sqrt{nT}}(\Delta_{1,\eta}(\tilde{\theta})+\Delta_{2,\eta}(\tilde{\theta}))+\frac{1}{\sqrt{nT}}I_{\eta\omega}(\tilde{\theta})I_{\omega\omega}^{-1}(\tilde{\theta})(\Delta_{1,\omega}(\tilde{\theta})+\Delta_{2,\omega}(\tilde{\theta})) \\
			&=\frac{1}{nT\widetilde{\sigma_\xi^2}}\left[\xi'_L(\tilde{\theta})J_LW_{1L}Y_L-tr(W_{1L}), Y_{L,-1}'J_L\xi_L(\tilde{\theta}), (W_{L,-1}Y_{L,-1})'J_L\xi_L(\tilde{\theta})\right]'\\
			&\hspace{4mm}-\frac{1}{\sqrt{nT}}\Big(-tr\left[W_{1L}(\tilde{\theta})\left(\frac{1}{T}1_T1_T'\otimes J_n\right)\right]-tr\left[W_{1L}(\tilde{\theta})\left(I_T \otimes \frac{1}{n}1_n1_n'\right)\right], \\
			&\hspace{20mm} tr\left[W_{2L}(\tilde{\theta})\left(\frac{1}{T}1_T1_T'\otimes J_n\right)\right],tr\left[W_{3L}(\tilde{\theta})\left(\frac{1}{T}1_T1_T'\otimes J_n\right)\right]\Big)',\\
			&\hspace{4mm}+\frac{1}{\sqrt{nT}}I_{\eta\omega}(\tilde{\theta})I_{\omega\omega}^{-1}(\tilde{\theta})	\begin{bmatrix} 0_{k_1 \times 1} \\ -\frac{1}{2\widetilde{\sigma_\xi^2}}\left(\sqrt{\frac{n}{T}}+\sqrt{\frac{T}{n}}\right)\end{bmatrix}.
		\end{align*}
	
		\subsection*{3. Proposition 4 proof}
		Let $\theta_0=(\delta_0',\eta_0',\omega_0')'$,
		$\theta^*=(0',0',\omega_0')' $, and
		$\tilde{\theta} = (0',0',\tilde{\omega}')'$, 
		where $\tilde{\omega}$ is the ML estimator. Consider $H_a^\delta: \delta_0=\frac{\zeta}{\sqrt{nT}}$ and $H_a^\eta: \eta_0=\frac{\nu}{\sqrt{nT}}.$ The first-order Taylor expansion of the score function, $L_{\delta}(\tilde{\theta})$, around $\theta_0$ under $H_a^\delta$ and $H_a^\eta$ is
		\begin{align}
			\label{eq14}
			\begin{split}
			\sqrt{nT}L_\delta(\tilde{\theta})=&\sqrt{nT}L_\delta(\theta_0)-\frac{\partial L_\delta(\theta_0)}{\partial \delta}\zeta-\frac{\partial L_\delta(\theta_0)}{\partial \eta}\nu\\
			&+\sqrt{nT}\frac{\partial L_\delta(\theta_0)}{\partial  \omega}(\tilde{\omega}-\omega_0) + o_p(1). 	
			\end{split}
		\end{align}

		\noindent Similarly the Taylor expansion of $L_{\omega}(\theta^*)$ around $\tilde{\theta}$ under $H_a^\delta$ \& $H_a^\eta$ is

	\begin{align}
		\label{eq15}
		\begin{split}
		\sqrt{nT}L_\omega(\theta^*)&=\sqrt{nT}L_\omega(\tilde{\theta})+\sqrt{nT}\frac{\partial L_\omega(\tilde{\theta})}{\partial \omega}(\omega_0-\tilde{\omega}) + o_p(1) \\
		&\overset{\text{a}}{=}I_{\omega\omega}\sqrt{nT}(\tilde{\omega}-\omega_0),
		\end{split}
	\end{align} 

	\noindent where `a' represents `asymptotically'. On the other hand, the Taylor expansion of $L_{\omega}(\theta^*)$ around $\theta_0$ under $H_a^\delta$ and $H_a^\eta$ is
	\begin{align}
		\label{eq16}
		\begin{split}
		\sqrt{nT}L_\omega(\theta^*)&=\sqrt{nT}L_\omega(\theta_0)-\frac{\partial L_\omega(\theta_0)}{\partial \delta}\zeta-\frac{\partial L_\omega(\theta_0)}{\partial \eta}\nu + o_p(1) \\
		&\overset{\text{a}}{=}\sqrt{nT}L_\omega(\theta_0)+I_{\omega\delta}\zeta+I_{\omega\eta}\nu.
		\end{split}
	\end{align}

	\noindent Since $\eqref{eq15}$ and $\eqref{eq16}$ are identical, 
	\begin{equation*}
	I_{\omega\omega}\sqrt{nT}(\tilde{\omega}-\omega_0)\overset{a}{=}\sqrt{nT}L_\omega(\theta_0)+I_{\omega\delta}\zeta+I_{\omega\eta}\nu_,
	\end{equation*}

	\noindent which implies
	\begin{equation}
		\label{eq17}
		\begin{split}	\sqrt{nT}(\tilde{\omega}-\omega_0)\overset{\text{a}}{=}I_{\omega\omega}^{-1}\sqrt{nT}L_\omega(\theta_0)+I_{\omega\omega}^{-1}I_{\omega\delta}\zeta+I_{\omega\omega}^{-1}I_{\omega\eta}\nu.
		\end{split}
	\end{equation}

	\noindent Using $\eqref{eq17}$ for $\eqref{eq14}$, 
	\begin{align*}
		\sqrt{nT}L_\delta(\tilde{\theta})&\overset{\text{a}}{=}\sqrt{nT}L_\delta(\theta_0)+I_{\delta\delta}\zeta+I_{\delta\eta}\nu-I_{\delta\omega}\left[I_{\omega\omega}^{-1}\sqrt{nT}L_\omega(\theta_0)+I_{\omega\omega}^{-1}I_{\omega\delta}\zeta+I_{\omega\omega}^{-1}I_{\omega\eta}\nu\right] \\
		&=\sqrt{nT}L_\delta(\theta_0)-I_{\delta\omega}I_{\omega\omega}^{-1}\sqrt{nT}L_\omega(\theta_0)+(I_{\delta\delta}-I_{\delta\omega}I_{\omega\omega}^{-1}I_{\omega\delta})\zeta+(I_{\delta\eta}-I_{\delta\omega}I_{\omega\omega}^{-1}I_{\omega\eta})\nu \\
		&=[I_{p}, -I_{\delta\omega}I_{\omega\omega}^{-1}]
		\times \begin{bmatrix}
		\sqrt{nT}L_\delta(\theta_0) \\
		\sqrt{nT}L_\omega(\theta_0) \\
		\end{bmatrix}
		+(I_{\delta\delta}-I_{\delta\omega}I_{\omega\omega}^{-1}I_{\omega\delta})\zeta 	+(I_{\delta\eta}-I_{\delta\omega}I_{\omega\omega}^{-1}I_{\omega\eta})\nu \\
		&:=[I_{p}, -I_{\delta\omega}I_{\omega\omega}^{-1}]
		\times \begin{bmatrix}
		\sqrt{nT}L_\delta(\theta_0) \\
		\sqrt{nT}L_\omega(\theta_0) \\
		\end{bmatrix}
		+I_{\delta\cdot \omega}\zeta+I_{\delta\eta\cdot\omega}\nu, 
	\end{align*}

	\noindent where $I_{\delta \cdot \omega}=I_{\delta\delta}-I_{\delta\omega}I_{\omega\omega}^{-1}I_{\omega\delta}$ and $I_{\delta\eta \cdot \omega}=I_{\delta\eta}-I_{\delta\omega}I_{\omega\omega}^{-1}I_{\omega\eta}.$ By Proposition 1,
	\begin{equation}
		 \label{eq18}
		\begin{bmatrix} \sqrt{nT}L_{\delta}(\theta_0) \\
		\sqrt{nT}L_{\omega}(\theta_0) \\
		\end{bmatrix} - \begin{bmatrix}
		\Delta_{1,\delta}(\theta_0)+\Delta_{2,\delta}(\theta_0) \\
		\Delta_{1,\omega}(\theta_0)+\Delta_{2,\omega}(\theta_0) \\
		\end{bmatrix} \xrightarrow{d} N\left(0, \begin{bmatrix} 
		I_{\delta\delta} & I_{\delta\omega} \\
		I_{\omega\delta} & I_{\omega\omega}\\
		\end{bmatrix} \right).
	\end{equation}

	Thus, $L_\delta(\tilde{\theta})$ is not centered around zero due to the bias terms and needs to be adjusted. I introduce the \textit{bias-corrected} score function at $\tilde{\theta}$ defined by
	\begin{equation*}
	\begin{cases}
	C_\delta(\tilde{\theta}):=L_\delta(\tilde{\theta})-\frac{1}{\sqrt{nT}}(\Delta_{1,\delta}(\tilde{\theta})+\Delta_{2,\delta}(\tilde{\theta}))+\frac{1}{\sqrt{nT}}I_{\delta \omega}(\tilde{\theta})I_{\omega\omega}^{-1}(\tilde{\theta})(\Delta_{1,\omega}(\tilde{\theta})+\Delta_{2,\omega}(\tilde{\theta})), \\
	C_\eta(\tilde{\theta}):=L_\eta(\tilde{\theta})-\frac{1}{\sqrt{nT}}(\Delta_{1,\eta}(\tilde{\theta})+\Delta_{2,\eta}(\tilde{\theta}))+\frac{1}{\sqrt{nT}}I_{\eta\omega}(\tilde{\theta})I_{\omega\omega}^{-1}(\tilde{\theta})(\Delta_{1,\omega}(\tilde{\theta})+\Delta_{2,\omega}(\tilde{\theta})),
	\end{cases}	
	\end{equation*}

	\noindent so that
	\begin{equation}
		\label{eq19}
		\begin{bmatrix}
		\sqrt{nT}C_\delta(\tilde{\theta})\\
		\sqrt{nT}C_\eta(\tilde{\theta})\end{bmatrix} 
		\xrightarrow{d}  
		N\left(\begin{bmatrix}
		I_{\delta \cdot \omega}\zeta + I_{\delta\eta \cdot \omega}\nu \\
		I_{\eta\delta \cdot \omega}\zeta+I_{\eta\cdot\omega}\nu
		\end{bmatrix},\begin{bmatrix}
		I_{\delta \cdot \omega} & I_{\delta\eta \cdot \omega} \\
		I_{\eta\delta\cdot\omega} & I_{\eta\cdot\omega} \end{bmatrix}\right). 
	\end{equation}
	\noindent Note that this result leads to the non-central $\chi^2$ distribution with over-rejection of size under the null ($H_0^\delta$) if there is a local misspecification in $\eta$ (i.e., $H_a^\eta$ holds.)

	One may notice that
	\begin{equation*}
	\sqrt{nT}C_\delta(\tilde{\theta})-I_{\delta \eta \cdot \omega}\nu \xrightarrow{d} N(I_{\delta \cdot \omega}\zeta, I_{\delta \cdot \omega}),
	\end{equation*}
	\noindent but $\nu$ is unknown. Using $\eqref{eq19}$ under $H_0^\delta$ and $H_a^\eta$, we may write the adjusted score function as
	\begin{equation*}
	C_\delta^*(\tilde{\theta})=C_\delta(\tilde{\theta})-I_{\delta\eta\cdot\omega}(\tilde{\theta})I_{\eta\cdot\omega}^{-1}(\tilde{\theta})C_\eta(\tilde{\theta}),
	\end{equation*}

	where we get the asymptotic distribution of $C_\delta^*(\tilde{\theta})$ as
	\begin{equation*}
	\sqrt{nT}C_\delta^*(\tilde{\theta})\xrightarrow{d} N((I_{\delta \cdot \omega}-I_{\delta \eta \cdot \omega}I_{\eta \cdot \omega}^{-1}I_{\eta\delta \cdot \omega})\zeta,I_{\delta \cdot \omega}-I_{\delta \eta \cdot \omega}I_{\eta \cdot \omega}^{-1}I_{\eta\delta \cdot \omega}).
	\end{equation*}

	Thus the adjusted or robust Rao's score test statistic is of the form
	\begin{equation*}
	RS_\delta^* (\tilde{\theta})=nTC_\delta^{*'}(\tilde{\theta})[Var(C_\delta^*(\tilde{\theta}))]^{-1}C_\delta^*(\tilde{\theta}) \xrightarrow{d} \chi^2_p(\varphi_2),
	\end{equation*}
	\noindent where $\varphi_2=\zeta'(I_{\delta\cdot\omega}-I_{\delta\eta\cdot\omega}I_{\eta\cdot\omega}^{-1}I_{\eta\delta\cdot\omega})\zeta. \qed$

\end{document}